\documentclass[12pt,a4paper]{article}
\usepackage{multirow,amssymb,enumitem,booktabs,subfig,xcolor,todonotes,microtype,setspace}
\usepackage[top=1.4in, bottom=1.4in, left=1.35in, right=1.35in]{geometry}
\usepackage{natbib}
\usepackage{url} 
\usepackage[pdftex,colorlinks=true,hypertexnames=false]{hyperref}
\definecolor{darkblue}{rgb}{0,0,.6}
\hypersetup{citecolor=darkblue,linkcolor=darkblue,urlcolor=darkblue}
\usepackage{amsmath}
\usepackage{amsfonts}
\usepackage{epsfig}
\usepackage{graphics}
\usepackage{palatino,eulervm}
\usepackage{mathrsfs}
\usepackage{graphicx}
\usepackage{lscape}
\usepackage{rotating}
\usepackage[linewidth=1pt]{mdframed}
\allowdisplaybreaks[4]
\DeclareMathOperator*{\argmin}{arg\,min}

\providecommand{\U}[1]{\protect\rule{.1in}{.1in}}

\graphicspath{{plots/}}

\usepackage{amsthm,thmtools}
\declaretheorem{theorem}
\declaretheorem{lemma}
\makeatletter
\def\th@newremark{\th@remark\thm@headfont{\bfseries}}
\makeatletter
\theoremstyle{newremark}
\newtheorem{remark}{Remark}

\newtheorem{assumption}{Assumption}
\declaretheoremstyle[
  spaceabove=6pt, spacebelow=6pt,
  headfont=\bfseries,
  notefont=\mdseries, notebraces={(}{)},
bodyfont=\normalfont,
  postheadspace=0.5em,
]{mystyle}

\usepackage{mathrsfs}

\usepackage{xr}
\begin{document}

\title{\Large\bf Factor Models of Matrix-Valued Time Series: \\ 
Nonstationarity and Cointegration}
\author{{\normalsize Degui Li\thanks{\footnotesize Faculty of Business Administration, Asia-Pacific Academy of Economics and Management, and Department of Economics, University of Macau.},\ \ \ Yayi Yan\thanks{\footnotesize School of Statistics and Data Science, Shanghai University of Finance and Economics. },\ \ \ Qiwei Yao\thanks{\footnotesize Department of Statistics, London School of Economics. }}\\
{\normalsize\em $^\ast$University of Macau, $^\dag$Shanghai University of Finance and Economics, $^\ddag$London School of Economics}}
\date{\small This version: \today}

\maketitle

\centerline{\bf Abstract}

\medskip

In this paper, we consider the nonstationary matrix-valued time series
with common stochastic trends. Unlike the traditional factor analysis
which flattens matrix observations into vectors, we adopt a matrix factor
model in order to fully explore the intrinsic matrix structure in the
data, allowing interaction between the row and column stochastic trends,
and subsequently improving the estimation convergence. It also reduces  the  computation complexity
in estimation.
 The main estimation methodology is built on the eigenanalysis of sample row and
column covariance matrices when the nonstationary matrix factors are of
full rank and the idiosyncratic components are temporally stationary, and
is further extended to tackle a more flexible setting when the matrix
factors are cointegrated and the idiosyncratic components may be
nonstationary. Under some mild conditions which allow the existence of
weak factors, we derive the convergence theory for the estimated factor
loading matrices and nonstationary factor matrices. In particular, the
developed methodology and theory are applicable to the general case of
heterogeneous strengths over weak factors. An easy-to-implement ratio
criterion is adopted to consistently estimate the size of latent factor
matrix. Both simulation and empirical studies are conducted to examine
the numerical performance of the developed model and methodology in
finite samples.

\medskip


\newpage


\section{Introduction}\label{sec1}
\renewcommand{\theequation}{1.\arabic{equation}}
\setcounter{equation}{0}

There has been increasing attention in modeling large-scale matrix-valued data over the past two decades with applications in various fields such as environmental science, economics, finance, public health and social networks. Typical examples include the international trade flow among a large number of countries \citep{CC23}, the Netflix movie-rating challenge with customers and movies being matrix rows and columns, respectively \citep{HTW15}, and friendships between pupils in schools \citep{CFIK20}. A traditional method for tackling matrix-valued data is to convert them into multivariate data, i.e., matrices are flattened into vectors, and then adopt statistical tools which have been well studied for multivariate data (with possible high dimension). However, this naive method neglects the intrinsic matrix nature of the data and may lose important information contained in the matrix structure, leading to sup-optimal estimation convergence theory \citep[e.g.,][]{CF23, HKTY23}.

\smallskip

There have been extensive studies on estimation and inference of
independent matrix-valued data, see, for example, \cite{NW11},
\cite{RT11}, \cite{LT12}, \cite{HTW15} and the references therein. In
practice, matrix-valued data are often collected over time with
significant serial correlation which may be useful to the subsequent
causality and forecasting analysis. There have been notable developments
in models, methodology and theory for matrix-variate time series.
\cite{CXY21} study the estimation of autoregressive models for matrix
time series via a bilinear form; \cite{CTC19}, \cite{WLC19},
\cite{CHYY23}, \cite{CF23}, \cite{HKTY23} and \cite{HYZC24} propose various matrix
factor models based on eigenanalysis using either
contemporaneous covariance or dynamic auto-covariance matrices; and
\cite{HCZY24} propose simultaneous decorrelation of matrix time series to
achieve an effective dimension reduction. 

\smallskip

Most of the aforementioned literature requires a stationarity assumption on matrix time series, facilitating theoretical derivation of standard estimation and inferential theory. However, this assumption is too restrictive and often rejected when we test some empirical matrix time series collected over a long time span. A transformation mechanism may be required to convert nonstationary matrix time series to stationary ones. For example, a first- or second-order difference is taken to transform the multinational macroeconomic time series to the stationary matrix-variate process \citep[e.g.,][]{CXY21, CF23, HKTY23}. In practice, the nonstationarity may be due to the existence of common stochastic trends. Conventional transformation mechanisms such as differencing can result in loss of important trending information. 

\smallskip

The present paper introduces a general matrix factor model framework for nonstationary matrix time series, where the common stochastic trending patterns of matrix observations are captured by the latent common component through a low-rank structure. In particular, we study the following two scenarios: (i) the nonstationary matrix factors are of full rank \footnote{Here, by ``full rank'' we mean that the covariance matrix of the (normalized) row or column factor process is of full rank, which indicates the absence of cointegration among the factors.} and the idiosyncratic components are temporally stationary, and (ii) the matrix factors are cointegrated and the idiosyncratic components may be nonstationary. The nonstationary matrix factors in scenario (i) capture the common stochastic trends whereas the cointegrated matrix factors in scenario (ii) are generated by the matrix error-correction model which extends the classic vector error-correction model to matrix time series. The developed matrix factor model is a natural extension of the conventional approximate factor model which has been extensively studied in the literature for stationary and nonstationary vector time series \citep[e.g.,][]{CR83, BN02, B04, BN04, FLM13, BLL21}. The advantage of the matrix factor model over the vector factor model (by flattening matrix observations into vector ones) is that the intrinsic matrix structure can be fully explored and the interaction between the row and column stochastic trends is allowed, which would lead to improvement of the estimation convergence for unknown elements in the matrix factor model. It also reduces the computation complexity 
in estimation (see the last paragraph in Section \ref{sec2.2} below).

\smallskip

For scenario (i) with full-rank I(1) (integrated of order one) matrix
factors and stationary idiosyncratic components, the main estimation
technique is built on the principal component analysis (PCA)
using either a sample row or a sample column
covariance matrix, which is called mPCA throughout the paper. This
estimation methodology is also valid when the nonnegative definite matrix
is constructed via aggregation of auto-covariance matrices over lags
\citep[e.g.,][]{LYB11, LY12, CTC19, WLC19}, in which case the
idiosyncratic components are often assumed to be temporally uncorrelated.
In the present paper, due to stronger signals from the stochastic trends
in common components, we relax this restriction, allowing the
idiosyncratic components in the matrix factor model to be serially
correlated. The convergence theory in Section \ref{sec3.2} reveals that
the developed estimators achieve the so-called ``super-fast" convergence
rates as in the recent work by \cite{CGGT25}. The remarkable rate
improvement is due to the following two reasons. First, the common
components in the proposed matrix factor model contain latent stochastic
trends, resulting in much stronger signals in the estimation equations
than those in the stationary setting. This significantly improve the
estimation convergence for the factor loading matrices compared with that
for stationary matrix time series \citep[e.g.,][]{WLC19, CF23}. Second,
the developed estimation methodology makes full use of the intrinsic
matrix structure and thus improves the convergence rates over the
traditional method via the vectorized factor model \citep[e.g.,][]{B04, BLL21}. 

\smallskip

For scanario (ii) with cointegrated matrix-valued factors and mixed stationary and nonstationary idiosyncratic components, we extend \cite{BN04}'s PANIC (Panel Analysis of Nonstationarity in Idiosyncratic and Common components) technique from vector time series to matrix time series and call the proposed method as mPANIC. Through the eigenanalysis of the differenced matrix time series, we obtain the row and column factor loading estimation which has typical stationary convergence rates as shown in Theorem \ref{thm:3.4}. Furthermore, assuming a ``sparsity" restriction which limits the number of nonstationary entries in the matrix idiosyncratic components, we derive the uniform convergence of the estimated cointegrated factor matrix.

\smallskip

Unlike the existing literature on matrix factor models with strong factors or loadings \citep[e.g.,][]{CF23, HKTY23, CGGT25}, we not only consider weak factors in the developed matrix factor model but also allow heterogeneous factor strength. The existence of weak factors requires an adaptive estimation formula with random normalizers for estimating the common factor matrix. As in \cite{BN23} for vectorized factor model estimation theory, the weak and varying factor strength further slows down both the mean squared and the uniform convergence rates to be developed in Sections \ref{sec3} and \ref{sec4} and makes it more challenging to correctly estimate the row and column factor numbers. Our asymptotic theory reveals that the proposed eigenvalue ratio criterion can consistently estimate the factor dimension without any restriction for the minimum factor strength in scenario (i), but requires the weakest factor strength to exceed a fixed threshold as in \cite{F22} to achieve consistency in scanario (ii). 

\smallskip

The finite-sample Monte-Carlo simulation study shows that both the mPCA and mPANIC estimates converge as the sample size increases, the mPCA (which directly tackles the nonstationary matrix time series) has faster convergence than the mPANIC, and the estimation errors increase as the factor strength decreases. The empirical application to international trade flows reveals that the estimated factor is connected to the so-called global demand index and the patterns of the estimated trending factor may be related to some major global events. In addition, the estimated row and column factor loading matrices can be used to capture the structure of bilateral trade linkages across some major economies.

\smallskip

The rest of the paper is organized as follows. Section \ref{sec2} introduces the matrix factor model with discussion on model identifiability, and the mPCA and mPANIC estimation methods. Section \ref{sec3} presents the convergence properties of the developed estimators with assumptions and remarks. They are extended to the case of heterogeneous factor strength in Section \ref{sec4}. Section \ref{sec5} reports both the simulation and empirical studies. Section \ref{sec6} concludes the paper. The supplemental document contains the proofs of the main asymptotic results and some technical lemmas with proofs. Throughout the paper, we let ${\sf vec}(\cdot)$ be the vectorization of a matrix and ${\sf vec}^{-1}(\cdot)$ inverse of the vectorization operator. We use $\otimes$ and $\odot$ to denote the Kronecker and Hadamard products, respectively. Let $\Vert\cdot\Vert$ be the Euclidean norm of a vector or operator norm of a matrix, and $\Vert\cdot\Vert_F$ the matrix Frobenius norm. Let $\psi_k(\mathbf{A})$ denote the $k$-th largest eigenvalue of a positive semidefinite matrix $\mathbf{A}$. For notational simplicity, we write ``$\max\{a,b\}$" as ``$a\vee b$", ``$\min\{a,b\}$" as ``$a\wedge b$", and ``with probability approaching one" as ``{\em w.p.a.1}".


\section{Model and methodology}\label{sec2}
\renewcommand{\theequation}{2.\arabic{equation}}
\setcounter{equation}{0}

In this section, we introduce the factor model for nonstationary matrix time series and then propose two types of PCA-based estimation techniques depending on whether the nonstationary matrix factors are of full rank or cointegrated, and whether the idiosyncratic components are stationary.

\subsection{Model formulation}\label{sec2.1}

Suppose that we observe a sequence of matrix time series ${\boldsymbol X}_t$, $t=1,\ldots,T$, where 
\[
{\boldsymbol X}_t=\left(
\begin{array}{ccc}
X_{t,11}&\ldots&X_{t,1p_2}\\
\vdots&\ddots&\vdots\\
X_{t,p_11}&\ldots&X_{t,p_1p_2}
\end{array}
\right)_{p_1\times p_2},
\]
where both the dimensions $p_1$ and $p_2$ diverge to infinity together with $T$. Consider the following matrix factor model formulation: 
\begin{equation}\label{eq2.1}
{\boldsymbol X}_t={\boldsymbol Z}_t+{\boldsymbol E}_t\ \ \text{with}\ \ {\boldsymbol Z}_t={\boldsymbol R}{\boldsymbol F}_t{\boldsymbol C}^{^\intercal},
\end{equation}
where ${\boldsymbol R}$ and ${\boldsymbol C}$ are row and column factor loading matrices with sizes $p_1\times r_1$ and $p_2\times r_2$, respectively, ${\boldsymbol F}_t\in{\mathscr R}^{r_1\times r_2}$ is a matrix of nonstationary factors, and ${\boldsymbol E}_t\in{\mathscr R}^{p_1\times p_2}$ is a matrix of idiosyncratic components. Throughout the paper, we assume that both $r_1$ and $r_2$ are fixed. 

\smallskip

It is well known that the latent factors ${\boldsymbol F}_t$ and loading matrices ${\boldsymbol R}$ and ${\boldsymbol C}$ are not identifiable, i.e., for any two invertible matrices ${\boldsymbol V}_1$ and ${\boldsymbol V}_2$ with sizes $r_1\times r_1$ and $r_2\times r_2$, respectively, model (\ref{eq2.1}) still holds when the triplet $({\boldsymbol R}, {\boldsymbol F}_t, {\boldsymbol C})$ are replaced by $({\boldsymbol R}{\boldsymbol V}_1, {\boldsymbol V}_1^{-1}{\boldsymbol F}_t{\boldsymbol V}_2^{-1}, {\boldsymbol C}{\boldsymbol V}_2^{^\intercal})$. However, the column spaces spanned by ${\boldsymbol R}$ and ${\boldsymbol C}$, ${\cal M}({\boldsymbol R})$ and ${\cal M}({\boldsymbol C})$, are uniquely determined. In fact, assuming that ${\boldsymbol R}$ and ${\boldsymbol C}$ are of full column rank, we have the following QR decomposition: ${\boldsymbol R}={\boldsymbol Q}_R{\boldsymbol W}_R$ and ${\boldsymbol C}={\boldsymbol Q}_C{\boldsymbol W}_C$, where ${\boldsymbol Q}_R$ and ${\boldsymbol Q}_C$ are $p_1\times r_1$ and $p_2\times r_2$ matrices with orthogonal columns, and ${\boldsymbol W}_R$ and ${\boldsymbol W}_C$ are $r_1\times r_1$ and $r_2\times r_2$ upper triangular matrices. Writing ${\boldsymbol W}_t={\boldsymbol W}_R{\boldsymbol F}_t{\boldsymbol W}_C^{^\intercal}$, a transformed factor matrix of integrated processes, by (\ref{eq2.1}), we readily have that
\[
{\boldsymbol X}_t={\boldsymbol Q}_R{\boldsymbol W}_t{\boldsymbol Q}_C^{^\intercal}+{\boldsymbol E}_t.
\]
It is clear that ${\cal M}({\boldsymbol R})$ is the same as ${\cal M}({\boldsymbol Q}_R)$ and ${\cal M}({\boldsymbol C})$ is the same as ${\cal M}({\boldsymbol Q}_C)$. 

\smallskip

On the other hand, by flattening matrix observations to vector ones, we can rewrite the matrix factor model (\ref{eq2.1}) as the following vectorized factor model:
\[
{\sf vec}({\boldsymbol X}_t)=\left({\boldsymbol C}\otimes{\boldsymbol R}\right){\sf vec}({\boldsymbol F}_t)+{\sf vec}({\boldsymbol E}_t).
\]
The conventional factor model has been extensively studied in the literature for vector time series \citep{CR83, BN02, FLM13}. In particular, model (\ref{eq2.1}) can be seen as a natural extension of the classic approximation factor model for vector nonstationary time series \citep[e.g.,][]{B04,BN04,BLL21}, fully exploring the intrinsic matrix structure and allowing interaction between the row and column stochastic trends in ${\boldsymbol F}_t$. It also extends factor models for stationary matrix time series \citep[e.g.,][]{WLC19, CXY21, CF23, CHYY23, HKTY23} to the nonstationary setting. 

\smallskip

We are interested in the following two scenarios for latent factors and idiosyncratic components: (i) the nonstationary matrix factors are of full rank and the idiosyncratic components are stationary over time, and (ii) the matrix factors are cointegrated and the idiosyncratic components are either stationary or nonstationary.  The nonstationarity of matrix time series in scenario (i) is fully captured by the common stochastic trends contained in matrix factors, whereas appropriate rotation of the cointegrated factors in scenario (ii) can separate out stationary and nonstationary factors. In the following two subsections, we provide detailed model settings and propose two types of PCA-based estimation techniques for these two scenarios, respectively.

\subsection{mPCA estimation}\label{sec2.2}

Assume that ${\boldsymbol F}_t$ is generated by an integrated I(1) matrix-valued process:
\begin{equation}\label{eq2.2}
{\boldsymbol F}_t={\boldsymbol F}_{t-1}+{\boldsymbol U}_t,
\end{equation}
where ${\boldsymbol U}_t\in{\mathscr R}^{r_1\times r_2}$ is a stationary and I(0) (integrated of order zero) matrix-valued process. In addition, the matrix idiosyncratic component ${\boldsymbol E}_t\in{\mathscr R}^{p_1\times p_2}$ is stationary and I(0). Under this model assumption, the matrix common component ${\boldsymbol Z}_t$ captures the stochastic trending patterns of matrix time series via the latent low-rank structure in (\ref{eq2.1}). Throughout this subsection, we assume that ${\boldsymbol F}_t$ is a matrix of full-rank I(1) random elements. Our main interest lies in estimation of the factor loading matrices ${\boldsymbol R}$ and ${\boldsymbol C}$ and construction of the matrix ${\boldsymbol F}_t$ for common stochastic trends. For this, we next introduce the mPCA approach via eigenanalysis of a nonnegative definite matrix which is either a sample row or column covariance matrix.

\smallskip

Assume $r_1$ and $r_2$ are known for the timebeing. Their consistent estimation can be constructed via a ratio criterion, see (\ref{eq2.5}) and (\ref{eq2.6}) below. Without loss of generality, assume that ${\boldsymbol X}_t$ has zero mean. We start with the estimation method for the row factor loading matrix ${\boldsymbol R}$ and define the sample row covariance matrix with size $p_1\times p_1$:
\begin{equation}\label{eq2.3}
\widehat{\boldsymbol\Omega}_{R}=\frac{1}{T}\sum_{t=1}^{T}{\boldsymbol X}_{t}{\boldsymbol X}_{t}^{^\intercal}.
\end{equation}
With the eigenanalysis of $\widehat{\boldsymbol\Omega}_{R}$, we obtain $\widehat{\boldsymbol R}=(\widehat{R}_{1},\ldots,\widehat{R}_{r_1})$ with $\widehat{R}_{k}$ being the eigenvector of $\widehat{\boldsymbol\Omega}_{R}$ corresponding to the $k$-th largest eigenvalue. The column space ${\cal M}({\boldsymbol R})$ is estimated by ${\cal M}(\widehat{\boldsymbol R})$. Equivalently $\widehat{\boldsymbol R}$ is the estimate of ${\boldsymbol R}$ subject to appropriate normalization and rotation (with the rotation matrix defined in Section \ref{sec3.1}). Defining the sample column covariance matrix:
\[
\widehat{\boldsymbol\Omega}_{C}=\frac{1}{T}\sum_{t=1}^{T}{\boldsymbol X}_{t}^{^\intercal}{\boldsymbol X}_{t},
\]
we may estimate the column factor loading matrix ${\boldsymbol C}$ in exactly the same way by conducting the eigenanalysis of $\widehat{\boldsymbol\Omega}_{C}$, and denote the resulting estimate by $\widehat{\boldsymbol C}$ which consists of the eigenvectors of $\widehat{\boldsymbol\Omega}_{C}$ corresponding to the $r_2$ largest eigenvalues. With the estimates $\widehat{\boldsymbol R}$ and $\widehat{\boldsymbol C}$, the matrix ${\boldsymbol F}_t$ of common stochastic trends can be estimated by 
\begin{equation}\label{eq2.4}
\widehat{\boldsymbol F}_t=\widetilde\lambda_{R,1}^{1/2}\widetilde{\boldsymbol V}_R^{-1/2}\widehat{\boldsymbol R}^{^\intercal} {\boldsymbol X}_t\widehat{\boldsymbol C}\widetilde{\boldsymbol V}_C^{-1/2},\ \ t=1,\ldots,T,\footnote{The normalization rate $\widetilde\lambda_{R,1}^{1/2}$ can be replaced by $\widetilde\lambda_{C,1}^{1/2}$ or $(\widetilde\lambda_{R,1}+\widetilde\lambda_{C,1})^{1/2}$, where $\widetilde\lambda_{C,1}$ is the maximum eigenvalue of $T^{-1}\widehat{\boldsymbol\Omega}_{C}$.}
\end{equation}
where $\widetilde\lambda_{R,1}$ is the maximum eigenvalue of $T^{-1}\widehat{\boldsymbol\Omega}_{R}$, $\widetilde{\boldsymbol V}_R$ is an $r_1\times r_1$ diagonal matrix with the diagonal entries being the $r_1$ largest eigenvalues of $T^{-1}\widehat{\boldsymbol\Omega}_R$ (in a descending order) whereas $\widetilde{\boldsymbol V}_C$ is an $r_2\times r_2$ diagonal matrix with the diagonal entries being the $r_2$ largest eigenvalues of $T^{-1}\widehat{\boldsymbol\Omega}_C$. The adaptive estimation formula in (\ref{eq2.4}) is due to possible existence of weak factors with the extra adaptive factors $\widetilde\lambda_{R,1}^{1/2}$, $\widetilde{\boldsymbol V}_R^{-1/2}$ and $\widetilde{\boldsymbol V}_C^{-1/2}$ related to the (varying) factor strength, see Assumptions \ref{ass:3.1}(ii) and \ref{ass:4.1}(i). When both the row and column factors are assumed to be strong, the estimation formula in (\ref{eq2.4}) can be simplified to $\widehat{\boldsymbol F}_t= (p_1p_2)^{-1/2} \widehat{\boldsymbol R}^{^\intercal} {\boldsymbol X}_t\widehat{\boldsymbol C}$.

\smallskip

The numbers of row and column factors, $r_1$ and $r_2$, are often unknown in practice. We next estimate them using the ratio criterion proposed by \cite{LY12} and \cite{AH13}. Let $\{\widehat{\lambda}_{R,i}:\ i=1,\ldots,p_1\}$ and $\{\widehat{\lambda}_{C,i}:\ i=1,\ldots,p_2\}$ be sets of eigenvalues (arranged in a descending order) of $\widehat{\boldsymbol\Omega}_{R}$ and $\widehat{\boldsymbol\Omega}_{C}$, respectively. We estimate $r_1$ and $r_2$ by 
\begin{equation}\label{eq2.5}
\widehat{r}_1=\argmin_{1\leq k\leq K_1}\left\vert \widehat{\lambda}_{R,k+1}/\widehat{\lambda}_{R,k}\right\vert
\end{equation}
and
\begin{equation}\label{eq2.6}
\widehat{r}_2=\argmin_{1\leq k\leq K_2}\left\vert \widehat{\lambda}_{C,k+1}/\widehat{\lambda}_{C,k}\right\vert,
\end{equation}
respectively, where $K_1$ and $K_2$ are user-specified upper bounds. 


In the construction of the mPCA above, we only need to conduct the eigenanalysis
for matrices of sizes $p_i \times p_i$ ($i=1, 2$). This is in contrast to the PCA
for flatterned ${\boldsymbol X}_t$ which involves an eigenanalysis
for a matrix of size $(p_1 p_2)\times (p_1 p_2)$. For moderately large $p_1$
and/or $p_2$, the computational complexity is substantially reduced.

\subsection{mPANIC estimation}\label{sec2.3}

We next remove the full-rank assumption on ${\boldsymbol F}_t$ and suppose that ${\boldsymbol F}_t$ is cointegrated and generated by the following error-correction model:
\begin{equation}\label{eq2.7}
\Delta{\boldsymbol F}_t={\boldsymbol A}_1{\boldsymbol F}_{t-1}{\boldsymbol A}_2^{^\intercal}+{\boldsymbol V}_t,
\end{equation}
where $\Delta{\boldsymbol F}_t={\boldsymbol F}_t-{\boldsymbol F}_{t-1}$,
${\boldsymbol
A}_1={\boldsymbol\alpha}_1{\boldsymbol\beta}_1^{^\intercal}$,
${\boldsymbol
A}_2={\boldsymbol\alpha}_2{\boldsymbol\beta}_2^{^\intercal}$,
${\boldsymbol\alpha}_j$ and ${\boldsymbol\beta}_j$ are full-rank
$r_j\times k_j$ matrices, $1\leq k_j\leq r_j$, $j=1,2$, and
$\{{\boldsymbol V}_t\}$ is a sequence of $r_1\times r_2$ white noise
matrices. Model (\ref{eq2.7}) is a natural extension of the vector
error-correction (VEC) model to matrix time series. \cite{LX24} and
\cite{HRW25} consider estimating the matrix error-correction model with
extra lag and constant terms for observed matrix time series. In
contrast, the factor process $\{{\boldsymbol F}_t\}$ is latent in the
present paper. Flattening the factor matrix to a vector, we may
rewrite (\ref{eq2.7}) as the following classic VEC model formulation:
\[
{\sf vec}(\Delta {\boldsymbol F}_t)=({\boldsymbol A}_2\otimes {\boldsymbol A}_1){\sf vec}({\boldsymbol F}_{t-1})+{\sf vec}({\boldsymbol V}_t)
\]
with
\[
({\boldsymbol A}_2\otimes {\boldsymbol A}_1)=\left({\boldsymbol\alpha}_2{\boldsymbol\beta}_2^{^\intercal}\right)\otimes \left({\boldsymbol\alpha}_1{\boldsymbol\beta}_1^{^\intercal}\right)=({\boldsymbol\alpha}_2\otimes {\boldsymbol\alpha}_1)({\boldsymbol\beta}_2\otimes {\boldsymbol\beta}_1)^{^\intercal}=:{\boldsymbol\alpha}{\boldsymbol\beta}^{^\intercal}.
\]
Letting ${\boldsymbol\beta}_\bot$ be the orthogonal complement of ${\boldsymbol\beta}$ and defining 
\[
{\boldsymbol P}=({\boldsymbol P}_1,\ {\boldsymbol P}_2),\quad {\boldsymbol P}_1={\boldsymbol\beta}_\bot\left({\boldsymbol\beta}_\bot^{^\intercal}{\boldsymbol\beta}_\bot\right)^{-1/2},\quad {\boldsymbol P}_2={\boldsymbol\beta}\left({\boldsymbol\beta}^{^\intercal}{\boldsymbol\beta}\right)^{-1/2},
\]
we may rotate ${\sf vec}({\boldsymbol F}_{t})$ via
\[
{\boldsymbol P}^{^\intercal}{\sf vec}({\boldsymbol F}_{t})=\left[
\begin{array}{c}
{\boldsymbol P}_1^{^\intercal}{\sf vec}({\boldsymbol F}_{t})\\
{\boldsymbol P}_2^{^\intercal}{\sf vec}({\boldsymbol F}_{t})
\end{array}
\right],
\]
to separate out the full-rank I(1) random vector ${\boldsymbol P}_1^{^\intercal}{\sf vec}({\boldsymbol F}_{t})$ and the stationary I(0) random vector ${\boldsymbol P}_2^{^\intercal}{\sf vec}({\boldsymbol F}_{t})$, see \cite{CP09}. Hence, the proposed structure is comparable to the matrix factor model setting in \cite{CGGT25} which includes both stationary and nonstationary matrix-valued factors and propose a projection-based three-stage estimation method.

\smallskip

Let $e_{t,(i,j)}$ be the $(i,j)$ entry of ${\boldsymbol E}_{t}$ and suppose that
\begin{equation}\label{eq2.8}
(1-\rho_{ij}L)e_{t,(i,j)}=\varepsilon_{t,(i,j)},\quad 1\leq i\leq p_1,\quad 1\leq j\leq p_2,
\end{equation} 
where $|\rho_{ij}|< 1$ or $\rho_{ij}=1$, $L$ denotes the lag operator, and $\{\varepsilon_{t,(i,j)}\}$ is a sequence of stationary I(0) random variables for each $(i,j)$. Note that $e_{t,(i,j)}$ is nonstationary if $\rho_{ij}=1$ and is stationary if $|\rho_{ij}|<1$. Similar assumptions can be found in \cite{BN04} and \cite{BLL21} for elements of idiosyncratic error vectors.

\smallskip

Neither the mPCA method proposed in Section \ref{sec2.2} nor the projection-based estimation in \cite{CGGT25} is applicable when ${\boldsymbol F}_t$ and ${\boldsymbol E}_t$ satisfy (\ref{eq2.7}) and (\ref{eq2.8}), respectively. We next adopt a matrix version of \cite{BN04}'s PANIC estimation technique (or mPANIC). Taking the first-order difference on both sides of (\ref{eq2.1}), we obtain
\begin{equation}\label{eq2.9}
\Delta{\boldsymbol X}_t={\boldsymbol R}(\Delta {\boldsymbol F}_t){\boldsymbol C}^{^\intercal}+\Delta {\boldsymbol E}_t,
\end{equation}
which becomes a factor model for stationary matrix-valued time series \citep[e.g.,][]{WLC19, CF23}. As in Section \ref{sec2.2},  we define the sample row covariance matrix using $\Delta{\boldsymbol X}_t$:
\begin{equation}\label{eq2.10}
\overline{\boldsymbol\Omega}_{R}=\frac{1}{T}\sum_{t=1}^{T}\Delta {\boldsymbol X}_{t} \Delta {\boldsymbol X}_{t}^{^\intercal},
\end{equation}
and then obtain $\overline{\boldsymbol R}=(\overline{R}_{1},\ldots,\overline{R}_{r_1})$ with $\overline{R}_{k}$ being the eigenvector of $\overline{\boldsymbol\Omega}_{R}$ corresponding to the $k$-th largest eigenvalue. The estimate of column factor loading matrix ${\boldsymbol C}$ is constructed in a similar way, denoted by $\overline{\boldsymbol C}=(\overline{C}_{1},\ldots,\overline{C}_{r_2})$. The cointegrated factor matrix ${\boldsymbol F}_t$ is estimated by
\begin{equation}\label{eq2.11}
\overline{\boldsymbol F}_t=\overline\lambda_{R,1}^{1/2}\overline{\boldsymbol V}_R^{-1/2}\overline{\boldsymbol R}^{^\intercal}{\boldsymbol X}_t\overline{\boldsymbol C}\,\overline{\boldsymbol V}_C^{-1/2},\ \ t=1,\ldots,T,
\end{equation}
where $\overline\lambda_{R,1}$ is the maximum eigenvalue of $\overline{\boldsymbol\Omega}_{R}$, $\overline{\boldsymbol V}_R$ is an $r_1\times r_1$ diagonal matrix with the diagonal entries being the $r_1$ largest eigenvalues of $\overline{\boldsymbol\Omega}_R$ (in a descending order) whereas $\overline{\boldsymbol V}_C$ is an $r_2\times r_2$ diagonal matrix with the diagonal entries being the $r_2$ largest eigenvalues of $\overline{\boldsymbol\Omega}_C$. Finally, the numbers of row and column factors, $r_1$ and $r_2$, are determined via the ratio criterion:
\begin{equation}\label{eq2.12}
\overline{r}_1=\argmin_{1\leq k\leq K_1}\left\vert \overline{\lambda}_{R,k+1}/\overline{\lambda}_{R,k}\right\vert
\end{equation}
and
\begin{equation}\label{eq2.13}
\overline{r}_2=\argmin_{1\leq k\leq K_2}\left\vert \overline{\lambda}_{C,k+1}/\overline{\lambda}_{C,k}\right\vert,
\end{equation}
where $\{\overline{\lambda}_{R,k}:\ k=1,\ldots,p_1\}$ and $\{\overline{\lambda}_{C,k}:\ k=1,\ldots,p_2\}$ are sets of eigenvalues (arranged in a descending order) of $\overline{\boldsymbol\Omega}_{R}$ and $\overline{\boldsymbol\Omega}_{C}$, respectively.


\section{Large-sample theory}\label{sec3}
\renewcommand{\theequation}{3.\arabic{equation}}
\setcounter{equation}{0}

In this section, we give some regularity conditions and establish the convergence properties for the mPCA and mPANIC estimators developed in Section \ref{sec2}. The factor strength is assumed to be homogeneous over (weak) factors in this section and the extension to the case of heterogeneous factor strength will be explored in Section \ref{sec4}.

\subsection{Convergence theory for mPCA estimation}\label{sec3.1}

We start with the introduction of some notation. Let 
\[
{\mathbf f}_t={\sf vec}({\boldsymbol F}_t),\ \ {\mathbf u}_t={\sf vec}({\boldsymbol U}_t),\ \ {\mathbf e}_t={\sf vec}({\boldsymbol E}_t).
\]
Denote the $i$-th row vectors of ${\boldsymbol R}$ and ${\boldsymbol C}$ by ${\boldsymbol R}_{i\bullet}$ and ${\boldsymbol C}_{i\bullet}$, respectively. Let $e_{t,(i,j)}$ be the $(i,j)$ entry of ${\boldsymbol E}_{t}$. We require the following conditions to derive the convergence properties of mPCA estimation.

\medskip

\renewcommand{\theassumption}{3.\arabic{assumption}}
\setcounter{assumption}{0}

\begin{assumption}\label{ass:3.1}

{\em (i)\ Let ${\mathbf u}_t=\sum_{j=0}^\infty {\boldsymbol A}_j {\boldsymbol\eta}_{t-j}$, where $\{{\boldsymbol A}_j\}$ is a sequence of $(r_1r_2)\times (r_1r_2)$ coefficient matrices and $\{{\boldsymbol\eta}_t\}$ is a sequence of i.i.d. random vectors with dimension $(r_1r_2)$, zero mean, positive definite covariance matrix denoted by ${\boldsymbol\Sigma}_\eta$ and finite fourth moment. In addition, 
\begin{equation}\label{eq3.1}
\sum_{j=0}^\infty j\Vert{\boldsymbol A}_j\Vert<\infty,\ \ \overline{\boldsymbol A}:=\sum_{j=0}^\infty {\boldsymbol A}_j \succ0,
\end{equation}
where ``$\succ0$" denotes positive definiteness.}

{\em (ii)\ There exist positive definite matrices ${\boldsymbol\Sigma}_R$ and ${\boldsymbol\Sigma}_C$ such that 
\begin{equation}\label{eq3.2}
\frac{1}{p_1^{\alpha_R}}{\boldsymbol R}^{^\intercal}{\boldsymbol R}\rightarrow{\boldsymbol\Sigma}_R\ \ \text{as}\ p_1\rightarrow\infty,
\end{equation}
and
\begin{equation}\label{eq3.3}
\frac{1}{p_2^{\alpha_C}}{\boldsymbol C}^{^\intercal}{\boldsymbol C}\rightarrow{\boldsymbol\Sigma}_C\ \ \text{as}\ p_2\rightarrow\infty,
\end{equation}
where $0<\alpha_R,\alpha_C\leq 1$. In addition, both $\Vert {\boldsymbol R}_{i\bullet}\Vert$ and $\Vert {\boldsymbol C}_{i\bullet}\Vert$ are bounded uniformly over $i$.}

{\em (iii)\ The eigenvalues of ${\boldsymbol\Sigma}_R^{1/2}[\int_0^1 {\boldsymbol W}(u){\boldsymbol\Sigma}_C{\boldsymbol W}(u)^{^\intercal}du]{\boldsymbol\Sigma}_R^{1/2}$ are positive, bounded and distinct with probability one, where ${\boldsymbol W}(\cdot)$ is an $r_1\times r_2$ matrix of Brownian motions with covariance of ${\sf vec}({\boldsymbol W}(\cdot))$ being $\overline{\boldsymbol A}{\boldsymbol\Sigma}_\eta \overline{\boldsymbol A}^{^\intercal}$. The same condition holds for ${\boldsymbol\Sigma}_C^{1/2}[\int_0^1 {\boldsymbol W}(u)^{^\intercal}{\boldsymbol\Sigma}_R{\boldsymbol W}(u)du]{\boldsymbol\Sigma}_C^{1/2}$.}

\end{assumption}

\medskip

\begin{assumption}\label{ass:3.2}

{\em (i)\ Let $\{{\boldsymbol E}_t\}$ be a sequence of zero-mean matrix random elements independent of $\{{\boldsymbol\eta}_t\}$, and $\max_{t}\max_{i,j}{\sf E}[e_{t,(i,j)}^4]<\infty$. }

{\em (ii)\ Letting $\rho_{R,t,(i_1,i_2)}=\frac{1}{p_2}\sum_{j=1}^{p_2}{\sf E}[e_{t,(i_1,j)}e_{t,(i_2,j)}]$ and $\rho_{C,t,(j_1,j_2)}=\frac{1}{p_1}\sum_{i=1}^{p_1}{\sf E}[e_{t,(i,j_1)}e_{t,(i,j_2)}]$,
{\small\[
\sum_{i_1=1}^{p_1}\sum_{i_2=1}^{p_1}\vert \rho_{R,t,(i_1,i_2)}\vert\leq c_1p_1,
\ \ 
\max_{1\leq i_1, i_2\leq p_1}{\sf E}\left\vert\sum_{s=t_1}^{t_2}\sum_{j=1}^{p_2} \left[e_{s,(i_1,j)}e_{s,(i_2,j)}-\rho_{R,s,(i_1,i_2)}\right]\right\vert^2\leq c_1p_2(t_2-t_1)
\]}
and
{\small\[
	\sum_{j_1=1}^{p_2}\sum_{j_2=1}^{p_2}\vert \rho_{C,t,(j_1,j_2)}\vert\leq c_1p_2,
	\ \ 
	\max_{1\leq j_1, j_2\leq p_2}{\sf E}\left\vert\sum_{s=t_1}^{t_2}\sum_{i=1}^{p_1}\left[e_{s,(i,j_1)}e_{s,(i,j_2)}-\rho_{C,s,(j_1,j_2)}\right]\right\vert^2\leq c_1p_1(t_2-t_1)
	\]}
for any $1\leq t\leq T$ and $1\leq t_1<t_2\leq T$, where $c_1$ is a positive constant.}

{\em (iii)\ Let
\[
\max_{1\leq j\leq p_2}{\sf E}\left\Vert p_1^{-\alpha_R/2}\sum_{i=1}^{p_1}e_{t,(i,j)}{\boldsymbol R}_{i\bullet}\right\Vert^4\leq c_2,\ \ \max_{1\leq j\leq p_2}{\sf E}\left\Vert p_1^{-\alpha_R/2}\sum_{s=t_1}^{t_2}\sum_{i=1}^{p_1}e_{s,(i,j)}{\boldsymbol R}_{i\bullet}\right\Vert^2\leq c_2(t_2-t_1)
\]
and
\[
\max_{1\leq i\leq p_1}{\sf E}\left\Vert p_2^{-\alpha_C/2}\sum_{j=1}^{p_2}e_{t,(i,j)}{\boldsymbol C}_{j\bullet}\right\Vert^4\leq c_2,\ \ \max_{1\leq i\leq p_1}{\sf E}\left\Vert p_2^{-\alpha_C/2}\sum_{s=t_1}^{t_2}\sum_{j=1}^{p_2}e_{s,(i,j)}{\boldsymbol C}_{j\bullet}\right\Vert^2\leq c_2(t_2-t_1)
\]
for any $1\leq t\leq T$ and $1\leq t_1<t_2\leq T$, where $c_2$ is a positive constant.}

\end{assumption}

\medskip

\begin{assumption}\label{ass:3.3}

{\em (i)\ As $p_1,p_2,T\to \infty$, 
\[
\frac{p_1^{1-\alpha_R}p_2^{1-\alpha_C}}{T} \to 0,\quad \frac{p_1^{2-\alpha_R}}{p_2^{\alpha_C}T^2}\to 0\quad \text{and}\quad \frac{p_2^{2-\alpha_C}}{p_1^{\alpha_R}T^2}\to 0.
\] }

{\em (ii)\ There exists a positive constant $c_3$ such that}
\begin{equation}\label{eq3.4}
\max_{1\leq t\leq T}{\sf E}\left\Vert \frac{{\boldsymbol R}^{^\intercal}{\boldsymbol E}_t{\boldsymbol C}}{p_1^{\alpha_R/2}p_2^{\alpha_C/2}}\right\Vert^4\leq c_3.
\end{equation}

\end{assumption}

\medskip

\renewcommand{\theremark}{3.\arabic{remark}}
\setcounter{remark}{0}

\begin{remark}\label{re:3.1}

(i) Assumption \ref{ass:3.1} imposes some fundamental conditions on the matrix factors and factor loading matrices. It follows from Assumption \ref{ass:3.1}(i) and the Beveridge-Nelson decomposition \citep[e.g.,][]{PS92} that 
\begin{equation}\label{eq3.5}
{\mathbf f}_t=\sum_{s=1}^t{\mathbf u}_s+{\mathbf f}_0=\overline{\boldsymbol A}\sum_{s=1}^t{\boldsymbol \eta}_s+{\mathbf f}_0+\widetilde{\mathbf u}_0-\widetilde{\mathbf u}_t,
\end{equation}
where $\widetilde{\mathbf u}_t=\sum_{j=0}^\infty\widetilde{\boldsymbol A}_j{\boldsymbol\eta}_{t-j}$ with $\widetilde{\boldsymbol A}_j=\sum_{k=j+1}^\infty{\boldsymbol A}_k$. By (\ref{eq3.1}), $\{\widetilde{\mathbf u}_t\}$ is a stationary vector linear process and $\overline{\boldsymbol A}{\boldsymbol\Sigma}_\eta \overline{\boldsymbol A}^{^\intercal}$ is of full rank. Hence, ${\boldsymbol F}_t$ is a matrix of full-rank integrated random elements. In fact, with (\ref{eq3.5}) and Gaussian approximation theorem in \cite{GZ09}, we may show that 
\begin{equation}\label{eq3.6}
\max_{1\leq t\leq T} \left\Vert \frac{1}{\sqrt{T}} {\mathbf f}_t-{\mathbf w}(t/T)\right\Vert=O_P(T^{-1/4}),\quad {\mathbf w}(\cdot)={\sf vec}({\boldsymbol W}(\cdot)).
\end{equation}
More details are provided in the proof of Lemma B.3 available in the supplement. Assumption \ref{ass:3.1}(ii) allows the existence of weak factors with $\alpha_R$ and $\alpha_C$ representing the strength of row and column factors, respectively \citep[e.g.,][]{LYB11, CTC19, WLC19}. When $\alpha_R=\alpha_C=1$, the row and column factors have impacts on the majority of matrix time series entries, and are thus called ``strong factors" \citep[e.g.,][]{CF23}. In addition, (\ref{eq3.2}) and (\ref{eq3.3})  indicate that the nonstationary factors have homogeneous strength. We will consider the extension to the setting with heterogeneous factor strength in Section \ref{sec4} below.

(ii) Assumption \ref{ass:3.2} contains some high-level conditions which are mild restrictions on temporal and cross-row (or column) correlation. Similar assumptions can also be found in \cite{CF23} and they can be seen as a matrix extension of those conditions used by \cite{BN02} and \cite{B04}. The restriction on temporal dependence can be justified if we impose a mixing dependence condition on $\{{\boldsymbol E}_t\}$. From Assumption \ref{ass:3.2}, the idiosyncratic components are allowed to be heteroskedatic over time. Assumption \ref{ass:3.2}(i) indicates that $\{{\boldsymbol E}_t\}$ and $\{{\boldsymbol F}_t\}$ are mutually independent, but this independence restriction can be relaxed at the cost of more lengthy proofs. 

(iii) Assumption \ref{ass:3.3}(i) imposes some mild conditions to restrict the relationship between the matrix sizes $p_1,p_2$ and time series length $T$. For the special case of $\alpha_R=\alpha_C=1$, the conditions can be simplified to 
\[
\frac{p_1}{p_2T^2}\rightarrow0\ \ \text{and}\ \ \frac{p_2}{p_1T^2}\rightarrow0,
\]  
which implies that $p_1$ and $p_2$ are allowed to be of order much higher than $T$. Assumption \ref{ass:3.3}(ii) is a high-level condition to further restrict the cross-column and cross-row correlation of idiosyncratic error matrices when the loadings (or factors) are weak, see Assumption 5.3 in \cite{CF23} for a similar restriction for strong factors, i.e., $\alpha_R=\alpha_C=1$.   

\end{remark}

\medskip

Due to the model identifiability issue discussed in Section \ref{sec2.1}, we can only consistently estimate the row/column factor loading matrices up to appropriate rotation. Define two rotation matrices:
\begin{eqnarray}
\widehat{\boldsymbol H}_R&=&\left(\frac{1}{p_2^{\alpha_C}T^2}\sum_{t=1}^{T}{\boldsymbol F}_{t}{\boldsymbol C}^{^\intercal}{\boldsymbol C}{\boldsymbol F}_{t}^{^\intercal}\right)\left(p_1^{-\alpha_R/2}{\boldsymbol R}^{^\intercal}\widehat{\boldsymbol R}\right)\widehat{\boldsymbol V}_R^{-1},\notag\\
\widehat{\boldsymbol H}_C&=&\left(\frac{1}{p_1^{\alpha_R}T^2}\sum_{t=1}^{T}{\boldsymbol F}_{t}^{^\intercal}{\boldsymbol R}^{^\intercal}{\boldsymbol R}{\boldsymbol F}_{t}\right)\left(p_2^{-\alpha_C/2}{\boldsymbol C}^{^\intercal}\widehat{\boldsymbol C}\right)\widehat{\boldsymbol V}_C^{-1},\notag
\end{eqnarray}
where $\widehat{\boldsymbol V}_R$ is an $r_1\times r_1$ diagonal matrix with the diagonal entries being the $r_1$ largest eigenvalues of $(p_1^{\alpha_R}p_2^{\alpha_C}T)^{-1}\widehat{\boldsymbol\Omega}_R$ (in a descending order) and $\widehat{\boldsymbol V}_C$ is an $r_2\times r_2$ diagonal matrix with the diagonal entries being the $r_2$ largest eigenvalues of $(p_1^{\alpha_R}p_2^{\alpha_C}T)^{-1}\widehat{\boldsymbol\Omega}_C$. Lemma \ref{le:B.5} in Appendix B of the supplement shows that both $\widehat{\boldsymbol H}_R$ and $\widehat{\boldsymbol H}_C$ are invertible {\em w.p.a.1}, and derives their respective limits. The following theorem gives the mean squared convergence rates for the (normalized) mPCA estimates.

\medskip

\renewcommand{\thetheorem}{3.\arabic{theorem}}
\setcounter{theorem}{0}

\begin{theorem}\label{thm:3.1}

Suppose that Assumptions \ref{ass:3.1}, \ref{ass:3.2} and \ref{ass:3.3}(i) are satisfied. Then we have
\begin{eqnarray}
&&p_1^{-\alpha_R/2}\left\Vert \widetilde{{\boldsymbol R}}-{\boldsymbol R}\widehat{{\boldsymbol H}}_R\right\Vert_F=O_P\left(p_1^{1/2-\alpha_R/2}p_2^{-\alpha_C/2}T^{-1}\left(1 + p_1^{-\alpha_R/2}p_2^{1-\alpha_C/2} \right)\right),\label{eq3.7}\\
&&p_2^{-\alpha_C/2}\left\Vert \widetilde{{\boldsymbol C}}-{\boldsymbol C}\widehat{\boldsymbol H}_C\right\Vert_F=O_P\left(p_2^{1/2-\alpha_C/2}p_1^{-\alpha_R/2}T^{-1}\left(1 + p_2^{-\alpha_C/2}p_1^{1-\alpha_R/2} \right)\right),\label{eq3.8}
\end{eqnarray}
where $\widetilde{{\boldsymbol R}}=p_1^{\alpha_R/2}\widehat{{\boldsymbol R}}$ and $\widetilde{{\boldsymbol C}}=p_2^{\alpha_C/2}\widehat{{\boldsymbol C}}$.

\end{theorem}

\medskip

\begin{remark}\label{re:3.2}

(i) Theorem \ref{thm:3.1} establishes the convergence rates for $\widetilde{{\boldsymbol R}}$ and $\widetilde{{\boldsymbol C}}$, which are multiplied by $p_1^{\alpha_R/2}$ and $p_2^{\alpha_C/2}$. This is due to the normalization conditions in (\ref{eq3.2}) and (\ref{eq3.3}) for possibly weak factors. It follows from Assumption \ref{ass:3.3}(i) that the convergence rates in (\ref{eq3.7}) and (\ref{eq3.8}) tend to zero. Hence $\widetilde{{\boldsymbol R}}$ and $\widetilde{{\boldsymbol C}}$ are consistent estimates of ${\boldsymbol R}$ and ${\boldsymbol C}$ up to the asymptotically invertible rotation matrices $\widehat{{\boldsymbol H}}_R$ and $\widehat{{\boldsymbol H}}_C$, respectively. As a result, the column spaces spanned by ${\boldsymbol R}$ and ${\boldsymbol C}$, ${\cal M}({\boldsymbol R})$ and ${\cal M}({\boldsymbol C})$, can be consistently estimated by ${\cal M}(\widehat{\boldsymbol R})$ and ${\cal M}(\widehat{\boldsymbol C})$.

(ii) The general convergence rates for the estimated row and column factor loading matrices in Theorem \ref{thm:3.1} depend on not only the dimension triplet $(p_1, p_2,T)$ but also the parameters $\alpha_R$ and $\alpha_C$ which measure the factor strength. When $\alpha_R=\alpha_C=1$, we may rewrite (\ref{eq3.7}) and (\ref{eq3.8}) as 
\begin{eqnarray}
p_1^{-1/2}\left\Vert \widetilde{{\boldsymbol R}}-{\boldsymbol R}\widehat{{\boldsymbol H}}_R\right\Vert_F&=&O_P\left(p_1^{-1/2}T^{-1}+p_2^{-1/2}T^{-1}\right),\notag\\
p_2^{-1/2}\left\Vert \widetilde{{\boldsymbol C}}-{\boldsymbol C}\widehat{{\boldsymbol H}}_C\right\Vert_F&=&O_P\left(p_1^{-1/2}T^{-1}+p_2^{-1/2}T^{-1}\right),\notag
\end{eqnarray}
which are comparable to those in \cite{CGGT25}. These convergence rates are significantly faster than typical convergence rates obtained by some existing literature on factor loading estimation for stationary matrix time series. For example, \cite{CF23} obtain $O_P([p_1\wedge (p_2T)]^{-1/2})$ and $O_P([p_2\wedge (p_1T)]^{-1/2})$ for the row and column factor loading matrix estimates, respectively. This is in fact similar to the so-called ``super-consistency" property achieved in the linear cointegrating model estimation theory \citep[e.g.,][]{Ph91} and the super-fast convergence rate is due to stronger trending signals contained in ${\boldsymbol F}_t$.
 
(iii) As in Section \ref{sec2.1}, by flattening the matrix observations into vectors, model (\ref{eq2.1}) can be re-written in the vectorized version, which has been considered by \cite{B04}, \cite{BN04} and \cite{BLL21}. However, the classic estimation theory based on the vectorized factor model neglects the matrix structure in the nonstationary data, resulting in inferior convergence rates. In fact, following the PCA methodology and theory in \cite{B04}, we may obtain the convergence rate $O_P(T^{-1})$ for the estimation of  ${\boldsymbol C}\otimes {\boldsymbol R}$, slower than the rate in (\ref{eq3.7}) and (\ref{eq3.8}) when $\alpha_R=\alpha_C=1$. This demonstrates the advantage of adopting the matrix version of the factor model, which makes the use of extra information through cross-row (or cross-column) aggregation.  
\end{remark}

\medskip

We next present the uniform convergence rate of the estimated common trends via mPCA.

\medskip

\begin{theorem}\label{thm:3.2}

Suppose that Assumptions \ref{ass:3.1}--\ref{ass:3.3} are satisfied. Then, we have the following uniform convergence theory:
\begin{equation}\label{eq3.9}
\max_{1\leq t\leq T}\left\|\widehat{{\boldsymbol F}}_t - \widehat\nu_{R,1}^{1/2}\widehat{\boldsymbol V}_R^{-1/2}\widehat{{\boldsymbol H}}_R^{-1}{\boldsymbol F}_t\left(\widehat{{\boldsymbol H}}_C^{-1}\right)^{^\intercal}\widehat{\boldsymbol V}_C^{-1/2}\right\|_F=O_P\left( \varphi_1(p_1,p_2,T)+\varphi_2(p_1,p_2,T)\right),
\end{equation}
where $\widehat\nu_{R,1}=p_1^{-\alpha_R}p_2^{-\alpha_C}\widetilde\lambda_{R,1}$ is positive and bounded w.p.a.1,
\begin{eqnarray}
\varphi_1(p_1,p_2,T)&=&p_1^{(1-2\alpha_R)/2}p_2^{1-\alpha_C}T^{-1/2} + p_2^{(1-2\alpha_C)/2}p_1^{1-\alpha_R}T^{-1/2},\notag\\
\varphi_2(p_1,p_2,T)&=&p_1^{-\alpha_R/2}p_2^{-\alpha_C/2}T^{1/4}.\notag
\end{eqnarray}
\end{theorem}

\medskip

\begin{remark}\label{re:3.3}
	
(i) Since the common components contain the stochastic trends, it is imperative to study the uniform convergence of $\widehat{\boldsymbol F}_t$ rather than the point-wise convergence (for each fixed $t$) as in Theorems 3 and 4 of \cite{CF23}. Note that 
{\small\begin{eqnarray}\label{eq3.10}
&&\widehat{\boldsymbol F}_t-\widetilde\lambda_{R,1}^{1/2}\widetilde{\boldsymbol V}_R^{-1/2}\widehat{\boldsymbol R}^{^\intercal} {\boldsymbol X}_t\widehat{\boldsymbol C}\widetilde{\boldsymbol V}_C^{-1/2}\notag\\
&=&\left[\widetilde\lambda_{R,1}^{1/2}\widetilde{\boldsymbol V}_R^{-1/2}\widehat{\boldsymbol R}^{^\intercal} {\boldsymbol Z}_t\widehat{\boldsymbol C}\widetilde{\boldsymbol V}_C^{-1/2}-\widehat\nu_{R,1}^{1/2}\widehat{\boldsymbol V}_R^{-1/2}\widehat{{\boldsymbol H}}_R^{-1}{\boldsymbol F}_t\left(\widehat{{\boldsymbol H}}_C^{-1}\right)^{^\intercal}\widehat{\boldsymbol V}_C^{-1/2}\right]+\widetilde\lambda_{R,1}^{1/2}\widetilde{\boldsymbol V}_R^{-1/2}\widehat{\boldsymbol R}^{^\intercal} {\boldsymbol E}_t\widehat{\boldsymbol C}\widetilde{\boldsymbol V}_C^{-1/2}.
\end{eqnarray}}
The rate $\varphi_1(p_1,p_2,T)$ is due to the uniform convergence for the first term on RHS of (\ref{eq3.10}), making use of Theorem \ref{thm:3.1}, whereas the rate $\varphi_2(p_1,p_2,T)$ is due to the uniform convergence of the second term on RHS of (\ref{eq3.10}) based on a combination of Theorem \ref{thm:3.1} and Lemma \ref{le:B.6} in Appendix B of the supplement.

(ii) Furthermore, the matrix ${\boldsymbol Z}_t$ of common components in (\ref{eq2.1}) can be estimated by 
\begin{equation}\label{eq3.11}
\widehat{\boldsymbol Z}_t=\widehat{\boldsymbol R}\left(\widehat{\boldsymbol R}^{^\intercal} {\boldsymbol X}_t\widehat{\boldsymbol C}\right)\widehat{\boldsymbol C}^{^\intercal},\ \ t=1,\ldots,T,
\end{equation}
as in \cite{WLC19}. Combining Theorems \ref{thm:3.1} and \ref{thm:3.2}, we may easily derive the uniform convergence rate for $\widehat{\boldsymbol Z}_t$.

\end{remark}

\medskip

We require a further condition to establish the consistency of eigenvalue ratio estimation.

\medskip

\begin{assumption}\label{ass:3.4}

{\em There exist $d_R\in(0,1] $ and $d_C\in(0,1]$ such that 
$$
\psi_{\lfloor d_R\underline{p}_1\rfloor}\left(\mathbb{E}_1\mathbb{E}_1^{^\intercal}/\overline{p}_1\right)\geq c_4+o_P(1)
$$
and
$$
\psi_{\lfloor d_C\underline{p}_2\rfloor}\left(\mathbb{E}_2\mathbb{E}_2^{^\intercal}/\overline{p}_2\right)\geq c_4+o_P(1)
$$
for some constant $c_4 > 0$, where $\mathbb{E}_1 = \left({\boldsymbol E}_1,...,{\boldsymbol E}_T\right)$, $\mathbb{E}_2 = \left({\boldsymbol E}_1^{^\intercal},...,{\boldsymbol E}_T^{^\intercal}\right)$, $\overline{p}_1=p_1\vee(Tp_2)$, $\underline{p}_1 = p_1 \wedge(Tp_2)$, $\overline{p}_2=p_2\vee(Tp_1)$ and $\underline{p}_2=p_2\wedge(Tp_1)$.}

\end{assumption}

\medskip

\begin{remark}\label{re:3.4}

Assumption~\ref{ass:3.4} is a high-level condition that ensures the $k$-th eigenvalues of $\widehat{\boldsymbol \Omega}_R$ (resp., $\widehat{\boldsymbol \Omega}_C$), $k > r_1$ (resp., $k > r_2$), remain bounded away from zero after suitable normalization. This condition can be verified under a separable error structure for ${\boldsymbol E}_t$, see Assumptions C–D in \citet{AH13} for vectorized factor models, and Lemma 1 in \citet{CL24} for tensor factor models (including matrix factor models as a special case).

\end{remark}

\medskip

The following theorem establishes the consistency property of the ratio criterion.

\medskip

\begin{theorem}\label{thm:3.3}

Suppose that Assumptions \ref{ass:3.1}--\ref{ass:3.4} are satisfied, and $r_1\wedge r_2\geq1$. Then, we have
\begin{equation}\label{eq3.12}
{\sf P}\left(\widehat{r}_1=r_1\right)\rightarrow1\quad \text{and} \quad {\sf P}\left(\widehat{r}_2=r_2\right)\rightarrow1
\end{equation} 
for any $K_1\in (r_1, \lfloor d_R\underline{p}_1\rfloor - 2r_1]$ and $K_2\in (r_2, \lfloor d_C\underline{p}_2\rfloor - 2r_2]$.
\end{theorem}

\medskip

\begin{remark}\label{re:3.5}

(i) The restriction of $r_1\wedge r_2\geq1$ implies the existence of common trends in the matrix factor model and thus avoids the possibility of spurious factor modelling of large-scale nonstationary time series \citep{OW21}. This is also consistent with the fact that the developed ratio criterion would select at least one factor \citep[e.g.,][]{AH13}. In addition, the pre-specified upper bounds $K_1$ and $K_2$ are allowed to diverge to infinity.

(ii) Unlike the stationary case \citep[e.g.,][]{F22}, consistent estimation of the number of factors in our framework does not require the factor strengths $\alpha_{R}$ and $\alpha_{C}$ to exceed a fixed threshold, as long as Assumption \ref{ass:3.3}(i) holds. Intuitively, this is due to the fact that our method directly targets the nonstationary matrix time series and strong signals in the common stochastic trends dominate those in the matrix idiosyncratic components, leading to much faster convergence as illustrated in Theorem \ref{thm:3.1} and Remark \ref{re:3.2}.

\end{remark}


\subsection{Convergence theory for mPANIC estimation}\label{sec3.2}

We next derive the convergence properties for the mPANIC estimator developed in Section \ref{sec2.3}.

\medskip

\begin{assumption}\label{ass:3.5}

{\em (i)\ The equation $\vert {\boldsymbol I}_{r_1r_2}-({\boldsymbol I}_{r_1r_2}+{\boldsymbol\alpha}{\boldsymbol\beta}^{^\intercal})z\vert=0$ has roots on or outside the unit circle, and the matrix ${\boldsymbol I}_{k_1k_2}+{\boldsymbol\beta}^{^\intercal}{\boldsymbol\alpha}$ has eigenvalues strictly smaller than one, where ${\boldsymbol\alpha}={\boldsymbol\alpha}_2\otimes {\boldsymbol\alpha}_1$ and ${\boldsymbol\beta}={\boldsymbol\beta}_2\otimes {\boldsymbol\beta}_1$.}

{\em (ii) Let $\{{\boldsymbol V}_t\}$ be a sequence of i.i.d. $r_1\times r_2$ matrix-valued random elements with mean zero and finite fourth moment.}

{\em (iii)\ The eigenvalues of ${\boldsymbol\Sigma}_R^{1/2}{\sf E}\left[ \Delta {\boldsymbol F}_t{\boldsymbol\Sigma}_C \Delta {\boldsymbol F}_t^{^\intercal}\right]{\boldsymbol\Sigma}_R^{1/2}$ are positive, bounded and distinct. The same condition holds for ${\boldsymbol\Sigma}_C^{1/2} {\sf E}\left[ \Delta {\boldsymbol F}^{^\intercal}_t{\boldsymbol\Sigma}_R \Delta {\boldsymbol F}_t\right] {\boldsymbol\Sigma}_C^{1/2}$.}

\end{assumption}

\medskip

\begin{assumption}\label{ass:3.6}

{\em (i)\ Let $\{\varepsilon_{t,(i,j)}\}$ be a sequence of zero-mean random elements independent of $\{{\boldsymbol V}_t\}$, and $\max_{t}\max_{i,j}{\sf E}[\varepsilon_{t,(i,j)}^4]<\infty$. }

{\em (ii)\ Assumption \ref{ass:3.2}(ii)(iii) continues to hold when $e_{t,(i,j)}$ is replaced by $\varepsilon_{t,(i,j)}$.}

\end{assumption}

\medskip

\begin{assumption}\label{ass:3.7}	
	{\em As $p_1,p_2\to \infty$, $p_1^{1/2-\alpha_R}p_2^{1-\alpha_C}+p_2^{1/2-\alpha_C}p_1^{1-\alpha_R} \to 0$.} 	
\end{assumption}

\medskip

\medskip

\begin{remark}\label{re:3.6}

Assumption \ref{ass:3.5}(i) is crucial to derive the classic Granger’s representation for latent factors \citep[e.g.,][]{J95, CP09, LX24}, see (\ref{eqB.20}) in the supplement. The independence condition on $\{{\boldsymbol V}_t\}$ in Assumption \ref{ass:3.5}(ii) may be replaced by stationary martingale differences. Assumption \ref{ass:3.5}(iii) is comparable to Assumption \ref{ass:3.1}(iii) and a similar condition has been commonly used in the literature on factor model estimation theory. Assumption \ref{ass:3.6} is analogous to Assumption \ref{ass:3.2}, containing some high-level restrictions on temporal, cross-row and cross-column dependence. Assumption \ref{ass:3.7} implies that $\alpha_R$ and $\alpha_C$ must be larger than $1/2$, which is more restrictive that that in Section \ref{sec3.1}.

\end{remark}

\medskip

As in Section \ref{sec3.1}, we define two rotation matrices:
\begin{eqnarray}
\overline{\boldsymbol H}_R&=&\left(\frac{1}{p_2^{\alpha_C}T}\sum_{t=1}^{T}\Delta{\boldsymbol F}_{t}{\boldsymbol C}^{^\intercal}{\boldsymbol C} \Delta{\boldsymbol F}_{t}^{^\intercal}\right)\left(p_1^{-\alpha_R/2}{\boldsymbol R}^{^\intercal}\overline{\boldsymbol R}\right)\check{\boldsymbol V}_R^{-1},\notag\\
\overline{\boldsymbol H}_C&=&\left(\frac{1}{p_1^{\alpha_R}T}\sum_{t=1}^{T}\Delta{\boldsymbol F}_{t}^{^\intercal}{\boldsymbol R}^{^\intercal}\Delta{\boldsymbol R}{\boldsymbol F}_{t}\right)\left(p_2^{-\alpha_C/2}{\boldsymbol C}^{^\intercal}\overline{\boldsymbol C}\right)\check{\boldsymbol V}_C^{-1},\notag
\end{eqnarray}
where $\check{\boldsymbol V}_R$ is an $r_1\times r_1$ diagonal matrix with the diagonal entries being the $r_1$ largest eigenvalues of $(p_1^{\alpha_R}p_2^{\alpha_C})^{-1}\overline{\boldsymbol\Omega}_R$ (in a descending order), and $\check{\boldsymbol V}_C$ is defined similarly with $r_1$ and $\overline{\boldsymbol\Omega}_R$ replaced by $r_2$ and $\overline{\boldsymbol\Omega}_C$, respectively. The following theorem gives the mean squared convergence rates for (normalized) mPANIC estimates of row and column factor loadings.

\medskip

\begin{theorem}\label{thm:3.4}

Suppose that Assumptions \ref{ass:3.1}(ii) and \ref{ass:3.5}--\ref{ass:3.7} are satisfied. Then we have
\begin{eqnarray}
&&p_1^{-\alpha_R/2}\left\Vert \check{{\boldsymbol R}}-{\boldsymbol R}\overline{{\boldsymbol H}}_R\right\Vert_F=O_P\left(p_1^{1/2-\alpha_R}p_2^{1-\alpha_C}+{p_1^{1-\alpha_R}}p_2^{1/2-\alpha_C}T^{-1/2}\right),\label{eq3.13}\\
&&p_2^{-\alpha_C/2}\left\Vert \check{{\boldsymbol C}}-{\boldsymbol C}\overline{\boldsymbol H}_C\right\Vert_F=O_P\left(p_2^{1/2-\alpha_C}p_1^{1-\alpha_R}+{p_2^{1-\alpha_C}}p_1^{1/2-\alpha_R}T^{-1/2}\right),\label{eq3.14}
\end{eqnarray}
where $\check{{\boldsymbol R}}=p_1^{\alpha_R/2}\overline{{\boldsymbol R}}$ and $\check{{\boldsymbol C}}=p_2^{\alpha_C/2}\overline{{\boldsymbol C}}$.

\end{theorem}

\medskip

\begin{remark}\label{re:3.7}

Since the mPANIC estimation of factor loadings is obtained via eigenanalysis of sample row and column covariance matrices with differenced matrix time series observations, the convergence rates in (\ref{eq3.13}) and (\ref{eq3.14}) are typical stationary rates, which are much slower than those in (\ref{eq3.7}) and (\ref{eq3.8}). For the special case of strong factors with $\alpha_R=\alpha_C=1$, the rates are simplified to 
\[
O_P\left(p_1^{-1/2}+(p_2T)^{-1/2}\right)\quad\text{and}\quad O_P\left(p_2^{-1/2}+(p_1T)^{-1/2}\right),
\] 
which are the same as those in \cite{CF23}.

\end{remark}

\medskip

Write
\begin{equation}\label{eq3.15}
{\boldsymbol E}_t={\boldsymbol E}_t^\circ+{\boldsymbol E}_t^\dag,\quad {\boldsymbol E}_t^\circ={\boldsymbol E}_t\odot {\boldsymbol\Gamma}^\circ,\quad {\boldsymbol E}_t^\dag={\boldsymbol E}_t\odot {\boldsymbol\Gamma}^\dag,
\end{equation}
and 
\[
{\boldsymbol\Gamma}^\circ=\left(\Gamma_{ij}^\circ\right)_{p_1\times p_2}\quad \text{with}\quad \Gamma_{ij}^\circ={\sf I}(|\rho_{ij}|<1)\quad \text{and}\quad{\boldsymbol\Gamma}^\dag=\left(\Gamma_{ij}^\dag\right)_{p_1\times p_2}\quad \text{with}\quad \Gamma_{ij}^\dag={\sf I}(\rho_{ij}=1),
\]
where ${\sf I}(\cdot)$ denotes the indicator function. It follows from (\ref{eq3.15}) that ${\boldsymbol E}_t$ is decomposed into a sum of I(0) matrix idiosyncratic component ${\boldsymbol E}_t^\circ$ and I(1) component ${\boldsymbol E}_t^\dag$. In addition, we impose the following structural restriction on ${\boldsymbol\Gamma}^\dag$, i.e.,
\begin{equation}\label{eq3.16}
{\boldsymbol\Gamma}^\dag=\left(
\begin{array}{ll}
{\boldsymbol\Gamma}_1^\dag& {\boldsymbol O}_{s_R\times (p_2-s_C)}\\
{\boldsymbol O}_{(p_1-s_R)\times s_C} & {\boldsymbol O}_{(p_1-s_R)\times(p_2-s_C)}
\end{array}
\right),
\end{equation}
where ${\boldsymbol\Gamma}_1^\dag$ is an $s_R\times s_C$ block matrix with all the entries being ones, and ${\boldsymbol O}_{k\times l}$ is a null matrix with size $k\times l$. The condition (\ref{eq3.16}) limits the number of nonstationary entries in the idiosyncratic matrix, which is similar to the commonly-used sparsity restriction in high-dimensional econometrics. 

\medskip

\begin{assumption}\label{ass:3.8}

{\em (i)\ As $p_1,p_2,T\to \infty$, 
\[
p_1^{2-2\alpha_R}p_2^{2-2\alpha_C}=O(T)
\]
and
\[
\left(\frac{s_p}{p_1^{\alpha_R}}\frac{s_C}{p_2^{\alpha_C}}\right)^{1/2}T^{1/4}=O(1).
\]
 } 

{\em (ii)\ There exists a positive constant $c_5$ such that
\begin{equation}\label{eq3.17}
{\sf E}\left\Vert \frac{{\boldsymbol R}^{^\intercal}{\boldsymbol E}_t^\circ{\boldsymbol C}}{p_1^{\alpha_R/2}p_2^{\alpha_C/2}}\right\Vert^4\leq c_5\quad \text{and}\quad {\sf E}\left\Vert \frac{{\boldsymbol E}_t^\dag}{s_R^{1/2}s_C^{1/2}}\right\Vert^4+{\sf E}\left\Vert \frac{{\boldsymbol R}^{^\intercal}{\boldsymbol E}_t^\dag{\boldsymbol C}}{s_R^{\alpha_R/2}s_C^{\alpha_C/2}}\right\Vert^4\leq c_5t^2
\end{equation}
for any $1\leq t\leq T$. }

{\em (iii)\ Letting $e_{t,(i,j)}^\dag$ be the $(i,j)$-entry of ${\boldsymbol E}_t^\dag$, there exists a positive constant $c_6$ such that
\[
\max_{1\leq i\leq s_R}{\sf E}\left\Vert s_C^{-\alpha_C/2}\sum_{j=1}^{p_2}e_{t,(i,j)}^\dag{\boldsymbol C}_{j\bullet}\right\Vert^4\leq c_6t^2
\]
and
\[
\max_{1\leq j\leq s_C}{\sf E}\left\Vert s_R^{-\alpha_R/2}\sum_{i=1}^{p_1}e_{t,(i,j)}^\dag{\boldsymbol R}_{i\bullet}\right\Vert^4\leq c_6t^2
\]
for any $1\leq t\leq T$.}

\end{assumption}

\medskip

\begin{remark}\label{re:3.8}

Assumption \ref{ass:3.8}(i) restricts the relationship between the dimensions $p_1$, $p_2$ and $T$, indicating that $s_R\ll p_1^{\alpha_R}$ and $s_C\ll p_2^{\alpha_C}$. When $\alpha_R=\alpha_C=1$, Assumption \ref{ass:3.8}(i) is simplified to 
\[
\left(\frac{s_p}{p_1}\frac{s_C}{p_2}\right)^{1/2}T^{1/4}=O(1).
\]
It follows from (\ref{eq3.16}) that only the upper-left $s_R\times s_C$ block matrix of ${\boldsymbol E}_t^\dag$ contains non-zero entries which are nonstationary I(1). Hence, the normalization rates $s_R^{1/2}$, $s_R^{\alpha_R/2}$, $s_C^{1/2}$ and $s_C^{\alpha_C/2}$ are used in the high-level conditions in Assumption \ref{ass:3.8}(ii)(iii). We may verify these conditions if $\varepsilon_{t,(i,j)}$ are further assumed to be i.i.d. or satisfy some weak dependence conditions.

\end{remark}

\medskip

We next state the uniform convergence property of the cointegrated matrix factor estimation. 

\medskip

\begin{theorem}\label{thm:3.5}

Suppose that Assumptions \ref{ass:3.1}(ii) and \ref{ass:3.5}--\ref{ass:3.8} are satisfied. The following uniform convergence property holds:
\begin{equation}\label{eq3.18}
\max_{1\leq t\leq T}\left\|\overline{{\boldsymbol F}}_t - \check{\nu}_{R,1}^{1/2}\check{\boldsymbol V}_R^{-1/2}\overline{{\boldsymbol H}}_R^{-1}{\boldsymbol F}_t\left(\overline{{\boldsymbol H}}_C^{-1}\right)^{^\intercal}\check{\boldsymbol V}_C^{-1/2}\right\|_F=O_P\left( \varpi_1(p_1,p_2,T)+\varpi_2(p_1,p_2,T)\right),
\end{equation}
where $\check{\nu}_{R,1}=p_1^{-\alpha_R}p_2^{-\alpha_C}\overline\lambda_{R,1}$ is positive and bounded w.p.a.1,
\begin{eqnarray}
\varpi_1(p_1,p_2,T)&=&T^{1/2}\left(p_1^{1/2-\alpha_R} p_2^{1-\alpha_C}+p_2^{1/2-\alpha_C}p_1^{1-\alpha_R}\right),\notag\\
\varpi_2(p_1,p_2,T)&=&\left(s_R^{1/2}/p_1\right)^{\alpha_R}\left(s_C^{1/2}/p_2\right)^{\alpha_C}T^{3/4}.\notag
\end{eqnarray}
 
\end{theorem}

\medskip

\begin{remark}\label{re:3.9}

With (\ref{eq2.11}) and (\ref{eq3.15}), we have 
\begin{eqnarray}\label{eq3.19}
&&\overline{{\boldsymbol F}}_t - \overline\lambda_{R,1}^{1/2}\overline{\boldsymbol V}_R^{-1/2}\overline{\boldsymbol R}^{^\intercal}{\boldsymbol X}_t\overline{\boldsymbol C}\,\overline{\boldsymbol V}_C^{-1/2}\notag\\
&=&\left[\overline\lambda_{R,1}^{1/2}\overline{\boldsymbol V}_R^{-1/2}\overline{\boldsymbol R}^{^\intercal}{\boldsymbol Z}_t\overline{\boldsymbol C}\,\overline{\boldsymbol V}_C^{-1/2}-\check{\nu}_{R,1}^{1/2}\check{\boldsymbol V}_R^{-1/2}\overline{{\boldsymbol H}}_R^{-1}{\boldsymbol F}_t\left(\overline{{\boldsymbol H}}_C^{-1}\right)^{^\intercal}\check{\boldsymbol V}_C^{-1/2}\right] \notag\\
&&+\overline\lambda_{R,1}^{1/2}\overline{\boldsymbol V}_R^{-1/2}\overline{\boldsymbol R}^{^\intercal}{\boldsymbol E}_t^\circ\overline{\boldsymbol C}\,\overline{\boldsymbol V}_C^{-1/2}+\overline\lambda_{R,1}^{1/2}\overline{\boldsymbol V}_R^{-1/2}\overline{\boldsymbol R}^{^\intercal}{\boldsymbol E}_t^\dag\overline{\boldsymbol C}\,\overline{\boldsymbol V}_C^{-1/2}.
\end{eqnarray}
The rate $\varpi_1(p_1,p_2,T)$ is due to the uniform convergence for the first term on RHS of (\ref{eq3.19}) and the factor loading estimation errors; the second term on RHS of (\ref{eq3.19}) is dominated by the rate $\varpi_1(p_1,p_2,T)$ as the stationary matrix idiosyncratic component ${\boldsymbol E}_t^\circ$ contains much weaker signal than the nonstationary common components; the rate $\varpi_2(p_1,p_2,T)$ is due to the uniform convergence of the third term on RHS of (\ref{eq3.19}) which contains the nonstationary matrix idiosyncratic component ${\boldsymbol E}_t^\dag$ satisfying the sparsity restriction.

\end{remark}

\begin{theorem}\label{thm:3.6}
Suppose that Assumptions \ref{ass:3.1}(ii), \ref{ass:3.5}--\ref{ass:3.7} are satisfied, $r_1\wedge r_2\geq1$, and Assumption \ref{ass:3.4} is valid when $e_{t,(i,j)}$ is replaced by $\varepsilon_{t,(i,j)}$. Then, we have
\[
{\sf P}\left(\overline{r}_1=r_1\right)\rightarrow1\quad \text{and} \quad {\sf P}\left(\overline{r}_2=r_2\right)\rightarrow1
\]
for any $K_1\in (r_1, \lfloor d_R\underline{p}_1\rfloor - 2r_1]$ and $K_2\in (r_2, \lfloor d_C\underline{p}_2\rfloor - 2r_2]$.
\end{theorem}



\section{Extension to heterogeneous factor strength}\label{sec4}
\renewcommand{\theequation}{4.\arabic{equation}}
\setcounter{equation}{0}

The convergence theory developed in Section \ref{sec3} assumes that the row/column (weak) factor strengths are homogeneous with the constant values $\alpha_R$ and $\alpha_C$, see Assumption \ref{ass:3.1}(ii). This assumption, however, may be too restrictive for real-world applications. In this section, we further extend the methodology and theory, allowing for heterogeneous factor strength. 

\subsection{Convergence of mPCA with heterogeneous factor strength}\label{sec4.1}

Due to heterogeneity in the weak factor strength, we replace Assumptions \ref{ass:3.1}(ii)(iii), \ref{ass:3.2}(iii) and \ref{ass:3.3} by the following conditions.

\medskip

\renewcommand{\theassumption}{4.\arabic{assumption}}
\setcounter{assumption}{0}

\begin{assumption}\label{ass:4.1}

{\em (i)\ There exist diagonal matrices ${\boldsymbol\Sigma}_R^\ast$ and ${\boldsymbol\Sigma}_C^\ast$ with positive diagonal entries such that 
\begin{equation}\label{eq4.1}
\mathbf{B}_{R}^{-1}{\boldsymbol R}^{^\intercal}{\boldsymbol R}\mathbf{B}_{R}^{-1}\rightarrow{\boldsymbol\Sigma}_R^\ast\ \ \text{as}\ p_1\rightarrow\infty,\ \ \mathbf{B}_{R} = {\sf diag}\left(p_1^{\alpha_{R,1}/2},...,p_1^{\alpha_{R,r_1}/2} \right),
\end{equation}
and
\begin{equation}\label{eq4.2}
\mathbf{B}_{C}^{-1}{\boldsymbol C}^{^\intercal}{\boldsymbol C}\mathbf{B}_{C}^{-1}\rightarrow{\boldsymbol\Sigma}_C^\ast\ \ \text{as}\ p_2\rightarrow\infty,\ \ \mathbf{B}_{C} = {\sf diag}\left(p_2^{\alpha_{C,1}/2},...,p_2^{\alpha_{C,r_2}/2} \right).
\end{equation}
In addition, both $\Vert {\boldsymbol R}_{i\bullet}\Vert$ and $\Vert {\boldsymbol C}_{i\bullet}\Vert$ are bounded uniformly over $i$.}
	
{\em (ii)\ The eigenvalues of $({\boldsymbol\Sigma}_R^{\ast})^{1/2}[\int_0^1 {\boldsymbol W}(u){\boldsymbol\Sigma}_C^{\ast,1}{\boldsymbol W}(u)^{^\intercal}du]({\boldsymbol\Sigma}_R^{\ast})^{1/2}$ are positive, bounded and distinct with probability one, where ${\boldsymbol W}(\cdot)$ is defined in Assumption \ref{ass:3.1}(iii) and ${\boldsymbol\Sigma}_C^{\ast,1} =\lim_{p_2\to \infty} p_2^{-\alpha_{C,1}}{\boldsymbol C}^{^\intercal}{\boldsymbol C}$. The same condition holds for $({\boldsymbol\Sigma}_C^{\ast})^{1/2}[\int_0^1 {\boldsymbol W}(u)^{^\intercal}{\boldsymbol\Sigma}_R^{\ast,1}{\boldsymbol W}(u)du]({\boldsymbol\Sigma}_C^{\ast})^{1/2}$ with ${\boldsymbol\Sigma}_R^{\ast,1} =\lim_{p_1\to \infty} p_1^{-\alpha_{R,1}}{\boldsymbol R}^{^\intercal}{\boldsymbol R}$.}
\end{assumption}

\medskip

\begin{assumption}\label{ass:4.2}

{\em (i)\ There exists a positive constant $c_7$ such that 
\[
\max_{1\leq i\leq p_1}{\sf E}\left\Vert \sum_{j=1}^{p_2}e_{t,(i,j)}{\boldsymbol C}_{j\bullet}\mathbf{B}_C^{-1}\right\Vert^4\leq c_7,\ \ \max_{1\leq i\leq p_1}{\sf E}\left\Vert \sum_{s=t_1}^{t_2}\sum_{j=1}^{p_2}e_{s,(i,j)}{\boldsymbol C}_{j\bullet}\mathbf{B}_C^{-1}\right\Vert^2\leq c_7(t_2-t_1)
\]
for any $1\leq t\leq T$ and $1\leq t_1<t_2\leq T$.}
	
{\em (ii)\ There exists a positive constant $c_8$ such that 
\[
\max_{1\leq j\leq p_2}{\sf E}\left\Vert \sum_{i=1}^{p_1}e_{t,(i,j)}{\boldsymbol R}_{i\bullet}\mathbf{B}_R^{-1}\right\Vert^4\leq c_8,\ \ \max_{1\leq j\leq p_2}{\sf E}\left\Vert \sum_{s=t_1}^{t_2}\sum_{i=1}^{p_1}e_{s,(i,j)}{\boldsymbol R}_{i\bullet}\mathbf{B}_R^{-1}\right\Vert^2\leq c_8(t_2-t_1)
\]
for any $1\leq t\leq T$ and $1\leq t_1<t_2\leq T$.}
	
{\em (iii)\ There exists a positive constant $c_9$ such that 
$$\max_{1\leq t\leq T}{\sf E}\left\| \mathbf{B}_R^{-1}\mathbf{R}^{^\intercal}{\boldsymbol E}_t\mathbf{C}\mathbf{B}_C^{-1} \right\|_F^4 \leq c_9.$$}

\end{assumption}

\medskip

\begin{assumption}\label{ass:4.3}

{\em As $p_1,p_2,T\to \infty$, $$\left(p_1^{\alpha_{R,1}-\alpha_{R,r_1}}\vee p_2^{\alpha_{C,1}-\alpha_{C,r_2}}\right)\frac{p_1^{1-\alpha_{R,1}}p_2^{1-\alpha_{C,1}}}{T} \to 0$$
and
$$p_1^{\alpha_{R,1}-\alpha_{R,r_1}} \frac{p_1^{2-\alpha_{R,1}}}{p_2^{\alpha_{C,1}}T^2}\to 0\quad \text{and}\quad p_2^{\alpha_{C,1}-\alpha_{C,r_2}}\frac{p_2^{2-\alpha_{C,1}}}{p_1^{\alpha_{R,1}}T^2}\to 0.$$}

\end{assumption}

\medskip

\renewcommand{\theremark}{4.\arabic{remark}}
\setcounter{remark}{0}

\begin{remark}\label{re:4.1}

Assumptions \ref{ass:4.1} and \ref{ass:4.2} are similar in spirit to Assumptions \ref{ass:3.1}(ii)(iii) and \ref{ass:3.2}(iii) in Section \ref{sec3.1}. In particular, Assumption \ref{ass:4.1}(i) characterizes heterogeneous factor strengths via the diagonal matrices $\mathbf{B}_{R}$ and $\mathbf{B}_{C}$. Assuming ${\boldsymbol\Sigma}_R^*$ and ${\boldsymbol\Sigma}_C^*$ to be diagonal allows for the identification of factors with different strength orders and facilitates the choice of suitable normalizers, ensuring that the sample eigenvalue matrices $\widehat{\boldsymbol V}_R^\ast$ and $\widehat{\boldsymbol V}_C^\ast$ (to be defined below) are positive definite. This condition is also adopted by some recent works on vector and matrix factor models with weak loadings (or factors), e.g., Assumption A2'(ii) in \cite{BN23} and Assumption (V1) in \cite{CL24WP}. As pointed out by \cite{CL24WP}, this requirement is not restrictive, since weak factor models with heterogeneous strengths and non-diagonal limiting matrices are observationally equivalent to those with diagonal ones. Assumption \ref{ass:4.3} requires a more stringent requirement on the relative growth rates of $p_1$, $p_2$, and $T$. In particular, the precise condition also  depends on the degree of heterogeneity between the strongest and weakest factor strengths, similar to Assumption A4$^\prime$ in \cite{BN23} for vectorized factor models.

\end{remark}

\medskip

Due to the heterogeneous factor strength, a new normalization matrix is required to properly scale the sample moments of row/column factor loadings, see Assumption \ref{ass:4.1}(i). Accordingly, the rotation matrices also need to be modified, i.e.,
\begin{eqnarray*}
\widehat{\boldsymbol H}_R^\ast&=&\left(\frac{1}{p_2^{\alpha_{C,1}}T^2}\sum_{t=1}^{T}{\boldsymbol F}_{t}{\boldsymbol C}^{^\intercal}{\boldsymbol C}{\boldsymbol F}_{t}^{^\intercal}\right)\left({\boldsymbol R}^{^\intercal}\widehat{\boldsymbol R}\mathbf{B}_R^{-1}\right)\left(\widehat{\boldsymbol V}_R^\ast\right)^{-1},\notag\\
\widehat{\boldsymbol H}_C^\ast&=&\left(\frac{1}{p_1^{\alpha_{R,1}}T^2}\sum_{t=1}^{T}{\boldsymbol F}_{t}^{^\intercal}{\boldsymbol R}^{^\intercal}{\boldsymbol R}{\boldsymbol F}_{t}\right)\left({\boldsymbol C}^{^\intercal}\widehat{\boldsymbol C}\mathbf{B}_C^{-1}\right)\left(\widehat{\boldsymbol V}_C^\ast\right)^{-1},
\end{eqnarray*}
where $\widehat{\boldsymbol V}_R^\ast = p_2^{-\alpha_{C,1}}\mathbf{B}_R^{-2}\widetilde{\boldsymbol V}_R$,  $\widehat{\boldsymbol V}_C^\ast = p_1^{-\alpha_{R,1}}\mathbf{B}_C^{-2}\widetilde{\boldsymbol V}_C$, $\widetilde{\boldsymbol V}_R$ and $\widetilde{\boldsymbol V}_C$ are defined as in \eqref{eq2.4}. Lemma \ref{le:B.15} in Appendix B of the supplement shows that both $\widehat{\boldsymbol H}_R$ and $\widehat{\boldsymbol H}_C$ are invertible {\em w.p.a.1}. The following theorem extends Theorems \ref{thm:3.1}--\ref{thm:3.3} to the case of varying factor strength. 

\medskip

\renewcommand{\thetheorem}{4.\arabic{theorem}}
\setcounter{theorem}{0}

\begin{theorem}\label{thm:4.1}

Suppose that Assumptions \ref{ass:3.1}(i), \ref{ass:3.2}(i)(ii) and \ref{ass:4.1}--\ref{ass:4.3} are satisfied. 

(i) The mPCA estimators of the factor loading matrices have the following mean squared convergence: 
\begin{eqnarray}
\left\Vert \left(\widetilde{\boldsymbol R}^\ast-{\boldsymbol R}\widehat{\boldsymbol H}_R^\ast\right)\mathbf{B}_R^{-1}\right\Vert_F
&=&O_P\left(\xi_R(p_1,p_2,T)\right),\label{eq4.3}\\
\left\Vert \left(\widetilde{\boldsymbol C}^\ast-{\boldsymbol C}\widehat{\boldsymbol H}_C^\ast\right)\mathbf{B}_C^{-1}\right\Vert_F
&=&O_P\left(\xi_C(p_1,p_2,T)\right),\label{eq4.4}
\end{eqnarray}
where $\widetilde{\boldsymbol R}^\ast = \widehat{\boldsymbol R}\mathbf{B}_R$, $\widetilde{\boldsymbol C}^\ast = \widehat{\boldsymbol C}\mathbf{B}_C$, and 
\begin{eqnarray}
\xi_R(p_1,p_2,T)&=&p_1^{\alpha_{R,1}-\alpha_{R,r_1}} p_1^{1/2-\alpha_{R,1}/2}p_2^{-\alpha_{C,1}/2}T^{-1}\left(1 + p_1^{-\alpha_{R,1}/2}p_2^{1-\alpha_{C,1}/2} \right),\notag\\
\xi_C(p_1,p_2,T)&=&p_2^{\alpha_{C,1}-\alpha_{C,r_2}} p_2^{1/2-\alpha_{C,1}/2}p_1^{-\alpha_{R,1}/2}T^{-1}\left(1+p_2^{-\alpha_{C,1}/2}p_1^{1-\alpha_{R,1}/2}\right).\notag
\end{eqnarray}

(ii) The factor matrix estimators have the following uniform convergence property:
\begin{equation}
\max_{1\leq t\leq T}\left\|\widehat{\mathbf{F}}_t - \widehat\nu_{R,1}^{\ast^{1/2}}\widehat{\boldsymbol V}_R^{\ast^{-1/2}}\widehat{{\boldsymbol H}}_R^{\ast^{-1}}{\boldsymbol F}_t\left(\widehat{{\boldsymbol H}}_C^{\ast^{-1}}\right)^{^\intercal}\widehat{\boldsymbol V}_C^{\ast^{-1/2}} \right\|_F
=O_P\left(\varphi_1^\ast(p_1,p_2,T) + \varphi_2^\ast(p_1,p_2,T)\right),\label{eq4.5}
\end{equation}
where $\widehat\nu_{R,1}^*=p_1^{-\alpha_{R,1}}p_2^{-\alpha_{C,1}}\widetilde\lambda_{R,1}$ is positive and bounded w.p.a.1,
\begin{eqnarray}
\varphi_1^\ast(p_1,p_2,T)&=&\left(p_1^{\alpha_{R,1}-\alpha_{R,r_1}}\vee p_2^{\alpha_{C,1}-\alpha_{C,r_2}}\right)\left(p_1^{(1-2\alpha_{R,1})/2}p_2^{1-\alpha_{C,1}}T^{-1/2} + p_2^{(1-2\alpha_{C,1})/2}p_1^{1-\alpha_{R,1}}T^{-1/2}\right),\notag\\
\varphi_2^\ast(p_1,p_2,T)&=&\left(p_1^{\alpha_{R,1}-\alpha_{R,r_1}} p_2^{\alpha_{C,1}-\alpha_{C,r_2}}\right)p_1^{-\alpha_{R,1}/2}p_2^{-\alpha_{C,1}/2}T^{1/4}.\notag
\end{eqnarray}

(iii) If, in addition, Assumption \ref{ass:3.4} is satisfied, $r_1\wedge r_2\geq1$,
\[
p_1^{\alpha_{R,1}-\alpha_{R,r_1}}\xi_R(p_1,p_2,T)\to 0\quad \text{and}\quad p_2^{\alpha_{C,1}-\alpha_{C,r_1}}\xi_C(p_1,p_2,T)\to 0,
\]
we have
$$
{\sf P}\left(\widehat{r}_1=r_1\right)\rightarrow1 \quad \text{and}\quad {\sf P}\left(\widehat{r}_2=r_2\right)\rightarrow1
$$
for any $K_1\in (r_1, \lfloor d_R\underline{p}_1\rfloor - 2r_1]$ and $K_2\in (r_2, \lfloor d_C\underline{p}_2\rfloor - 2r_2]$.

\end{theorem}

\medskip

\begin{remark}\label{re:4.2}

(i) When the factor strength is homogeneous, Theorem \ref{thm:4.1}(i) coincides exactly with the convergence property in Theorem \ref{thm:3.1}. In the presence of heterogeneous factor strengths, the mean squared convergence for factor loading estimators slows down, scaling the homogeneous convergence rate by a multiplicative factor that depends on the discrepancy between the strongest and weakest factor strengths. When estimating the row factor loadings, only the strongest column factor strength, $\alpha_{C,1}$, enters the convergence rate $\xi_R(p_1,p_2,T)$. Other weaker column factor strengths do not affect the rate and hence play no role in the average estimation error. The finding is the same for column factor loading estimation convergence.

(ii) The rate $\varphi_1^\ast(p_1,p_2,T)$ in (\ref{eq4.5}) arises from the uniform convergence of the first term (by slightly modifying some notation) on the right-hand side of (\ref{eq3.10}), using Theorem \ref{thm:4.1}(i), while $\varphi_2^\ast(p_1,p_2,T)$ is determined by the second term, using both Theorem \ref{thm:4.1}(i) and Lemma \ref{le:B.16} in Appendix B. In contrast to the case with homogeneous factor strength, $\varphi_1^\ast(p_1,p_2,T)$ depends on $p_1^{\alpha_{R,1}-\alpha_{R,r_1}} \vee p_2^{\alpha_{C,1}-\alpha_{C,r_2}}$ and $\varphi_2^\ast(p_1,p_2,T)$ depends on the product $p_1^{\alpha_{R,1}-\alpha_{R,r_1}} p_2^{\alpha_{C,1}-\alpha_{C,r_2}}$, both slowing down the uniform convergence rates.

(iii) Similar to Theorem \ref{thm:3.3}, consistent estimation of the number of factors in the case of heterogeneous factor strengths does not require the minimum factor strengths $\alpha_{R,r_1}$ and $\alpha_{C,r_2}$ to exceed a fixed threshold as in \cite{F22}. Nevertheless, weaker factor strength does require a larger sample size $T$ to achieve the consistency.

\end{remark}

\subsection{Convergence of mPANIC with heterogeneous factor strength}\label{sec4.2}

Similarly to Section \ref{sec4.1}, we replace Assumptions \ref{ass:3.5}(iii), \ref{ass:3.7}, and \ref{ass:3.8} by the following assumptions.

\medskip

\begin{assumption}\label{ass:4.4}	

{\em (i)\ As $p_1,p_2,T\to \infty$, 
\begin{eqnarray}
&&\left(p_1^{\alpha_{R,1}-\alpha_{R,r_1}}\vee p_2^{\alpha_{C,1}-\alpha_{C,r_2}}\right)\left(	p_1^{1/2-\alpha_{R,1}}p_2^{1-\alpha_{C,1}}+p_2^{1/2-\alpha_{C,1}}p_1^{1-\alpha_{R,1}}\right) \to 0,\notag\\
&&\left(p_1^{\alpha_{R,1}-\alpha_{R,r_1}} p_2^{\alpha_{C,1}-\alpha_{C,r_2}}\right)^2	p_1^{2-2\alpha_{R,1}}p_2^{2-2\alpha_{C,1}}=O(T),\notag\\
&&\left(p_1^{\alpha_{R,1}-\alpha_{R,r_1}} p_2^{\alpha_{C,1}-\alpha_{C,r_2}}\right)^{1/2}	\left(\frac{s_p}{p_1^{\alpha_{R,1}}}\frac{s_C}{p_2^{\alpha_{C,1}}}\right)^{1/2}T^{1/4}=O(1).\notag
\end{eqnarray}
} 
	
{\em (ii)\ There exists a positive constant $c_{10}$ such that
\begin{equation}\label{eq4.6}
{\sf E}\left\Vert \mathbf{B}_R^{-1}{\boldsymbol R}^{^\intercal}{\boldsymbol E}_t^\circ{\boldsymbol C}\mathbf{B}_C^{-1}\right\Vert^4\leq c_{10}\quad \text{and}\quad {\sf E}\left\Vert \frac{{\boldsymbol E}_t^\dag}{s_R^{1/2}s_C^{1/2}}\right\Vert^4+{\sf E}\left\Vert\mathbf{S}_R^{-1} {\boldsymbol R}^{^\intercal}{\boldsymbol E}_t^\dag{\boldsymbol C}\mathbf{S}_C^{-1}\right\Vert^4\leq c_{10}t^2
\end{equation}
for any $1\leq t\leq T$, in which $\mathbf{S}_R = \mathrm{diag}\left(s_R^{\alpha_{R,1}/2},\ldots,s_R^{\alpha_{R,r_1}/2} \right)$ and $\mathbf{S}_C = \mathrm{diag}\left(s_C^{\alpha_{C,1}/2},\ldots,s_C^{\alpha_{C,r_2}/2} \right)$. }
	
{\em (iii)\ Letting $e_{t,(i,j)}^\dag$ be the $(i,j)$-entry of ${\boldsymbol E}_t^\dag$, there exists a positive constant $c_{11}$ such that
\[
\max_{1\leq i\leq s_R}{\sf E}\left\Vert \sum_{j=1}^{p_2}e_{t,(i,j)}^\dag{\boldsymbol C}_{j\bullet}\mathbf{S}_C^{-1}\right\Vert^4\leq c_{11}t^2\quad \text{and}\quad\max_{1\leq j\leq s_C}{\sf E}\left\Vert \sum_{i=1}^{p_1}e_{t,(i,j)}^\dag{\boldsymbol R}_{i\bullet}\mathbf{S}_R^{-1}\right\Vert^4\leq c_{11}t^2
\]
for any $1\leq t\leq T$.}
		
{\em (iv)\ The eigenvalues of ${\boldsymbol\Sigma}_R^{\ast^{1/2}}{\sf E}\left[ \Delta {\boldsymbol F}_t{\boldsymbol\Sigma}_C^{\ast,1} \Delta {\boldsymbol F}_t^{^\intercal}\right]{\boldsymbol\Sigma}_R^{\ast^{1/2}}$ are positive, bounded and distinct. The same condition holds for ${\boldsymbol\Sigma}_C^{\ast^{1/2}} {\sf E}\left[ \Delta {\boldsymbol F}^{^\intercal}_t{\boldsymbol\Sigma}_R^{\ast,1} \Delta {\boldsymbol F}_t\right] {\boldsymbol\Sigma}_C^{\ast^{1/2}}$. }

\end{assumption}

\medskip

\begin{remark}\label{re:4.3}

Assumption \ref{ass:4.4}(i) imposes some restrictions on the relative growth rates of $p_1,p_2,T$, where $p_1^{\alpha_{R,1}-\alpha_{R,r_1}}\vee p_2^{\alpha_{C,1}-\alpha_{C,r_2}}$ or $p_1^{\alpha_{R,1}-\alpha_{R,r_1}} p_2^{\alpha_{C,1}-\alpha_{C,r_2}}$ is involved to accommodate the discrepancy between the strongest and weakest factor strengths. These conditions indicate that both $\alpha_{R,r_1}$ and $\alpha_{C,r_2}$ need to be greater than $0.5$ as discussed in Remark \ref{re:3.6}. The high-level conditions in Assumption \ref{ass:4.4}(ii)(iii) extend Assumption \ref{ass:3.8}(ii)(iii) to the case of heterogeneous factor strength whereas Assumption \ref{ass:4.4}(iv) is comparable to Assumption \ref{ass:3.5}(iii).

\end{remark}

\medskip

Define
\begin{eqnarray*}
\overline{\boldsymbol H}_R^\ast&=&\left(\frac{1}{p_2^{\alpha_{C,1}}T}\sum_{t=1}^{T}\Delta {\boldsymbol F}_{t}{\boldsymbol C}^{^\intercal}{\boldsymbol C}\Delta{\boldsymbol F}_{t}^{^\intercal}\right) \left({\boldsymbol R}^{^\intercal}\overline{\boldsymbol R}\mathbf{B}_R^{-1}\right)\check{\boldsymbol V}_R^{\ast^{-1}},\notag\\
\overline{\boldsymbol H}_C^\ast&=&\left(\frac{1}{p_1^{\alpha_{R,1}}T}\sum_{t=1}^{T}\Delta{\boldsymbol F}_{t}^{^\intercal}{\boldsymbol R}^{^\intercal}{\boldsymbol R}\Delta{\boldsymbol F}_{t}\right)\left({\boldsymbol C}^{^\intercal}\overline{\boldsymbol C}\mathbf{B}_C^{-1}\right)\check{\boldsymbol V}_C^{\ast^{-1}},
\end{eqnarray*}
where $\check{\boldsymbol V}_R^\ast = p_2^{-\alpha_{C,1}}\mathbf{B}_R^{-2}\overline{\boldsymbol V}_R$,  $\check{\boldsymbol V}_C^\ast = p_1^{-\alpha_{R,1}}\mathbf{B}_C^{-2}\overline{\boldsymbol V}_C$, $\overline{\boldsymbol V}_R$ and $\overline{\boldsymbol V}_C$ are defined as in \eqref{eq2.11}. The following theorem extends Theorems \ref{thm:3.4}--\ref{thm:3.6} to the case of heterogeneous factor strength.

\medskip

\begin{theorem}\label{thm:4.2}

Suppose that Assumptions \ref{ass:3.5}(i)(ii), \ref{ass:3.6}, \ref{ass:4.1}(i) and \ref{ass:4.4} are satisfied. 

(i) The mPANIC estimators of the factor loading matrices have the following mean squared convergence: 
\begin{eqnarray}
\left\Vert \left(\check{{\boldsymbol R}}^\ast-{\boldsymbol R}\overline{{\boldsymbol H}}_R^\ast\right)\mathbf{B}_R^{-1}\right\Vert_F&=&O_P\left(\xi_R^\dag(p_1,p_2,T)\right),\label{eq4.7}\\
\left\Vert \left(\check{{\boldsymbol C}}^\ast-{\boldsymbol C}\overline{\boldsymbol H}_C^\ast\right)\mathbf{B}_C^{-1}\right\Vert_F&=&O_P\left(\xi_C^\dag(p_1,p_2,T)\right),\label{eq4.8}
\end{eqnarray}
where $\check{{\boldsymbol R}}^\ast=\overline{{\boldsymbol R}}\mathbf{B}_R$, $\check{{\boldsymbol C}}^\ast=\overline{{\boldsymbol C}}\mathbf{B}_C$,
\begin{eqnarray}
\xi_R^\dag(p_1,p_2,T)&=&p_1^{\alpha_{R,1}-\alpha_{R,r_1}}\left(p_1^{1/2-\alpha_{R,1}}p_2^{1-\alpha_{C,1}}+p_1^{1-\alpha_{R,1}}p_2^{1/2-\alpha_{C,1}}T^{-1/2}\right),\notag\\
\xi_C^\dag(p_1,p_2,T)&=&p_2^{\alpha_{C,1}-\alpha_{C,r_2}}\left(p_2^{1/2-\alpha_{C,1}}p_1^{1-\alpha_{R,1}}+p_2^{1-\alpha_{C,1}}p_1^{1/2-\alpha_{R,1}}T^{-1/2}\right).\notag
\end{eqnarray}

(ii) The factor matrix estimators have the following uniform convergence property:
\begin{equation}\label{eq4.9}
\max_{1\leq t\leq T}\left\|\overline{{\boldsymbol F}}_t - \check{\nu}_{R,1}^{\ast^{1/2}}\check{\boldsymbol V}_R^{\ast^{-1/2}}\overline{{\boldsymbol H}}_R^{\ast^{-1}}{\boldsymbol F}_t\left(\overline{{\boldsymbol H}}_C^{\ast^{-1}}\right)^{^\intercal}\check{\boldsymbol V}_C^{\ast^{-1/2}}\right\|_F=O_P\left(\varpi_1^\ast(p_1,p_2,T)+\varpi_2^\ast(p_1,p_2,T)\right),
\end{equation}
where $\check{\nu}_{R,1}^\ast=p_1^{-\alpha_{R,1}}p_2^{-\alpha_{C,1}}\overline\lambda_{R,1}$ is positive and bounded w.p.a.1,
\begin{eqnarray}
\varpi_1^\ast(p_1,p_2,T)&=&\left(p_1^{\alpha_{R,1}-\alpha_{R,r_1}}\vee p_2^{\alpha_{C,1}-\alpha_{C,r_2}}\right)\left(p_1^{1/2-\alpha_{R,1}} p_2^{1-\alpha_{C,1}}+p_2^{1/2-\alpha_{C,1}}p_1^{1-\alpha_{R,1}}\right)T^{1/2},\notag\\
\varpi_2^\ast(p_1,p_2,T)&=&\left(p_1^{\alpha_{R,1}-\alpha_{R,r_1}} p_2^{\alpha_{C,1}-\alpha_{C,r_2}}\right)\left(\frac{s_R^{1/2}}{p_1}\right)^{\alpha_{R,1}}\left(\frac{s_C^{1/2}}{p_2}\right)^{\alpha_{C,1}}T^{3/4}.\notag
\end{eqnarray}

(iii) If, in addition, Assumption \ref{ass:3.4} continues to hold when $e_{t,(i,j)}$ is replaced by $\varepsilon_{t,(i,j)}$ and $r_1\wedge r_2\geq1$, 
\[
p_1^{\alpha_{R,1}-\alpha_{R,r_1}}\xi_R^\dag(p_1,p_2,T)\to 0\quad \text{and}\quad p_2^{\alpha_{C,1}-\alpha_{C,r_1}}\xi_C^\dag(p_1,p_2,T)\to 0,
\]
we have
$$
{\sf P}\left(\overline{r}_1=r_1\right)\rightarrow1 \quad \text{and}\quad {\sf P}\left(\overline{r}_2=r_2\right)\rightarrow1
$$
for any $K_1\in (r_1, \lfloor d_R\underline{p}_1\rfloor - 2r_1]$ and $K_2\in (r_2, \lfloor d_C\underline{p}_2\rfloor - 2r_2]$.

\end{theorem}

\medskip

\begin{remark}\label{re:4.4}

As discussed in Remark \ref{re:4.2}(i)(ii), the mean squared convergence rates in Theorem \ref{thm:4.2}(i) and the uniform convergence rate in Theorem \ref{thm:4.2}(ii) are slower than those in Theorems \ref{thm:3.4} and \ref{thm:3.5} due to varying (weak) factor strengths. In addition, since the mPANIC estimation is built on the eigenanalysis of stationary matrix time series (after taking the first-order difference), the rates in (\ref{eq4.7})--(\ref{eq4.9}) are also slower than those in (\ref{eq4.3})--(\ref{eq4.5}). The extra condition in Theorem \ref{thm:4.2}(iii) indicates that when $\alpha_{R,1}=\alpha_{C,1} = 1$, the weakest factor strength must exceed $0.75$ to ensure that the proposed ratio criterion can consistently estimate the number of weak factors in mPANIC. In contrast, \cite{F22} requires the factor
strength to be greater than $0.5$. This discrepancy arises because \cite{F22} assumes the factor loadings are sparse and the rotation matrix is an identity matrix, which can be achieved when the covariance matrices of both the factor loadings and common factors are diagonal. By exploiting this additional structure, specifically, through the use of partial sums of eigenvectors to amplify the eigen-gap, their eigenvalue ratio estimate accommodates factors with strength as low as $0.5$. In our setting, where the factors follow a matrix error correction process, such mutual orthogonality among the factors is not achievable.

\end{remark}


\section{Numerical studies}\label{sec5}
\renewcommand{\theequation}{5.\arabic{equation}}
\setcounter{equation}{0}

In this section we provide both Monte-Carlo simulation and empirical studies to examine the finite-sample performance of the proposed model and methods.


\subsection{Simulation studies}\label{sec5.1}

To generate common factors, we adopt model \eqref{eq2.2} for integrated full-rank $\boldsymbol{F}_t$, and \eqref{eq2.7} for cointegrated $\boldsymbol{F}_t$, where we set $r_1=r_2=2$ and the initial value $\boldsymbol{F}_0 = 0$. In \eqref{eq2.2}, entries of $\boldsymbol{U}_t$ are mutually independent AR(1) processes defined by
$$
u_{t,(i,j)} = 0.3 u_{t-1,(i,j)} + \varepsilon_{t,(i,j)}^u \quad \text{with}\quad u_{0,(i,j)}=0,\quad \varepsilon_{t,(i,j)}^u\stackrel{i.i.d.}\sim {\sf N}(0,1).
$$
In (\ref{eq2.7}), entries of $\boldsymbol{V}_t$ are mutually independent sequences of i.i.d. standard normal variables, ${\boldsymbol\alpha}_1 = [-0.1,0.1]^{^\intercal}$, ${\boldsymbol\beta}_1 = [1,-1]^{^\intercal}$, ${\boldsymbol\alpha}_2 = [0.1,-0.1]^{^\intercal}$, ${\boldsymbol\beta}_2 = [1,-1]^{^\intercal}$. To generate factor loadings, we set $\mathbf{R} = \mathbf{U}_R\mathbf{B}_R$ and $\mathbf{C} = \mathbf{V}_C\mathbf{B}_C$, where $\mathbf{U}_R$ and $\mathbf{V}_C$ are random orthonormal matrices of dimensions $p_1 \times r_1$ and $p_2 \times r_2$, respectively, which are obtained via the QR decomposition of random matrices with i.i.d. standard normal entries. We consider three settings for the factor loadings with different levels of factor strength:
\begin{description}

\item (i) $\alpha_{R,1} = \alpha_{C,1} =\alpha_{R,2} = \alpha_{C,2} = 1$;

\item (ii) $\alpha_{R,1} = \alpha_{C,1} = 1$, $\alpha_{R,2} = \alpha_{C,2} = 0.8$;

\item (iii) $\alpha_{R,1} = \alpha_{C,1} = 1$, $\alpha_{R,2} = \alpha_{C,2} = 0.6$. 

\end{description}

\noindent Weak factors (or factor loadings) exist in cases (ii) and (iii). The idiosyncratic error matrix $\boldsymbol{E}_t$ is generated by
$$
\boldsymbol{E}_t = 0.3 \boldsymbol{E}_{t-1} + {\boldsymbol\Gamma}_R^{1/2} {\boldsymbol \Xi}_t {\boldsymbol\Gamma}_C^{1/2},$$
where $\boldsymbol{E}_0 = {\bf O}_{p_1\times p_2}$, $\{{\boldsymbol \Xi}_t\}$ is a sequence of matrices containing i.i.d. standard normal elements, ${\boldsymbol\Gamma}_R$ and  ${\boldsymbol\Gamma}_C$ are $p_1\times p_1$ and $p_2\times p_2$ matrices with the $(i,j)$-entry being $0.5^{|i-j|}$.

\smallskip

For each setting, we repeat the simulation $1000$ times. Since both common factors and factor loadings are estimated only up to some rotation matrices, we report the root mean squared error between the estimated and true projection matrices (onto the corresponding column spaces). For instance, we compute the root mean squared error of row factor loading matrix estimation by
$$
\text{RMSE}(\widehat{\mathbf{R}}) = \frac{1}{1000}\sum_{i=1}^{1000} \left\|\widehat{\mathbf{R}}^{(i)}\widehat{\mathbf{R}}^{(i)^\intercal} - \mathbf{U}_R\mathbf{U}_R^{^\intercal} \right\|_F,
$$
where $\widehat{\mathbf{R}}^{(i)}$ denotes the estimate of $\mathbf{R}$ in the $i$-th replication, and $\mathbf{U}_R$ the matrix of left singular vectors of $\mathbf{R}$. $\text{RMSE}(\widehat{\mathbf{C}})$ and $\text{RMSE}(\widehat{\mathbf{F}})$ are defined similarly. When implementing the eigenvalue ratio estimators \eqref{eq2.5}, \eqref{eq2.6} \eqref{eq2.12} and \eqref{eq2.13}, we set $K_1=K_2=10$.

\smallskip

Table \ref{tab:1} presents the simulation results for both mPCA and mPANIC under various settings, where $\text{Mean}(\widehat{r}_1)$ and $\text{Mean}(\widehat{r}_2)$ represent the average estimated number of row and column factors over 1000 simulation replications, $\text{CP}(\widehat{r}_1)$ and $\text{CP}(\widehat{r}_2)$ denote the corresponding proportions of correct estimations. Several key insights can be drawn from the simulation results. First, we observe that estimation errors for factor loadings and common factors decrease as the sample size increases, confirming the consistency of both the mPCA and mPANIC methods. Second, mPCA consistently yields smaller estimation errors than mPANIC, as it directly targets the nonstationary time series, resulting in faster convergence. Third, as the weakest factor strength (i.e., $\alpha_{R,2}$ and $\alpha_{C,2}$) decreases, estimation errors increase and the accuracy in estimating the number of factors deteriorates. When factor strengths are $1$ and $0.8$ in case (ii), both the mPCA and mPANIC methods perform well in recovering the true number of factors, with $\text{CP}(\widehat{r}_1)$ and $\text{CP}(\widehat{r}_2)$ close to 1. However, when the weakest factor strength is $0.6$ in case (iii), $\text{CP}(\widehat{r}_1)$ and $\text{CP}(\widehat{r}_2)$ under mPANIC are close to 0, whereas mPCA achieves a higher CP of around 0.9, which is consistent with the implication of Theorems \ref{thm:4.1}(iii) and \ref{thm:4.2}(iii). In fact, to ensure consistent selection of the factor number in mPANIC, the current simulation design requires both $\alpha_{R,2}$ and $\alpha_{C,2}$ to be larger than $0.75$.

\begin{table}
	\centering
	\caption{Simulation results}
	\begin{tabular}{l| ccccccc}	
		\hline\hline
				\multicolumn{8}{c}{Panel A: mPCA estimation} \\
				\hline
				&$\text{RMSE}(\widehat{\mathbf{R}})$ &$\text{RMSE}(\widehat{\mathbf{C}})$ &$\text{RMSE}(\widehat{\mathbf{F}})$ &$\text{Mean}(\widehat{r}_1)$ &$\text{CP}(\widehat{r}_1)$ &$\text{Mean}(\widehat{r}_2)$ &$\text{CP}(\widehat{r}_2)$\\
								\hline
		& \multicolumn{7}{c}{ $\alpha_{R,1} = \alpha_{C,1} = 1$, $\alpha_{R,2} = \alpha_{C,2} = 1$} \\
		 $T=50,\,p_1=p_2=30$& 0.008  & 0.008  & 0.041  & 2.000  & 1.000  & 1.998  & 0.998  \\
		 $T=100,\,p_1=p_2=30$& 0.004  & 0.004  & 0.028  & 2.000  & 1.000  & 2.000  & 1.000  \\
		$T=100,\,p_1=p_2=60$ & 0.003  & 0.003  & 0.014  & 2.000  & 1.000  & 2.000  & 1.000  \\
		& \multicolumn{7}{c}{$\alpha_{R,1} = \alpha_{C,1} = 1$, $\alpha_{R,2} = \alpha_{C,2} = 0.8$} \\
		$T=50,\,p_1=p_2=30$ &     0.012  & 0.012  & 0.061  & 1.986  & 0.986  & 1.980  & 0.980  \\
	    $T=100,\,p_1=p_2=30$ &   0.006  & 0.006  & 0.042  & 1.998  & 0.998  & 1.994  & 0.994  \\
	    $T=100,\,p_1=p_2=60$ &   0.004  & 0.004  & 0.023  & 1.998  & 0.998  & 1.999  & 0.999  \\
		& \multicolumn{7}{c}{$\alpha_{R,1} = \alpha_{C,1} = 1$, $\alpha_{R,2} = \alpha_{C,2} = 0.6$} \\
		$T=50,\,p_1=p_2=30$ &    0.018  & 0.019  & 0.100  & 1.867  & 0.867  & 1.873  & 0.873  \\
		$T=100,\,p_1=p_2=30$ &    0.009  & 0.009  & 0.069  & 1.946  & 0.946  & 1.943  & 0.943  \\
		$T=100,\,p_1=p_2=60$ &    0.007  & 0.007  & 0.043  & 1.943  & 0.943  & 1.943  & 0.943  \\
		\hline
		\multicolumn{8}{c}{Panel B: mPANIC estimation} \\
		\hline
		& \multicolumn{7}{c}{ $\alpha_{R,1} = \alpha_{C,1} = 1$, $\alpha_{R,2} = \alpha_{C,2} = 1$} \\
	    $T=50,\,p_1=p_2=30$ &     0.076  & 0.076  & 0.055  & 2.000  & 1.000  & 2.000  & 1.000  \\
		$T=100,\,p_1=p_2=30$ &     0.065  & 0.065  & 0.041  & 2.000  & 1.000  & 2.000  & 1.000  \\
		$T=100,\,p_1=p_2=60$ &     0.038  & 0.038  & 0.020  & 2.000  & 1.000  & 2.000  & 1.000  \\
		& \multicolumn{7}{c}{$\alpha_{R,1} = \alpha_{C,1} = 1$, $\alpha_{R,2} = \alpha_{C,2} = 0.8$} \\
		$T=50,\,p_1=p_2=30$ &     0.143  & 0.142  & 0.087  & 1.985  & 0.985  & 1.993  & 0.993  \\
		$T=100,\,p_1=p_2=30$ &     0.126  & 0.127  & 0.063  & 1.999  & 0.999  & 1.999  & 0.999  \\
		$T=100,\,p_1=p_2=60$ &     0.079  & 0.080  & 0.033  & 2.000  & 1.000  & 2.000  & 1.000  \\
		& \multicolumn{7}{c}{$\alpha_{R,1} = \alpha_{C,1} = 1$, $\alpha_{R,2} = \alpha_{C,2} = 0.6$} \\
		$T=50,\,p_1=p_2=30$ &     0.319  & 0.319  & 0.173  & 1.044  & 0.044  & 1.049  & 0.049  \\
		$T=100,\,p_1=p_2=30$ &     0.291  & 0.293  & 0.124  & 1.013  & 0.013  & 1.015  & 0.015  \\
		$T=100,\,p_1=p_2=60$ &     0.203  & 0.201  & 0.072  & 1.012  & 0.012  & 1.007  & 0.007  \\
			\hline\hline
	\end{tabular}\label{tab:1}
\end{table}


\subsection{An empirical application}\label{sec5.2}

In this section, we apply the proposed model and methodology to analyze international trade flows. By modeling possibly nonstationary trade volumes directly, we aim to examine the evolution pattern of global trade and the international trade network, and further assess the impact of major events such as the global financial crisis and the COVID-19 pandemic. Our empirical analysis is based on monthly bilateral export volumes of commodity goods among 23 countries and regions, obtained from the International Monetary Fund’s Direction of Trade Statistics. The data cover free-on-board export values denominated in US dollars over a 273-month period from January 2000 to September 2022. The raw trade volume data are log-transformed, whereby the first differences correspond to the growth rates of international trade flows. The countries and regions include Australia (AUS), Canada (CAN), Mainland China (CHN), Denmark (DNK), Finland (FIN), France (FRA), Germany (DEU), Hong Kong China (HKG), Indonesia (IDN), Ireland (IRL), Italy (ITA), Japan (JPN), Korea (KOR), Malaysia (MYS), Mexico (MEX), Netherlands (NLD), New Zealand (NZL), Singapore (SGP), Spain (ESP), Sweden (SWE), Thailand (THA), United Kingdom (GBR), and United States (USA). 

\smallskip

\begin{figure}[htb!]
	\centering
	{\includegraphics[width=14cm]{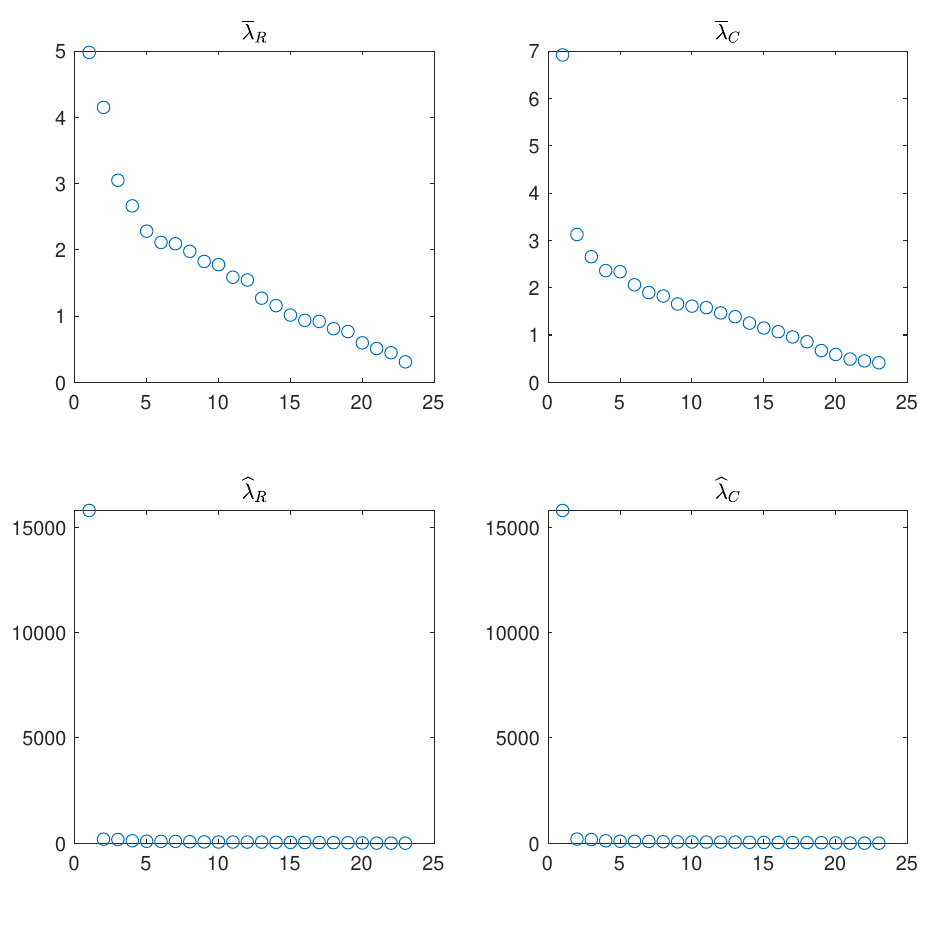}}
	\caption{Sorted eigenvalues of the sample row (left panels) and column (right panels) covariance matrices corresponding to models \eqref{eq2.2} (bottom panels) and \eqref{eq2.7} (top panels).}\label{Fig1}
\end{figure}

We first estimate the size of matrix factors. The proposed eigenvalue ratio criterion yields $\widehat{r}_1 = \widehat{r}_2 = 1$ under model \eqref{eq2.2}, and $\overline{r}_1 = 2,\ \overline{r}_2 = 1$ under model \eqref{eq2.7}. Figure \ref{Fig1} displays the sorted eigenvalues of the sample row (left) and column (right) covariance matrices for model \eqref{eq2.7} (top panels) and model \eqref{eq2.2} (bottom panels). As shown in Figure \ref{Fig1}, the eigenvalue gaps are more pronounced under model \eqref{eq2.2}, facilitating the identification of the number of factors. In contrast, the eigenvalues under model \eqref{eq2.7}, where the data are differenced, are more tightly clustered, making it more challenging to determine the true factor dimensionality. Furthermore, for the two estimated factors under model \eqref{eq2.7}, we apply the model-free method proposed in \cite{ZRY19} to test for cointegration. The result, however, indicates that these two factors are not cointegrated. Therefore, we proceed with model \eqref{eq2.2} in the subsequent analysis.

\smallskip

\begin{figure}[htb!]
	\centering
	{\includegraphics[width=14cm]{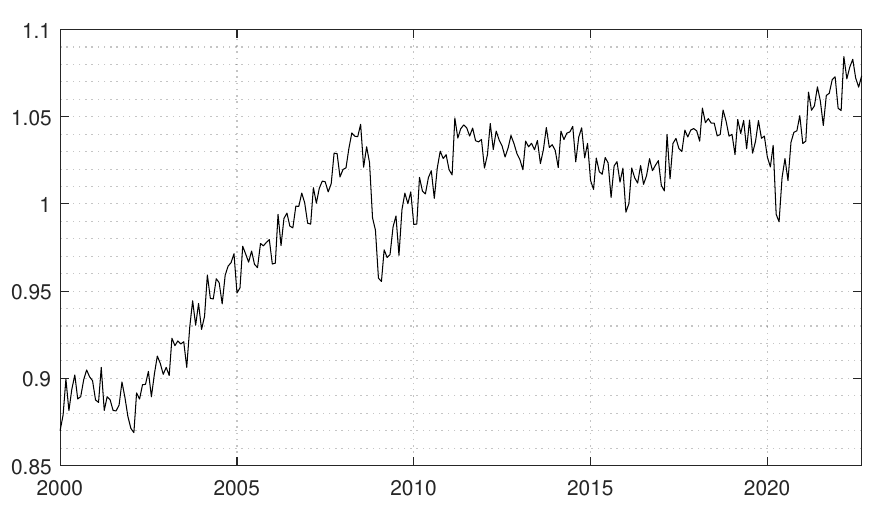}}
	\caption{Time series plots of estimated common factor under model \eqref{eq2.2}.}\label{Fig2}
\end{figure}

Figure \ref{Fig2} plots the time series of the estimated common factor. From a modeling perspective, the strong trending pattern in the factor reflects its nonstationary feature. Model \eqref{eq2.2} explicitly accommodates integrated components, thereby capturing persistent global movements in international trade. In this context, the estimated factor can be interpreted as a global demand index: bilateral trade volumes fall worldwide when it drops; and international trade expands globally when it rises. Thus, the model effectively identifies the dominant source of co-movement in international trade flows, highlighting its relevance for capturing global shocks in a parsimonious and interpretable manner. Furthermore, the dynamic behavior of the estimated factor aligns closely with several major global events. The plot shows a persistent upward trend beginning in 2002, which may be attributed to China's accession to the World Trade Organization and the subsequent acceleration in its trade activities with major economies (e.g., \citealp{HO16}). Notably, it exhibits a sharp drop around 2009, corresponding to the so-called ``Great Trade Collapse'' due to the 2008 Global Financial Crisis. The estimated factor also drops sharply in early 2020 due to the outbreak of COVID-19 pandemic, followed by a quick rebound, reflecting the V-shaped pattern of global trade disruption and recovery. In addition, a smaller but visible decline is observed around 2015--2016, coinciding with the documented slowdown in global trade during that period (e.g., \citealp{CMR20}).

\smallskip

\begin{figure}[htb!]
	\centering
	{\includegraphics[width=14cm]{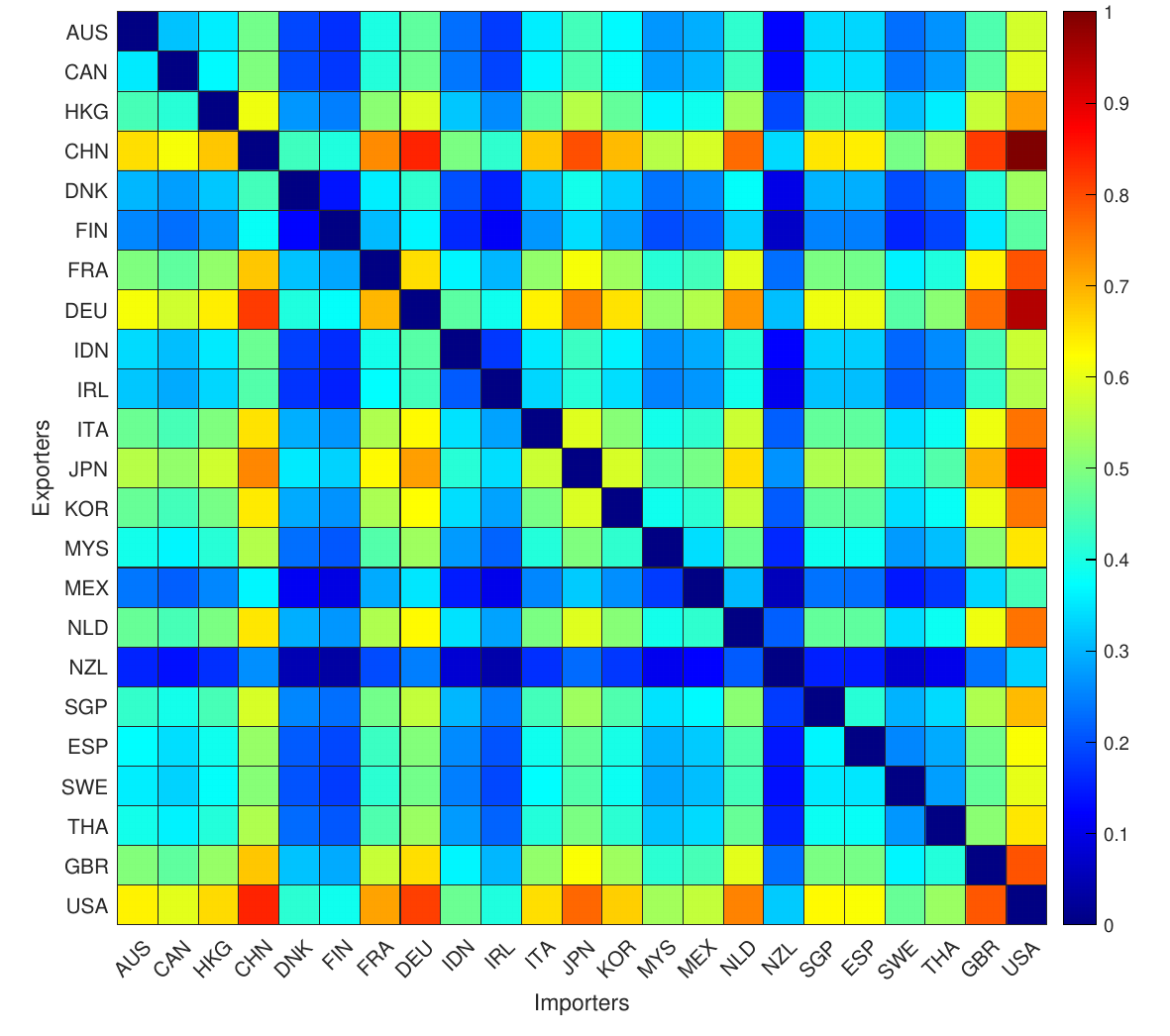}}

 	\caption{Heatmap of the normalized trade network matrix ``$\widehat{{\boldsymbol R}}\widehat{{\boldsymbol C}}^{^\intercal}$''.}\label{Fig3}
\end{figure}

Figure \ref{Fig3} displays the heatmap of the normalized matrix $\widehat{\boldsymbol R}\widehat{\boldsymbol C}^{^\intercal}$, where each element has been rescaled to lie between 0 and 1 for ease of comparison. This matrix captures the structure of bilateral trade linkages across 23 economies, conditional on the global common factor estimated from monthly international trade data. Each entry $(i,j)$ in $\widehat{\boldsymbol R}\widehat{\boldsymbol C}^{^\intercal}$ reflects the relative intensity of trade between exporter $i$ and importer $j$ in response to global shocks, such as aggregate demand or systemic disruptions. As $\widehat{\boldsymbol R}$ and $\widehat{\boldsymbol C}$ represent country-specific sensitivities to the global factor, this matrix effectively summarizes how global economic conditions propagate through the trade network. The heatmap reveals that the strongest bilateral linkage arises between China (CHN) and the United States (USA), highlighting their pivotal roles as global trade hubs. This is consistent with their status as the two largest trading nations over the past decades.

\section{Conclusion}\label{sec6}
\renewcommand{\theequation}{6.\arabic{equation}}
\setcounter{equation}{0}

We have introduced a general matrix factor model to tackle large-scale trending matrix time series, making full case of the intrinsic matrix data structure and aiming to improve the estimation convergence. This is substantially different from the conventional vectorized factor model framework which transforms the matrix time series observations into vectors and often results in loss of sample information. We propose two types of PCA-based estimation techniques: mPCA and mPANIC, according to different nonstationary features in matrix common and idiosyncratic components. Under various model assumptions, we establish the convergence theory for the estimated factor loading matrices and nonstationary factor matrices and the consistency of the factor number estimation via the eigenvalue ratio criterion. In particular, we allow the existence of weak factors in the proposed factor model structure, where the factor strength can be heterogeneous. The obtained mean squared convergence rates (for factor loading matrix estimation) and uniform convergence rates (for factor matrix estimation) depend on the matrix size $(p_1,p_2)$, time series length $T$ and the weak factor strength. The Monte-Carlo simulation study justifies the estimation convergence in finite samples. The empirical application to international trade flows shows that the estimated factor can be viewed as the global trade index and the estimated row and column loading matrices are useful to describe bilateral trade linkages between major economies (such as China and the United States).


\section*{Supplement}

The supplemental document contains proofs of the main theorems (in Appendix A) and some technical lemmas with proofs (in Appendix B).


\newpage

\begin{center}
	{\Large{Supplement to \textquotedblleft Factor Models of Matrix-Valued Time Series: \\ 
			Nonstationarity and Cointegration\textquotedblright }} 
	
	{\small \medskip }
	
	{\small \textsc{Degui Li}$^{*}$, \textsc{Yayi Yan}$^{\dag}$, \textsc{Qiwei Yao}$^{\ddag}$ }
	
	{\small \medskip }
	
	{\small\em $^{*}$University of Macau, $^{\dag}$ Shanghai University of
		Finance and Economics, $^{\ddag}$London School of Economics}
	\bigskip
\end{center}

This supplemental file contains two appendices. Specifically, we provide the detailed proofs of the main theoretical results in Appendix A and present some technical lemmas together with their proofs in Appendix B. Throughout the proofs, we let $C$ denote a generic positive constant whose value may change from line to line.

\section*{Appendix A:\ Proofs of the main results}
\renewcommand{\theequation}{A.\arabic{equation}}
\setcounter{equation}{0}

\noindent{\bf Proof of Theorem \ref{thm:3.1}}.\ \ By the definition of $\widehat{\mathbf R}$ and $\widehat{\mathbf V}_R$, we have
\begin{equation}\label{eqA.1}
	\left(\frac{1}{p_1^{\alpha_R}p_2^{\alpha_C}T}\widehat{\boldsymbol\Omega}_R\right)\widehat{\mathbf R} =\widehat{\mathbf R}\widehat{\mathbf V}_R.
\end{equation}
By (\ref{eq2.1}) and (\ref{eq2.3}), we readily have that
\begin{eqnarray}
\widehat{\boldsymbol\Omega}_R&=&\frac{1}{T}\sum_{t=1}^{T}\left({\boldsymbol Z}_{t}{\boldsymbol Z}_{t}^{^\intercal}+{\boldsymbol Z}_{t}{\boldsymbol E}_{t}^{^\intercal}+{\boldsymbol E}_{t}{\boldsymbol Z}_{t}^{^\intercal}+{\boldsymbol E}_{t}{\boldsymbol E}_{t}^{^\intercal}\right)\notag\\
&=&\frac{1}{T}\sum_{t=1}^{T}{\boldsymbol R}{\boldsymbol F}_{t}{\boldsymbol C}^{^\intercal}{\boldsymbol C}{\boldsymbol F}_{t}^{^\intercal}{\boldsymbol R}^{^\intercal}+{\boldsymbol\Lambda}_{1}+{\boldsymbol\Lambda}_{2}+{\boldsymbol\Lambda}_{3},\label{eqA.2}
\end{eqnarray}
where
\[
{\boldsymbol\Lambda}_{1}=\frac{1}{T}\sum_{t=1}^{T}{\boldsymbol Z}_{t}{\boldsymbol E}_{t}^{^\intercal},\ \ {\boldsymbol\Lambda}_{2}=\frac{1}{T}\sum_{t=1}^{T}{\boldsymbol E}_{t}{\boldsymbol Z}_{t}^{^\intercal},\ \ 
{\boldsymbol\Lambda}_{3}=\frac{1}{T}\sum_{t=1}^{T}{\boldsymbol E}_{t}{\boldsymbol E}_{t}^{^\intercal}.
\]
It follows from \eqref{eqA.1}--\eqref{eqA.2} and Lemma \ref{le:B.4} that
\begin{eqnarray*}
p_1^{\alpha_R/2}\widehat{{\boldsymbol R}}-{\boldsymbol R}\widehat{{\boldsymbol H}}_R&=&\frac{1}{p_1^{\alpha_R}p_2^{\alpha_C}T}(\boldsymbol{\Lambda}_{1}+\boldsymbol{\Lambda}_{2}+\boldsymbol{\Lambda}_{3})\left(p_1^{\alpha_R/2}\widehat{{\boldsymbol R}}\right)\widehat{{\boldsymbol V}}_{R}^{-1}.
\end{eqnarray*}

By Lemma \ref{le:B.1}, we readily have that
\begin{equation}\label{eqA.3}
\left\|\frac{1}{p_1^{\alpha_R}p_2^{\alpha_C}T}\boldsymbol{\Lambda}_{3} \right\|_F=O_P\left(p_1^{\frac{1}{2}-\alpha_R}p_2^{1-\alpha_C}T^{-1} + p_1^{1-\alpha_R}p_2^{\frac{1}{2}-\alpha_C}T^{-\frac{3}{2}}\right).
\end{equation}
Using the fact $\left\|{\boldsymbol R} \right\|_F = O(p_1^{\alpha_R/2})$ from Assumption \ref{ass:3.1}(ii) and Lemma \ref{le:B.2}, we have
\begin{equation}\label{eqA.4}
\left\|\frac{1}{p_1^{\alpha_R}p_2^{\alpha_C}T}(\boldsymbol{\Lambda}_{1}+\boldsymbol{\Lambda}_{2})\right\|_F = \frac{1}{p_1^{\alpha_R}p_2^{\alpha_C}T^2} \times O(p_1^{\alpha_R/2}) \times O_P\left(p_1^{1/2} p_2^{\alpha_C/2}T\right) = O_P\left(p_1^{\frac{1-\alpha_R}{2}}p_2^{-\frac{\alpha_C}{2}}T^{-1}\right).
\end{equation}
As $\left\|\widehat{{\boldsymbol R}} \right\|_F = O_P(1)$, by (\ref{eqA.3}), (\ref{eqA.4}), Lemma \ref{le:B.4} and $p_1^{1-\alpha_R}p_2^{1-\alpha_C}=o(T)$ in Assumption \ref{ass:3.3}(i), we complete the proof of (\ref{eq3.7}). The proof of (\ref{eq3.8}) is analogous and thus omitted here. \hfill$\blacksquare$

\bigskip

\noindent{\bf Proof of Theorem \ref{thm:3.2}}.\ \ Let $\widetilde{\boldsymbol R}$ and $\widetilde{\boldsymbol C}$ be defined as in (\ref{eq3.7}) and (\ref{eq3.8}). Note that 
\begin{eqnarray*}
\widehat{{\boldsymbol F}}_t&=&\widetilde\lambda_{R,1}^{1/2}\widetilde{\boldsymbol V}_R^{-1/2}\widehat{\boldsymbol R}^{^\intercal} {\boldsymbol X}_t\widehat{\boldsymbol C}\widetilde{\boldsymbol V}_C^{-1/2}\\
&=&\widehat\nu_{R,1}^{1/2} \widehat{\boldsymbol V}_R^{-1/2} \widehat{{\boldsymbol H}}_R^{-1}{\boldsymbol F}_t\left(\widehat{{\boldsymbol H}}_C^{-1}\right)^{^\intercal}\widehat{\boldsymbol V}_C^{-1/2}+\widetilde\lambda_{R,1}^{1/2}\widetilde{\boldsymbol V}_R^{-1/2}\left[\widehat{{\boldsymbol R}}^{^\intercal}\left({\boldsymbol R}-\widetilde{{\boldsymbol R}}\widehat{{\boldsymbol H}}_R^{-1}\right){\boldsymbol F}_t\left({\boldsymbol C}-\widetilde{{\boldsymbol C}}\widehat{{\boldsymbol H}}_C^{-1}\right)^{^\intercal}\widehat{{\boldsymbol C}}+\right.\\
&&\left.\widehat{{\boldsymbol R}}^{^\intercal} \left( {\boldsymbol R} - \widetilde{{\boldsymbol R}} \widehat{{\boldsymbol H}}_R^{-1} \right){\boldsymbol F}_t \left( \widehat{{\boldsymbol H}}_C^{-1}\right)^{^\intercal}\left(\widetilde{\mathbf C}^{^\intercal}\widehat{\mathbf C}\right)  + \left(\widehat{\mathbf R}^{^\intercal}\widetilde{\mathbf R}\right)\widehat{{\boldsymbol H}}_R^{-1} {\boldsymbol F}_t \left( {\boldsymbol C} - \widehat{{\boldsymbol C}} \widehat{{\boldsymbol H}}_C^{-1} \right)^{^\intercal} \widehat{{\boldsymbol C}}\right]\widetilde{\boldsymbol V}_C^{-1/2} +\\
&& \left(\widetilde\lambda_{R,1}^{1/2}p_1^{-\alpha_R/2}p_2^{-\alpha_C/2}\right)\widetilde{\boldsymbol V}_R^{-1/2}\left[\left( \widetilde{{\boldsymbol R}} - {\boldsymbol R} \widehat{{\boldsymbol H}}_R \right)^{^\intercal}  {\boldsymbol E}_t\left( \widetilde{{\boldsymbol C}} - {\boldsymbol C} \widehat{{\boldsymbol H}}_C \right)
+\left( \widetilde{{\boldsymbol R}} - {\boldsymbol R} \widehat{{\boldsymbol H}}_R \right)^{^\intercal}{\boldsymbol E}_t {\boldsymbol C} \widehat{{\boldsymbol H}}_C+\right. \\
&& \left.\widehat{{\boldsymbol H}}_R^{^\intercal} {\boldsymbol R}^{^\intercal}{\boldsymbol E}_t \left( \widetilde{{\boldsymbol C}} - {\boldsymbol C} \widehat{{\boldsymbol H}}_C\right)+\widehat{{\boldsymbol H}}_R^{^\intercal} {\boldsymbol R}^{^\intercal} {\boldsymbol E}_t {\boldsymbol C} \widehat{{\boldsymbol H}}_C\right]\widetilde{\boldsymbol V}_C^{-1/2}.
\end{eqnarray*}
By Lemma \ref{le:B.4}, we may show that 
\begin{equation}\label{eqA.5}
\left\vert \widehat\nu_{R,1}-\nu_{R,1}\right\vert=o_P(1),
\end{equation}
where $\nu_{R,1}$ denotes the maximum eigenvalue of ${\boldsymbol\Sigma}_R^{1/2}[\int_0^1 {\boldsymbol W}(u){\boldsymbol\Sigma}_C{\boldsymbol W}(u)^{^\intercal}du]{\boldsymbol\Sigma}_R^{1/2}$, which is positive and bounded with probability one due to Assumption \ref{ass:3.1}(iii). This indicates that $\widetilde\lambda_{R,1}$ is of order $p_1^{\alpha_R}p_2^{\alpha_C}$ {\em w.p.a.1}.

As $\max_{1\leq t\leq T}\|{\boldsymbol F}_t\|_F = O_P(\sqrt{T})$ (see the proof of Lemma \ref{le:B.3}), using Assumption \ref{ass:3.3}(i), Theorem \ref{thm:3.1} and Lemma \ref{le:B.5} , we have
\begin{eqnarray}
	&&\max_{1 \leq t \leq T} \left\| \widetilde\lambda_{R,1}^{1/2}\widetilde{\boldsymbol V}_R^{-1/2}\widehat{{\boldsymbol R}}^{^\intercal} \left( {\boldsymbol R} - \widetilde{{\boldsymbol R}} \widehat{{\boldsymbol H}}_R^{-1} \right) 	{\boldsymbol F}_t \left( {\boldsymbol C} - \widetilde{{\boldsymbol C}} \widehat{{\boldsymbol H}}_C^{-1}\right)^{^\intercal} \widehat{{\boldsymbol C}}\widetilde{\boldsymbol V}_C^{-1/2}
	\right\|_F \notag\\
	&=&O_P\left(p_1^{\alpha_R/2}p_2^{\alpha_C/2}\right)O_P\left(p_1^{-\alpha_R/2}p_2^{-\alpha_C/2}\right)O_P\left(p_1^{1/2}p_2^{-\alpha_C/2}T^{-1}\left(1 + p_1^{-\alpha_R/2}p_2^{1-\alpha_C/2} \right)\right)\notag\\
	&&O_P\left(T^{1/2}\right)O_P\left(p_2^{1/2}p_1^{-\alpha_R/2}T^{-1}\left(1 + p_2^{-\alpha_C/2}p_1^{1-\alpha_R/2} \right)\right)O_P\left(p_1^{-\alpha_R/2}p_2^{-\alpha_C/2}\right)\notag\\
	&=&o_P\left(p_1^{(1-\alpha_R)/2} p_2^{-\alpha_C/2}T^{-1/2}+ p_1^{(1-2\alpha_R)/2}p_2^{1-\alpha_C}T^{-1/2} \right)=o_P\left(\varphi_1(p_1,p_2,T)\right),\label{eqA.6}\\
	&&\max_{1 \leq t \leq T} 
	\left\| \widetilde\lambda_{R,1}^{1/2}\widetilde{\boldsymbol V}_R^{-1/2}\widehat{{\boldsymbol R}}^{^\intercal} \left( {\boldsymbol R} - \widetilde{{\boldsymbol R}} \widehat{{\boldsymbol H}}_R^{-1} \right){\boldsymbol F}_t \left( \widehat{{\boldsymbol H}}_C^{-1}\right)^{^\intercal}\left(\widetilde{\mathbf C}^{^\intercal}\widehat{\mathbf C}\right)\widetilde{\boldsymbol V}_C^{-1/2} \right\|_F\notag\\
	&=&O_P\left(p_1^{\alpha_R/2}p_2^{\alpha_C/2}\right) O_P\left(p_1^{-\alpha_R/2}p_2^{-\alpha_C/2}\right)O_P\left(p_1^{1/2}p_2^{-\alpha_C/2}T^{-1}\left(1 + p_1^{-\alpha_R/2}p_2^{1-\alpha_C/2} \right)\right)\notag\\
	&&O_P\left(T^{1/2}\right)O_P\left(p_2^{\alpha_C/2}\right)O_P\left(p_1^{-\alpha_R/2}p_2^{-\alpha_C/2}\right)\notag\\
	&=&O_P\left(p_1^{(1-\alpha_R)/2} p_2^{-\alpha_C/2}T^{-1/2}+ p_1^{(1-2\alpha_R)/2}p_2^{1-\alpha_C}T^{-1/2} \right)=O_P\left(\varphi_1(p_1,p_2,T)\right),\label{eqA.7}\\
	&&\max_{1 \leq t \leq T} 
	\left\| \widetilde\lambda_{R,1}^{1/2} \widetilde{\boldsymbol V}_R^{-1/2}\left(\widehat{\mathbf R}^{^\intercal}\widetilde{\mathbf R}\right)\widehat{{\boldsymbol H}}_R^{-1} {\boldsymbol F}_t \left( {\boldsymbol C} - \widehat{{\boldsymbol C}} \widehat{{\boldsymbol H}}_C^{-1} \right)^{^\intercal} \widehat{{\boldsymbol C}}\widetilde{\boldsymbol V}_C^{-1/2}	\right\| _F\notag\\
	&=&O_P\left(p_1^{\alpha_R/2}p_2^{\alpha_C/2}\right)O_P\left(p_1^{-\alpha_R/2}p_2^{-\alpha_C/2}\right)O_P\left(p_1^{\alpha_R/2}\right)O_P\left(T^{1/2}\right)\notag\\
	&&O_P\left(p_2^{1/2}p_1^{-\alpha_R/2}T^{-1}\left(1 + p_2^{-\alpha_C/2}p_1^{1-\alpha_R/2} \right)\right)O_P\left(p_1^{-\alpha_R/2}p_2^{-\alpha_C/2}\right)\notag\\
	&=&O_P\left(p_2^{(1-\alpha_C)/2} p_1^{-\alpha_R/2}T^{-1/2}+ p_2^{(1-2\alpha_C)/2}p_1^{1-\alpha_R}T^{-1/2} \right)=O_P\left(\varphi_1(p_1,p_2,T)\right).\label{eqA.8}
\end{eqnarray}
Meanwhile, by using Assumption \ref{ass:3.3}(i), Theorem \ref{thm:3.1} and Lemmas \ref{le:B.5} and \ref{le:B.6}, we have
\begin{eqnarray}
&&\max_{1 \leq t \leq T} \left\| \left(\widetilde\lambda_{R,1}^{1/2}p_1^{-\alpha_R/2}p_2^{-\alpha_C/2}\right)\widetilde{\boldsymbol V}_R^{-1/2}\left( \widetilde{{\boldsymbol R}} - {\boldsymbol R} \widehat{{\boldsymbol H}}_R \right)^{^\intercal}  {\boldsymbol E}_t\left( \widetilde{{\boldsymbol C}} - {\boldsymbol C} \widehat{{\boldsymbol H}}_C \right)
\widetilde{\boldsymbol V}_C^{-1/2}	\right\|_F \notag\\
&=& O_P\left(p_1^{-\alpha_R/2}p_2^{-\alpha_C/2}\right)O_P\left(p_1^{1/2}p_2^{-\alpha_C/2}T^{-1}\left(1 + p_1^{-\alpha_R/2}p_2^{1-\alpha_C/2} \right)\right)\notag\\
	&&O_P\left((p_1p_2)^{1/2}T^{1/4}\right)O_P\left(p_2^{1/2}p_1^{-\alpha_R/2}T^{-1}\left(1 + p_2^{-\alpha_C/2}p_1^{1-\alpha_R/2} \right)\right)O_P\left(p_1^{-\alpha_R/2}p_2^{-\alpha_C/2}\right)\notag\\
	&=&O_P\left(p_1^{-\alpha_R/2}p_2^{-\alpha_C/2}T^{1/4}\right)O_P\left(p_1^{1-\alpha_R/2}p_2^{-\alpha_C/2}T^{-1}+ p_1^{1-\alpha_R}p_2^{1-\alpha_C}T^{-1}\right)\notag\\
	&&O_P\left(p_2^{1-\alpha_C/2}p_1^{-\alpha_R/2}T^{-1}+ p_2^{1-\alpha_C}p_1^{1-\alpha_R}T^{-1}\right)\notag\\
	&=&o_P\left(p_1^{-\alpha_R/2}p_2^{-\alpha_C/2}T^{1/4}\right)=o_P\left(\varphi_2(p_1,p_2,T)\right),\label{eqA.9}\\
&&\max_{1 \leq t \leq T} \left\| \left(\widetilde\lambda_{R,1}^{1/2}p_1^{-\alpha_R/2}p_2^{-\alpha_C/2}\right)\widetilde{\boldsymbol V}_R^{-1/2} \left( \widetilde{{\boldsymbol R}} - {\boldsymbol R} \widehat{{\boldsymbol H}}_R \right)^{^\intercal}{\boldsymbol E}_t {\boldsymbol C} \widehat{{\boldsymbol H}}_C\widetilde{\boldsymbol V}_C^{-1/2}\right\|_F\notag\\
&=& O_P\left(p_1^{-\alpha_R/2}p_2^{-\alpha_C/2}\right)O_P\left(p_1^{1/2}p_2^{-\alpha_C/2}T^{-1}\left(1 + p_1^{-\alpha_R/2}p_2^{1-\alpha_C/2} \right)\right)\notag\\
&&O_P\left(p_1^{1/2}p_2^{\alpha_C/2}T^{1/4}\right)O_P\left(p_1^{-\alpha_R/2}p_2^{-\alpha_C/2}\right)\notag\\
&=&O_P\left(p_1^{-\alpha_R/2}p_2^{-\alpha_C/2}T^{1/4}\right)O_P\left(p_1^{1-\alpha_R/2}p_2^{-\alpha_C/2}T^{-1}+ p_1^{1-\alpha_R}p_2^{1-\alpha_C}T^{-1}\right)\notag\\
&=&o_P\left(p_1^{-\alpha_R/2}p_2^{-\alpha_C/2}T^{1/4}\right)=o_P\left(\varphi_2(p_1,p_2,T)\right),\label{eqA.10}\\
&&\max_{1 \leq t \leq T} 
	\left\|\left(\widetilde\lambda_{R,1}^{1/2}p_1^{-\alpha_R/2}p_2^{-\alpha_C/2}\right)\widetilde{\boldsymbol V}_R^{-1/2}\widehat{{\boldsymbol H}}_R^{^\intercal} {\boldsymbol R}^{^\intercal}{\boldsymbol E}_t \left( \widetilde{{\boldsymbol C}} - {\boldsymbol C} \widehat{{\boldsymbol H}}_C\right)\widetilde{\boldsymbol V}_C^{-1/2}
	\right\|_F\notag \\
	&=&O_P\left(p_1^{-\alpha_R/2}p_2^{-\alpha_C/2}\right)O_P\left(p_1^{\alpha_R/2}p_2^{1/2}T^{1/4}\right)\notag\\
	&&O_P\left(p_2^{1/2}p_1^{-\alpha_R/2}T^{-1}\left(1 + p_2^{-\alpha_C/2}p_1^{1-\alpha_R/2} \right)\right)O_P\left(p_1^{-\alpha_R/2}p_2^{-\alpha_C/2}\right)\notag\\
&=&O_P\left(p_1^{-\alpha_R/2}p_2^{-\alpha_C/2}T^{1/4}\right)O_P\left(p_2^{1-\alpha_C/2}p_1^{-\alpha_R/2}T^{-1}+ p_2^{1-\alpha_C}p_1^{1-\alpha_R}T^{-1}\right)\notag\\
&=&o_P\left(p_1^{-\alpha_R/2}p_2^{-\alpha_C/2}T^{1/4}\right)=o_P\left(\varphi_2(p_1,p_2,T)\right),\label{eqA.11}\\
&&\max_{1 \leq t \leq T} \left\|  \left(\widetilde\lambda_{R,1}^{1/2}p_1^{-\alpha_R/2}p_2^{-\alpha_C/2}\right) \widetilde{\boldsymbol V}_R^{-1/2} \widehat{{\boldsymbol H}}_R^{^\intercal} {\boldsymbol R}^{^\intercal} {\boldsymbol E}_t {\boldsymbol C} \widehat{{\boldsymbol H}}_C\widetilde{\boldsymbol V}_C^{-1/2}\right\|_F =O_P\left(\varphi_2(p_1,p_2,T)\right). \label{eqA.12}
\end{eqnarray}
Combining (\ref{eqA.7})--(\ref{eqA.12}), we readily have \eqref{eq3.9}. \hfill$\blacksquare$

\bigskip

\noindent{\bf Proof of Theorem \ref{thm:3.3}}.\ \ By Lemma \ref{le:B.4}, we may show there exists a positive constant $\epsilon_0$ such that
\begin{equation}\label{eqA.13}
	{\sf P}\left(\left\vert\frac{\widehat\lambda_{R,k+1}}{\widehat\lambda_{R,k}}\right\vert>\epsilon_0\right)\rightarrow1,\ \ k=0,\ldots,r_1-1.
\end{equation}
On the other hand, for $k=r_1$, following the proof of Lemma \ref{le:B.4}, we have $\widehat\lambda_{R,r_1+1}=o_P(\widehat\lambda_{R,r_1})$ and thus
\begin{equation}\label{eqA.14}
	{\sf P}\left(\left\vert\frac{\widehat\lambda_{R,r_1+1}}{\widehat\lambda_{R,r_1}}\right\vert \geq \epsilon_0\right)\rightarrow0.
\end{equation}

We next consider $k\geq r_1+1$. Define 
$$
\mathbf{R}^* = \mathbf{R} + \mathbb{E}_1\mathbb{F}_C\left(\mathbb{F}_C^{^\intercal}\mathbb{F}_C\right)^{-1}
$$
where
$$
\mathbb{E}_1 = \left(\mathbf{E}_1,\mathbf{E}_2,...,\mathbf{E}_T\right) \quad \text{and} \quad \mathbb{F}_C^{^\intercal} = \left(\mathbf{F}_1\mathbf{C}^{^\intercal},\mathbf{F}_2\mathbf{C}^{^\intercal},...,\mathbf{F}_T\mathbf{C}^{^\intercal}\right).
$$
Then, we have
$$
\widehat{\pmb{\Omega}}_R = \mathbf{R}^*\left(\frac{1}{T}\sum_{t=1}^{T}{\boldsymbol F}_{t}{\boldsymbol C}^{^\intercal}{\boldsymbol C}{\boldsymbol F}_{t}^{^\intercal}\right)\mathbf{R}^{*^\intercal} + \frac{1}{T}\mathbb{E}_1 \left(\mathbf{I}_{Tp_2} - \mathbb{F}_C\left(\mathbb{F}_C^{^\intercal}\mathbb{F}_C\right)^{-1}\mathbb{F}_C^{^\intercal}\right) \mathbb{E}_1^{^{\intercal}}.
$$
For any $1\leq j\leq \underline{p}_1-2r_1$ with $\underline{p}_1=p_1\wedge(Tp_2)$, by using Lemmas A.6 and A.5 in \cite{AH13}, we may show that
\begin{eqnarray*}
	&&\psi_{r_1+j}\left(\frac{1}{T}\mathbb{E}_1 \left(\mathbf{I}_{Tp_2} - \mathbb{F}_C\left(\mathbb{F}_C^{^\intercal}\mathbb{F}_C\right)^{-1}\mathbb{F}_C^{^\intercal}\right) \mathbb{E}_1^{^\intercal} \right) \leq \psi_{r_1+j}\left(\widehat{\pmb{\Omega}}_R \right) \notag\\
	&\leq& \psi_{r_1+j}\left(\frac{1}{T}\mathbb{E}_1 \left(\mathbf{I}_{Tp_2} - \mathbb{F}_C\left(\mathbb{F}_C^{^\intercal}\mathbb{F}_C\right)^{-1}\mathbb{F}_C^{^\intercal}\right) \mathbb{E}_1^{^\intercal} \right) + \psi_{r_1+1}\left(\mathbf{R}^*\left(\frac{1}{T}\sum_{t=1}^{T}{\boldsymbol F}_{t}{\boldsymbol C}^{^\intercal}{\boldsymbol C}{\boldsymbol F}_{t}^{^\intercal}\right)\mathbf{R}^{*^\intercal} \right) \notag\\
	&=&\psi_{r_1+j}\left(\frac{1}{T}\mathbb{E}_1 \left(\mathbf{I}_{Tp_2} - \mathbb{F}_C\left(\mathbb{F}_C^{^\intercal}\mathbb{F}_C\right)^{-1}\mathbb{F}_C^{^\intercal}\right) \mathbb{E}_1^{^\intercal} \right) \leq \psi_{j}\left(\frac{1}{T}\mathbb{E}_1\mathbb{E}_1^{^\intercal} \right).
\end{eqnarray*}
Using Lemmas A.6 and A.5 in \cite{AH13} again, we have 
$$
\psi_{2r_1+j}\left(\frac{1}{T}\mathbb{E}_1\mathbb{E}_1^{^\intercal} \right)\leq\psi_{r_1+j}\left(\frac{1}{T}\mathbb{E}_1 \left(\mathbf{I}_{Tp_2} - \mathbb{F}_C\left(\mathbb{F}_C^{^\intercal}\mathbb{F}_C\right)^{-1}\mathbb{F}_C^{^\intercal}\right) \mathbb{E}_1^{^\intercal} \right).
$$
Hence, for any $1\leq j\leq \underline{p_1} -2r_1$,
$$
\psi_{2r_1+j}\left(\frac{1}{T}\mathbb{E}_1\mathbb{E}_1^{^\intercal} \right)\leq \psi_{r_1+j}\left(\widehat{\pmb{\Omega}}_R \right) \leq \psi_{j}\left(\frac{1}{T}\mathbb{E}_1\mathbb{E}_1^{^\intercal} \right),
$$
which, together with Assumption \ref{ass:3.4}, leads to
\begin{equation}\label{eqA.15}
	{\sf P}\left(\left\vert\frac{\widehat\lambda_{R,k+1}}{\widehat\lambda_{R,k}}\right\vert>\epsilon_0
	\right)\rightarrow1
\end{equation}
for $k=r_1+1,\ldots,K_1$, where $K_1\in (r_1, \lfloor d_R\underline{p}_1\rfloor - 2r_1]$. Combining (\ref{eqA.13})--(\ref{eqA.15}), we prove the consistency property for $\widehat{r}_1$. The consistency of $\widehat{r}_2$ can be proved in exactly the same way. \hfill$\blacksquare$

\bigskip

\noindent{\bf Proof of Theorem \ref{thm:3.4}}.\ \ It follows from the definition of $\overline{\mathbf R}$ that
\begin{equation}\label{eqA.16}
	\left(\frac{1}{p_1^{\alpha_R}p_2^{\alpha_C}}\overline{\boldsymbol\Omega}_R\right)\overline{\mathbf R} =\overline{\mathbf R}\,\check{\boldsymbol V}_R,
\end{equation}
where $\check{\boldsymbol V}_R$ is an $r_1\times r_1$ diagonal matrix with the diagonal entries being the $r_1$ largest eigenvalues of $(p_1^{\alpha_R}p_2^{\alpha_C})^{-1}\overline{\boldsymbol\Omega}_R$ (in a descending order). By (\ref{eq2.9}), we have
\begin{equation}\label{eqA.17}
\frac{1}{p_1^{\alpha_R}p_2^{\alpha_C}}\overline{\boldsymbol\Omega}_R=\frac{1}{p_1^{\alpha_R}p_2^{\alpha_C}T}\sum_{t=1}^{T}{\boldsymbol R}(\Delta{\boldsymbol F}_{t}){\boldsymbol C}^{^\intercal}{\boldsymbol C}(\Delta{\boldsymbol F}_{t})^{^\intercal}{\boldsymbol R}^{^\intercal}+\frac{1}{p_1^{\alpha_R}p_2^{\alpha_C}}\left({\boldsymbol\Pi}_{1}+{\boldsymbol\Pi}_{2}+{\boldsymbol\Pi}_{3}\right),
\end{equation}
where
\begin{eqnarray}
{\boldsymbol\Pi}_{1}&=&\frac{1}{T}\sum_{t=1}^{T}{\boldsymbol R}(\Delta{\boldsymbol F}_{t}){\boldsymbol C}^{^\intercal}(\Delta{\boldsymbol E}_{t})^{^\intercal},\notag\\
{\boldsymbol\Pi}_{2}&=&\frac{1}{T}\sum_{t=1}^{T}(\Delta{\boldsymbol E}_{t}){\boldsymbol C}(\Delta{\boldsymbol F}_{t})^{^\intercal}{\boldsymbol R}^{^\intercal},\notag\\ 
{\boldsymbol\Pi}_{3}&=&\frac{1}{T}\sum_{t=1}^{T}(\Delta{\boldsymbol E}_{t})(\Delta{\boldsymbol E}_{t})^{^\intercal}.\notag
\end{eqnarray}
By virtue of \eqref{eqA.16}, \eqref{eqA.17} and Lemma \ref{le:B.9} that
\[
p_1^{\alpha_R/2}\overline{{\boldsymbol R}}-{\boldsymbol R}\overline{{\boldsymbol H}}_R=\frac{1}{p_1^{\alpha_R}p_2^{\alpha_C}}(\boldsymbol{\Pi}_{1}+\boldsymbol{\Pi}_{2}+\boldsymbol{\Pi}_{3})\left(p_1^{\alpha_R/2}\overline{{\boldsymbol R}}\right)\check{\boldsymbol V}_{R}^{-1}.
\]
By Lemma \ref{le:B.7}, we have
\begin{equation}\label{eqA.18}
\frac{1}{p_1^{\alpha_R}p_2^{\alpha_C}}\left\| \boldsymbol{\Pi}_{3} \right\|_F=O_P\left(p_1^{\frac{1}{2}-\alpha_R}p_2^{1-\alpha_C} + p_1^{1-\alpha_R}p_2^{\frac{1}{2}-\alpha_C}T^{-1/2}\right).
\end{equation}
As $\left\|{\boldsymbol R} \right\|_F = O(p_1^{\alpha_R/2})$, by Lemma \ref{le:B.8}, we have
\begin{equation}\label{eqA.19}
\frac{1}{p_1^{\alpha_R}p_2^{\alpha_C}}\left\|\boldsymbol{\Pi}_{1}+\boldsymbol{\Pi}_{2}\right\|_F = \frac{1}{p_1^{\alpha_R}p_2^{\alpha_C}T} \times O(p_1^{\alpha_R/2}) \times O_P\left(p_1^{1/2} p_2^{\alpha_C/2}T^{1/2}\right) = O_P\left(p_1^{\frac{1-\alpha_R}{2}}p_2^{-\frac{\alpha_C}{2}}T^{-1/2}\right).
\end{equation}
Combining (\ref{eqA.17})--(\ref{eqA.19}) and Lemma \ref{le:B.9}, we complete the proof of (\ref{eq3.13}). The proof of (\ref{eq3.14}) is analogous and thus omitted. \hfill$\blacksquare$

\bigskip

\noindent{\bf Proof of Theorem \ref{thm:3.5}}.\ \ The proof is similar to the proof of Theorem \ref{thm:3.2} in general. By (\ref{eq2.11}), we readily have that
\begin{eqnarray*}
\overline{{\boldsymbol F}}_t&=&\check{\nu}_{R,1}^{1/2}\check{\boldsymbol V}_R^{-1/2}\overline{{\boldsymbol H}}_R^{-1} {\boldsymbol F}_t\left(\overline{{\boldsymbol H}}_C^{-1}\right)^{^\intercal}\check{\boldsymbol V}_C^{-1/2}+\overline\lambda_{R,1}^{1/2}\overline{\boldsymbol V}_R^{-1/2}\left[\overline{{\boldsymbol R}}^{^\intercal}\left({\boldsymbol R}-\check{{\boldsymbol R}}\overline{{\boldsymbol H}}_R^{-1}\right) {\boldsymbol F}_t\left({\boldsymbol C}-\check{{\boldsymbol C}}\overline{{\boldsymbol H}}_C^{-1}\right)^{^\intercal}\overline{{\boldsymbol C}}\right.\\
&&\left.+\overline{{\boldsymbol R}}^{^\intercal} \left( {\boldsymbol R} - \check{{\boldsymbol R}} \overline{{\boldsymbol H}}_R^{-1} \right){\boldsymbol F}_t \left( \overline{{\boldsymbol H}}_C^{-1}\right)^{^\intercal}\left(\check{\mathbf C}^{^\intercal}\overline{\mathbf C}\right)  + \left(\overline{\mathbf R}^{^\intercal}\check{\mathbf R}\right)\overline{{\boldsymbol H}}_R^{-1} {\boldsymbol F}_t \left( {\boldsymbol C} - \check{{\boldsymbol C}} \overline{{\boldsymbol H}}_C^{-1} \right)^{^\intercal} \overline{{\boldsymbol C}}\right] \overline{\boldsymbol V}_C^{-1/2}\\
&&+ \left(\overline\lambda_{R,1}^{1/2}p_1^{-\alpha_R/2}p_2^{-\alpha_C/2}\right)\overline{\boldsymbol V}_R^{-1/2}\left[\left( \check{{\boldsymbol R}} - {\boldsymbol R} \overline{{\boldsymbol H}}_R \right)^{^\intercal} {\boldsymbol E}_t\left( \check{{\boldsymbol C}} - {\boldsymbol C} \overline{{\boldsymbol H}}_C \right)
+  \left( \check{{\boldsymbol R}} - {\boldsymbol R} \overline{{\boldsymbol H}}_R \right)^{^\intercal}{\boldsymbol E}_t {\boldsymbol C} \overline{{\boldsymbol H}}_C\right. \\
&& \left.+\overline{{\boldsymbol H}}_R^{^\intercal} {\boldsymbol R}^{^\intercal} {\boldsymbol E}_t \left( \check{{\boldsymbol C}} - {\boldsymbol C} \overline{{\boldsymbol H}}_C\right)+\overline{{\boldsymbol H}}_R^{^\intercal} {\boldsymbol R}^{^\intercal} {\boldsymbol E}_t {\boldsymbol C} \overline{{\boldsymbol H}}_C\right]\overline{\boldsymbol V}_C^{-1/2},
\end{eqnarray*}
where $\check{\boldsymbol R}$ and $\check{\boldsymbol C}$ be defined in Theorem \ref{thm:3.4} and $\check{\nu}_{R,1}=p_1^{-\alpha_R}p_2^{-\alpha_C}\overline\lambda_{R,1}$.

It follows from Lemma \ref{le:B.9} that 
\begin{equation}\label{eqA.20}
\left\vert \check{\nu}_{R,1}-\nu_{R,1}^\circ\right\vert=o_P(1),
\end{equation}
where $\nu_{R,k}^\circ$ denotes the largest eigenvalue of  ${\boldsymbol\Sigma}_R^{1/2}{\sf E}\left[ \Delta {\boldsymbol F}_t{\boldsymbol\Sigma}_C \Delta {\boldsymbol F}_t^{^\intercal}\right]{\boldsymbol\Sigma}_R^{1/2}$ and is positive and bounded due to Assumption \ref{ass:3.5}(iii).

By (\ref{eqB.21}), Lemma \ref{le:B.10} and Theorem \ref{thm:3.4}, using $p_1^{1/2-\alpha_R}p_2^{1-\alpha_C}+p_2^{1/2-\alpha_C}p_1^{1-\alpha_R}=o(1)$ in Assumption \ref{ass:3.7}, we readily have that
\begin{eqnarray}
	&&\max_{1 \leq t \leq T} \left\| \overline\lambda_{R,1}^{1/2}\overline{\boldsymbol V}_R^{-1/2}\overline{{\boldsymbol R}}^{^\intercal}\left({\boldsymbol R}-\check{{\boldsymbol R}}\overline{{\boldsymbol H}}_R^{-1}\right) {\boldsymbol F}_t\left({\boldsymbol C}-\check{{\boldsymbol C}}\overline{{\boldsymbol H}}_C^{-1}\right)^{^\intercal}\overline{{\boldsymbol C}}\, \overline{\boldsymbol V}_C^{-1/2}
	\right\|_F \notag\\
	&=&O_P\left(p_1^{\alpha_R/2}p_2^{\alpha_C/2}\right)O_P\left(p_1^{-\alpha_R/2}p_2^{-\alpha_C/2}\right)O_P\left(p_1^{1/2-\alpha_R/2}p_2^{1-\alpha_C}+p_1^{1-\alpha_R/2}p_2^{1/2-\alpha_C}T^{-1/2}\right)\notag\\
	&&O_P\left(T^{1/2}\right)O_P\left(p_2^{1/2-\alpha_C/2}p_1^{1-\alpha_R}+p_2^{1-\alpha_C/2}p_1^{1/2-\alpha_R}T^{-1/2}\right)O_P\left(p_1^{-\alpha_R/2}p_2^{-\alpha_C/2}\right)\notag\\
	&=&o_P\left(T^{1/2}\left(p_1^{1/2-\alpha_R} p_2^{1-\alpha_C}+ p_1^{1-\alpha_R}p_2^{1/2-\alpha_C}\right) \right)=o_P\left(\varpi_1(p_1,p_2,T)\right),\label{eqA.21}\\
	&&\max_{1 \leq t \leq T} \left\|\overline\lambda_{R,1}^{1/2}\overline{\boldsymbol V}_R^{-1/2}\overline{{\boldsymbol R}}^{^\intercal} \left( {\boldsymbol R} - \check{{\boldsymbol R}} \overline{{\boldsymbol H}}_R^{-1} \right) {\boldsymbol F}_t \left( \overline{{\boldsymbol H}}_C^{-1}\right)^{^\intercal}\left(\check{\mathbf C}^{^\intercal}\overline{\mathbf C}\right) \overline{\boldsymbol V}_C^{-1/2} \right\|_F\notag\\
	&=&O_P\left(p_1^{\alpha_R/2}p_2^{\alpha_C/2}\right) O_P\left(p_1^{-\alpha_R/2}p_2^{-\alpha_C/2}\right)O_P\left(p_1^{1/2-\alpha_R/2}p_2^{1-\alpha_C}+p_1^{1-\alpha_R/2}p_2^{1/2-\alpha_C}T^{-1/2}\right)\notag\\
	&&O_P\left(T^{1/2}\right)O_P\left(p_2^{\alpha_C/2}\right)O_P\left(p_1^{-\alpha_R/2}p_2^{-\alpha_C/2}\right)\notag\\
	&=&O_P\left(T^{1/2}\left(p_1^{1/2-\alpha_R} p_2^{1-\alpha_C}+ p_1^{1-\alpha_R}p_2^{1/2-\alpha_C}T^{-1/2}\right) \right)=O_P\left(\varpi_1(p_1,p_2,T)\right),\label{eqA.22}\\
	&&\max_{1 \leq t \leq T} \left\|\overline\lambda_{R,1}^{1/2}\overline{\boldsymbol V}_R^{-1/2} \left(\overline{\mathbf R}^{^\intercal}\check{\mathbf R}\right)\overline{{\boldsymbol H}}_R^{-1} {\boldsymbol F}_t \left( {\boldsymbol C} - \check{{\boldsymbol C}} \overline{{\boldsymbol H}}_C^{-1} \right)^{^\intercal} \overline{{\boldsymbol C}}\overline{\boldsymbol V}_C^{-1/2}\right\| _F\notag\\
	&=&O_P\left(p_1^{\alpha_R/2}p_2^{\alpha_C/2}\right)O_P\left(p_1^{-\alpha_R/2}p_2^{-\alpha_C/2}\right)O_P\left(p_2^{\alpha_R/2}\right)\notag\\
	&&O_P\left(T^{1/2}\right)O_P\left(p_2^{1/2-\alpha_C/2}p_1^{1-\alpha_R}+p_2^{1-\alpha_C/2}p_1^{1/2-\alpha_R}T^{-1/2}\right)O_P\left(p_1^{-\alpha_R/2}p_2^{-\alpha_C/2}\right)\notag\\
	&=&O_P\left(T^{1/2}\left(p_2^{1/2-\alpha_C} p_1^{1-\alpha_R}+ p_2^{1-\alpha_C}p_1^{1/2-\alpha_R}T^{-1/2}\right) \right)=O_P\left(\varpi_1(p_1,p_2,T)\right).\label{eqA.23}
\end{eqnarray}

Similarly to the proofs of (\ref{eqA.9})--(\ref{eqA.12}), using Theorem \ref{thm:3.4}, Assumption \ref{ass:3.6}(ii) and $p_1^{2-2\alpha_R}p_2^{2-2\alpha_C}=O(T)$ in Assumption \ref{ass:3.8}(i), we may show that
\begin{eqnarray}
&&\max_{1 \leq t \leq T} \left\| \left(\overline\lambda_{R,1}^{1/2}p_1^{-\alpha_R/2}p_2^{-\alpha_C/2}\right)\overline{\boldsymbol V}_R^{-1/2}\left( \check{{\boldsymbol R}} - {\boldsymbol R} \overline{{\boldsymbol H}}_R \right)^{^\intercal} {\boldsymbol E}_t^\circ\left( \check{{\boldsymbol C}} - {\boldsymbol C} \overline{{\boldsymbol H}}_C \right)\overline{\boldsymbol V}_C^{-1/2}\right\|_F\notag\\
&=&O_P\left( p_1^{1/2-\alpha_R/2}p_2^{1/2-\alpha_C/2}T^{-1/4}\right)o_P\left(\varpi_1(p_1,p_2,T)\right) \notag\\
&=&o_P\left(\varpi_1(p_1,p_2,T)\right),\label{eqA.24}\\
&&\max_{1 \leq t \leq T} \left\| \left(\overline\lambda_{R,1}^{1/2}p_1^{-\alpha_R/2}p_2^{-\alpha_C/2}\right) \overline{\boldsymbol V}_R^{-1/2} \left( \check{{\boldsymbol R}} - {\boldsymbol R} \overline{{\boldsymbol H}}_R \right)^{^\intercal} {\boldsymbol E}_t^\circ {\boldsymbol C} \overline{{\boldsymbol H}}_C\overline{\boldsymbol V}_C^{-1/2}\right\|_F\notag \\
&=& O_P(p_2^{-\alpha_C/2}p_1^{1/2-\alpha_R/2}T^{-1/4})O_P\left(\varpi_1(p_1,p_2,T)\right) \notag\\
&=&o_P\left(\varpi_1(p_1,p_2,T)\right),\label{eqA.25}\\
&&\max_{1 \leq t \leq T} \left\| \left(\overline\lambda_{R,1}^{1/2}p_1^{-\alpha_R/2}p_2^{-\alpha_C/2}\right)\overline{\boldsymbol V}_R^{-1/2}\overline{{\boldsymbol H}}_R^{^\intercal} {\boldsymbol R}^{^\intercal}  {\boldsymbol E}_t^\circ \left( \check{{\boldsymbol C}} - {\boldsymbol C} \overline{{\boldsymbol H}}_C\right)\overline{\boldsymbol V}_C^{-1/2}\right\|_F\notag\\
&=& O_P(p_1^{-\alpha_R/2}p_2^{1/2-\alpha_C/2}T^{-1/4})O_P\left(\varpi_1(p_1,p_2,T)\right) \notag\\
&=&o_P\left(\varpi_1(p_1,p_2,T)\right),\label{eqA.26}\\ 
&&\max_{1 \leq t \leq T} \left\| \left(\overline\lambda_{R,1}^{1/2}p_1^{-\alpha_R/2}p_2^{-\alpha_C/2}\right)\overline{\boldsymbol V}_R^{-1/2}\overline{{\boldsymbol H}}_R^{^\intercal} {\boldsymbol R}^{^\intercal} {\boldsymbol E}_t^\circ {\boldsymbol C} \overline{{\boldsymbol H}}_C\overline{\boldsymbol V}_C^{-1/2}\right\|_F \notag\\ &=&O_P\left(p_1^{-\alpha_R/2}p_2^{-\alpha_C/2}T^{1/4}\right)=o_P\left(\varpi_1(p_1,p_2,T)\right).\label{eqA.27}
\end{eqnarray}
On the other hand, by Assumption \ref{ass:3.8}(i), Theorem \ref{thm:3.4} and Lemmas \ref{le:B.10} and \ref{le:B.11}, we have
\begin{eqnarray}
&&\max_{1 \leq t \leq T} \left\| \left(\overline\lambda_{R,1}^{1/2}p_1^{-\alpha_R/2}p_2^{-\alpha_C/2}\right)\overline{\boldsymbol V}_R^{-1/2}\left( \check{{\boldsymbol R}} - {\boldsymbol R} \overline{{\boldsymbol H}}_R \right)^{^\intercal} {\boldsymbol E}_t^\dag\left( \check{{\boldsymbol C}} - {\boldsymbol C} \overline{{\boldsymbol H}}_C \right)\overline{\boldsymbol V}_C^{-1/2}
	\right\|_F \notag\\
&=& O_P\left(p_1^{-\alpha_R/2}p_2^{-\alpha_C/2}\right)O_P\left(p_1^{1/2-\alpha_R/2}p_2^{1-\alpha_C}+p_1^{1-\alpha_R/2}p_2^{1/2-\alpha_C}T^{-1/2}\right)O_P\left([s_{R}s_C]^{1/2}T^{3/4}\right)\notag\\
	&&O_P\left(p_2^{1/2-\alpha_C/2}p_1^{1-\alpha_R}+p_2^{1-\alpha_C/2}p_1^{1/2-\alpha_R}T^{-1/2}\right)O_P\left(p_1^{-\alpha_R/2}p_2^{-\alpha_C/2}\right)\notag\\
	&=&O_P\left(\left(\frac{s_p}{p_1^{\alpha_R}}\frac{s_C}{p_2^{\alpha_C}}\right)^{1/2}T^{1/4}\right)O_P\left(p_1^{1/2-\alpha_R}p_2^{1-\alpha_C}+p_1^{1-\alpha_R}p_2^{1/2-\alpha_C}T^{-1/2}\right)\notag\\
	&&O_P\left(T^{1/2}\left(p_2^{1/2-\alpha_C} p_1^{1-\alpha_R}+ p_2^{1-\alpha_C}p_1^{1/2-\alpha_R}T^{-1/2}\right) \right)\notag\\
	&=&o_P\left(T^{1/2}\left(p_2^{1/2-\alpha_C} p_1^{1-\alpha_R}+ p_2^{1-\alpha_C/2}p_1^{1/2-\alpha_R}T^{-1/2}\right) \right)=o_P\left(\varpi_1(p_1,p_2,T)\right),\label{eqA.28}\\
&&\max_{1 \leq t \leq T} \left\| \left(\overline\lambda_{R,1}^{1/2}p_1^{-\alpha_R/2}p_2^{-\alpha_C/2}\right) \overline{\boldsymbol V}_R^{-1/2} \left( \check{{\boldsymbol R}} - {\boldsymbol R} \overline{{\boldsymbol H}}_R \right)^{^\intercal} {\boldsymbol E}_t^\dag {\boldsymbol C} \overline{{\boldsymbol H}}_C\overline{\boldsymbol V}_C^{-1/2}\right\|_F\notag\\
&=& O_P\left(p_1^{-\alpha_R/2}p_2^{-\alpha_C/2}\right)O_P\left(p_1^{1/2-\alpha_R/2}p_2^{1-\alpha_C}+p_1^{1-\alpha_R/2}p_2^{1/2-\alpha_C}T^{-1/2}\right)\notag\\ &&O_P\left(s_R^{1/2}s_C^{\alpha_C/2}T^{3/4}\right)O_P\left(p_1^{-\alpha_R/2}p_2^{-\alpha_C/2}\right)\notag\\
&=&O_P\left(\left(\frac{s_p}{p_1^{\alpha_R}}\frac{s_C^{\alpha_C}}{p_2^{2\alpha_C}}\right)^{1/2}T^{1/4}\right)O_P\left(T^{1/2}\left(p_1^{1/2-\alpha_R} p_2^{1-\alpha_C}+ p_1^{1-\alpha_R/2}p_2^{1/2-\alpha_C}T^{-1/2}\right) \right)\notag\\
&=&O_P\left(T^{1/2}\left(p_1^{1/2-\alpha_R} p_2^{1-\alpha_C}+ p_1^{1-\alpha_R/2}p_2^{1/2-\alpha_C}T^{-1/2}\right) \right)=O_P\left(\varpi_1(p_1,p_2,T)\right),\label{eqA.29}\\
&&\max_{1 \leq t \leq T} \left\| \left(\overline\lambda_{R,1}^{1/2}p_1^{-\alpha_R/2}p_2^{-\alpha_C/2}\right)\overline{\boldsymbol V}_R^{-1/2}\overline{{\boldsymbol H}}_R^{^\intercal} {\boldsymbol R}^{^\intercal} {\boldsymbol E}_t^\dag \left( \check{{\boldsymbol C}} - {\boldsymbol C} \overline{{\boldsymbol H}}_C\right)\overline{\boldsymbol V}_C^{-1/2}\right\|_F\notag \\
&=&O_P\left(p_1^{-\alpha_R/2}p_2^{-\alpha_C/2}\right)O_P\left(s_R^{\alpha_R/2}s_C^{1/2}T^{3/4}\right)\notag\\
&&O_P\left(p_2^{1/2-\alpha_C/2}p_1^{1-\alpha_R}+p_2^{1-\alpha_C/2}p_1^{1/2-\alpha_R}T^{-1/2}\right)O_P\left(p_1^{-\alpha_R/2}p_2^{-\alpha_C/2}\right)\notag\\
&=&O_P\left(\left(\frac{s_p^{\alpha_R}}{p_1^{2\alpha_R}}\frac{s_C}{p_2^{\alpha_C}}\right)^{1/2}T^{1/4}\right)O_P\left(T^{1/2}\left(p_2^{1/2-\alpha_C} p_1^{1-\alpha_R}+ p_2^{1-\alpha_C/2}p_1^{1/2-\alpha_R}T^{-1/2}\right) \right)\notag\\
&=&O_P\left(T^{1/2}\left(p_2^{1/2-\alpha_C} p_1^{1-\alpha_R}+ p_2^{1-\alpha_C}p_1^{1/2-\alpha_R}T^{-1/2}\right) \right)=O_P\left(\varpi_1(p_1,p_2,T)\right),\label{eqA.30}\\
&&\max_{1 \leq t \leq T} \left\| \left(\overline\lambda_{R,1}^{1/2}p_1^{-\alpha_R/2}p_2^{-\alpha_C/2}\right)\overline{\boldsymbol V}_R^{-1/2}\overline{{\boldsymbol H}}_R^{^\intercal} {\boldsymbol R}^{^\intercal} {\boldsymbol E}_t^\dag {\boldsymbol C} \overline{{\boldsymbol H}}_C\overline{\boldsymbol V}_C^{-1/2}\right\|_F \notag\\
&=& O_P\left(p_1^{-\alpha_R/2}p_2^{-\alpha_C/2}\right)O_P\left(s_R^{\alpha_R/2}s_C^{\alpha_C/2}T^{3/4}\right)O_P\left(p_1^{-\alpha_R/2}p_2^{-\alpha_C/2}\right)\notag\\
&=&O_P\left(\varpi_2(p_1,p_2,T)\right). \label{eqA.31}
\end{eqnarray}
By virtue of (\ref{eq3.15}) and (\ref{eqA.21})--(\ref{eqA.31}), we complete the proof of \eqref{eq3.18}. \hfill$\blacksquare$

\bigskip

\noindent{\bf Proof of Theorem \ref{thm:3.6}}.\ \ The proof is the same as that of Theorem \ref{thm:3.3}. \hfill$\blacksquare$

\bigskip

\noindent{\bf Proof of Theorem \ref{thm:4.1}}.\ \ (i) Note that 
\[
\left(\frac{1}{p_2^{\alpha_{C,1}}T}\widehat{\boldsymbol\Omega}_R\right)\widehat{\mathbf R}\mathbf{B}_{R}^{-2} =\widehat{\mathbf R}\widehat{\boldsymbol V}_R^\ast.
\] 
Letting $\boldsymbol{\Lambda}_{i}, i=1,2,3$, be defined as in the proof of Theorem \ref{thm:3.1}, by \eqref{eqA.1}--\eqref{eqA.2} and Lemma \ref{le:B.14}, we have
\begin{eqnarray*}
\left(\widetilde{\mathbf{R}}^\ast-\mathbf{R}\widehat{\mathbf{H}}_R^\ast\right)\mathbf{B}_R^{-1}&=& \frac{1}{p_2^{\alpha_{C,1}}T}(\boldsymbol{\Lambda}_{1}+\boldsymbol{\Lambda}_{2}+\boldsymbol{\Lambda}_{3}) \widehat{\mathbf{R}} \mathbf{B}_R^{-2} \left(\widehat{\boldsymbol V}_{R}^\ast\right)^{-1}.
\end{eqnarray*}

Note that $\|\mathbf{B}_R^{-1}\|_F = O(p_1^{-\alpha_{R,r_1}/2})$ and $\|\widehat{\mathbf{R}} \|_F = O(1)$, as $p_1^{1-\alpha_{R,1}}p_2^{1-\alpha_{C,1}}=O(T)$ implied by Assumption \ref{ass:4.3}, to complete the proof of (\ref{eq4.3}), it is sufficient to show that
\begin{eqnarray}
&&\left\|\frac{1}{p_2^{\alpha_{C,1}}T}(\boldsymbol{\Lambda}_{1}+\boldsymbol{\Lambda}_{2})\right\|_F= O_P\left(p_1^{1/2+\alpha_{R,1}/2}p_2^{- \alpha_{C,1}/2}T^{-1}\right),\label{eqA.32}\\
&&\left\|\frac{1}{p_2^{\alpha_{C,1}}T}\boldsymbol{\Lambda}_{3} \right\|_F=O_P\left(p_1^{1/2}p_2^{1-\alpha_{C,1}}T^{-1} + p_1 p_2^{1/2-\alpha_{C,1}}T^{-3/2}\right).\label{eqA.33}
\end{eqnarray}
Using $\left\|\mathbf{R}\right\|_F = O(p_1^{\alpha_{R,1}/2})$ and Lemma \ref{le:B.12}, we readily have that
$$
\left\|\frac{1}{p_2^{\alpha_{C,1}}T}(\boldsymbol{\Lambda}_{1}+\boldsymbol{\Lambda}_{2})\right\|_F = \frac{1}{p_2^{\alpha_{C,1}}T^2} \times O(p_1^{\alpha_{R,1}/2}) \times O_P\left(p_1^{1/2} p_2^{\alpha_{C,1}/2}T\right) = O_P\left(p_1^{1/2+\alpha_{R,1}/2}p_2^{- \alpha_{C,1}/2}T^{-1}\right),
$$
proving (\ref{eqA.32}). With Lemma \ref{le:B.1}, we can prove (\ref{eqA.33}). 

The proof of (\ref{eq4.4}) is analogous. We thus complete the proof of Theorem \ref{thm:4.1}(i).

\smallskip

(ii) Write
{\small\begin{eqnarray*}
\widehat{\mathbf{F}}_t&=&\widetilde{\lambda}_{R,1}^{1/2}\widetilde{\boldsymbol V}_{R}^{-1/2}\mathbf{B}_R\widehat{\mathbf{H}}_R^{\ast^{-1}}\mathbf{F}_t\left(\widehat{\mathbf{H}}_C^{\ast^{-1}}\right)^{^\intercal}\mathbf{B}_C\widetilde{\boldsymbol V}_{C}^{-1/2}+\widetilde{\lambda}_{R,1}^{1/2}\widetilde{\boldsymbol V}_{R}^{-1/2}\left[\widehat{\mathbf{R}}^{^\intercal}\left(\mathbf{R}-\widehat{\mathbf{R}}\mathbf{B}_R\widehat{\mathbf{H}}_R^{\ast^{-1}}\right)\mathbf{F}_t\left(\mathbf{C}-\widehat{\mathbf{C}}\mathbf{B}_C\widehat{\mathbf{H}}_C^{\ast^{-1}}\right)^{^\intercal}\widehat{\mathbf{C}}\right.\\
&&\left.+\widehat{\mathbf{R}}^{^\intercal}\left(\mathbf{R}-\widehat{\mathbf{R}}\mathbf{B}_R\widehat{\mathbf{H}}_R^{\ast^{-1}}\right)\mathbf{F}_t\left(\widehat{\mathbf{H}}_C^{\ast^{-1}}\right)^{^\intercal}\mathbf{B}_C+\mathbf{B}_R\widehat{\mathbf{H}}_R^{\ast^{-1}} \mathbf{F}_t \left(\mathbf{C}-\widehat{\mathbf{C}}\mathbf{B}_C\widehat{\mathbf{H}}_C^{\ast^{-1}}\right)^{^\intercal}\widehat{\mathbf{C}} \right]\widetilde{\boldsymbol V}_{C}^{-1/2}\\
&&+ \widetilde{\lambda}_{R,1}^{1/2}\widetilde{\boldsymbol V}_{R}^{-1/2}\left[\left( \widehat{\mathbf{R}} - \mathbf{R} \widehat{\mathbf{H}}_R^\ast\mathbf{B}_R^{-1} \right)^{^\intercal} \mathbf{E}_t\left( \widehat{\mathbf{C}} - \mathbf{C} \widehat{\mathbf{H}}_C^\ast\mathbf{B}_C^{-1}\right)
+ \left( \widehat{\mathbf{R}} - \mathbf{R} \widehat{\mathbf{H}}_R^\ast\mathbf{B}_R^{-1}\right)^{^\intercal} \mathbf{E}_t \mathbf{C} \widehat{\mathbf{H}}_C^\ast\mathbf{B}_C^{-1} \right.\\
&& + \left. \mathbf{B}_R^{-1}\left(\widehat{\mathbf{H}}_R^\ast\right)^{^\intercal} \mathbf{R}^{^\intercal} \mathbf{E}_t \left( \widehat{\mathbf{C}} - \mathbf{C} \widehat{\mathbf{H}}_C^\ast\mathbf{B}_C^{-1}\right)+ \mathbf{B}_R^{-1}\left(\widehat{\mathbf{H}}_R^\ast\right)^{^\intercal} \mathbf{R}^{^\intercal} \mathbf{E}_t \mathbf{C} \widehat{\mathbf{H}}_C^\ast\mathbf{B}_C^{-1} \right]\widetilde{\boldsymbol V}_{C}^{-1/2}.
\end{eqnarray*}}

By the definitions of $\widetilde{\boldsymbol V}_{R}$, $\widetilde{\boldsymbol V}_{C}$, $\widehat{\boldsymbol V}_{R}^\ast$ and $\widehat{\boldsymbol V}_{C}^\ast$, we have
\begin{equation}\label{eqA.34}
\widetilde{\lambda}_{R,1}^{1/2}\widetilde{\boldsymbol V}_{R}^{-1/2}\mathbf{B}_R\widehat{\mathbf{H}}_R^{\ast^{-1}}\mathbf{F}_t\left(\widehat{\mathbf{H}}_C^{\ast^{-1}}\right)^{^\intercal}\mathbf{B}_C\widetilde{\boldsymbol V}_{C}^{-1/2} = \widehat\nu_{R,1}^{\ast^{1/2}}\widehat{\boldsymbol V}_R^{\ast^{-1/2}}\widehat{{\boldsymbol H}}_R^{\ast^{-1}}{\boldsymbol F}_t\left(\widehat{{\boldsymbol H}}_C^{\ast^{-1}}\right)^{^\intercal}\widehat{\boldsymbol V}_C^{\ast^{-1/2}},
\end{equation}
where $\widehat\nu_{R,1}^{\ast} = p_1^{-\alpha_{R,1}}p_2^{-\alpha_{C,1}}\widetilde{\lambda}_{R,1}$ is the first diagonal element of $\widehat{\boldsymbol V}_R^{\ast}$, which is positive and bounded {\em w.p.a.1} according to Lemma \ref{le:B.14}. 

Using (\ref{eqA.32}) and (\ref{eqA.33}) in the proof of Theorem \ref{thm:4.1}(i), we have
\begin{eqnarray}
\widehat{\mathbf{R}}\mathbf{B}_R\widehat{\mathbf{H}}_R^{\ast^{-1}}-\mathbf{R}&=&\frac{1}{p_2^{\alpha_{C,1}}T}(\boldsymbol{\Lambda}_{1}+\boldsymbol{\Lambda}_{2}+\boldsymbol{\Lambda}_{3}) \widehat{\mathbf{R}} \mathbf{B}_R^{-1} \widehat{\boldsymbol V}_{R}^{\ast^{-1}}\widehat{\mathbf{H}}_R^{\ast^{-1}}\notag\\
&=&O_P\left(p_1^{\alpha_{R,1}-\alpha_{R,r_1}/2} p_1^{1/2-\alpha_{R,1}/2}p_2^{-\alpha_{C,1}/2}T^{-1}\left(1 + p_1^{-\alpha_{R,1}/2}p_2^{1-\alpha_{C,1}/2} \right)\right),\label{eqA.35}
\end{eqnarray}
and similarly
\begin{equation}\label{eqA.36}
\widehat{\mathbf{C}}\mathbf{B}_C\widehat{\mathbf{H}}_C^{\ast^{-1}}-\mathbf{C} =O_P\left(p_2^{\alpha_{C,1}-\alpha_{C,r_2}/2} p_2^{1/2-\alpha_{C,1}/2}p_1^{-\alpha_{R,1}/2}T^{-1}\left(1+p_2^{-\alpha_{C,1}/2}p_1^{1-\alpha_{R,1}/2}\right)\right).
\end{equation}
With (\ref{eqA.35}), (\ref{eqA.36}), Lemma \ref{le:B.14}, the fact that $\widetilde\lambda_{R,1}$ is of order $p_1^{\alpha_{R,1}}p_2^{\alpha_{C,1}}$ {\em w.p.a.1} and $\max_{1\leq t\leq T}\|\mathbf{F}_t\|_F = O_P(T^{1/2})$ as well as Assumption \ref{ass:4.3}, we have
\begin{eqnarray*}
&&\max_{1 \leq t \leq T} \left\|\widetilde{\lambda}_{R,1}^{1/2}\widetilde{\boldsymbol V}_{R}^{-1/2} \widehat{\mathbf{R}}^{^\intercal}\left(\mathbf{R}-\widehat{\mathbf{R}}\mathbf{B}_R\widehat{\mathbf{H}}_R^{\ast^{-1}}\right)\mathbf{F}_t\left(\mathbf{C}-\widehat{\mathbf{C}}\mathbf{B}_C\widehat{\mathbf{H}}_C^{\ast^{-1}}\right)^{^\intercal}\widehat{\mathbf{C}}\widetilde{\boldsymbol V}_{C}^{-1/2}\right\|_F \\
&=& O_P\left(p_1^{\alpha_{R,1}-\alpha_{R,r_1}} p_1^{1/2-\alpha_{R,1}/2}p_2^{-\alpha_{C,1}/2}T^{-1}\left(1 + p_1^{-\alpha_{R,1}/2}p_2^{1-\alpha_{C,1}/2} \right)\right) O_P(T^{1/2})\\
&&O_P\left(p_2^{\alpha_{C,1}-\alpha_{C,r_2}} p_2^{1/2-\alpha_{C,1}/2}p_1^{-\alpha_{R,1}/2}T^{-1}\left(1+p_2^{-\alpha_{C,1}/2}p_1^{1-\alpha_{R,1}/2}\right)\right)\notag\\
&=& o_P\left(\varphi_1^\ast(p_1,p_2,T)\right),\\
&&\max_{1 \leq t \leq T} \left\| \widehat{\mathbf{R}}^{^\intercal}\left(\mathbf{R}-\widehat{\mathbf{R}}\mathbf{B}_R\widehat{\mathbf{H}}_R^{\ast^{-1}}\right)\mathbf{F}_t\left(\widehat{\mathbf{H}}_C^{\ast^{-1}}\right)^{^\intercal}\mathbf{B}_C\right\|_F\\
&=& O_P\left(p_1^{\alpha_{R,1}-\alpha_{R,r_1}} p_1^{1/2-\alpha_{R,1}/2}p_2^{-\alpha_{C,1}/2}T^{-1/2}\left(1 + p_1^{-\alpha_{R,1}/2}p_2^{1-\alpha_{C,1}/2} \right)\right) =O_P\left(\varphi_1^\ast(p_1,p_2,T)\right),\\
&&\max_{1 \leq t \leq T} \left\|\mathbf{B}_R\widehat{\mathbf{H}}_R^{\ast^{-1}} \mathbf{F}_t \left(\mathbf{C}-\widehat{\mathbf{C}}\mathbf{B}_C\widehat{\mathbf{H}}_C^{\ast^{-1}}\right)^{^\intercal}\widehat{\mathbf{C}}\right\| _F\\
&=&O_P\left(p_2^{\alpha_{C,1}-\alpha_{C,r_2}} p_2^{1/2-\alpha_{C,1}/2}p_1^{-\alpha_{R,1}/2}T^{-1/2}\left(1+p_2^{-\alpha_{C,1}/2}p_1^{1-\alpha_{R,1}/2}\right)\right)=O_P\left(\varphi_1^\ast(p_1,p_2,T)\right).
\end{eqnarray*}

With (\ref{eqA.35}), (\ref{eqA.36}), Lemmas \ref{le:B.14}--\ref{le:B.16} and Assumption \ref{ass:4.3}, we have
\begin{eqnarray*}
&&\max_{1 \leq t \leq T} \left\|\widetilde{\lambda}_{R,1}^{1/2}\widetilde{\boldsymbol V}_{R}^{-1/2}\left( \widehat{\mathbf{R}} - \mathbf{R} \widehat{\mathbf{H}}_R^\ast\mathbf{B}_R^{-1} \right)^{^\intercal} \mathbf{E}_t\left( \widehat{\mathbf{C}} - \mathbf{C} \widehat{\mathbf{H}}_C^\ast\mathbf{B}_C^{-1}\right)\widetilde{\boldsymbol V}_{C}^{-1/2} \right\|_F \\
&=& O_P\left(p_1^{1/2-\alpha_{R,r_1}/2}p_2^{1/2-\alpha_{C,r_2}/2}T^{1/4}\right)O_P\left(p_1^{\alpha_{R,1}-\alpha_{R,r_1}} p_1^{1/2-\alpha_{R,1}/2}p_2^{-\alpha_{C,1}/2}T^{-1}\left(1 + p_1^{-\alpha_{R,1}/2}p_2^{1-\alpha_{C,1}/2} \right)\right)\notag\\
&&O_P\left(p_2^{\alpha_{C,1}-\alpha_{C,r_2}} p_2^{1/2-\alpha_{C,1}/2}p_1^{-\alpha_{R,1}/2}T^{-1}\left(1+p_2^{-\alpha_{C,1}/2}p_1^{1-\alpha_{R,1}/2}\right)\right)\notag \\
&=& O_P\left(p_1^{\alpha_{R,1}/2-\alpha_{R,r_1}}p_2^{\alpha_{C,1}/2-\alpha_{C,r_2}}T^{1/4}\right)O_P\left(p_1^{1-\alpha_{R,r_1}/2}p_2^{-\alpha_{C,1}/2}T^{-1}\left(1 + p_1^{-\alpha_{R,1}/2}p_2^{1-\alpha_{C,1}/2} \right)\right)\notag\\
&&O_P\left( p_2^{1-\alpha_{C,r_2}/2}p_1^{-\alpha_{R,1}/2}T^{-1}\left(1+p_2^{-\alpha_{C,1}/2}p_1^{1-\alpha_{R,1}/2}\right)\right)\\
&=&o_P\left(p_1^{\alpha_{R,1}/2-\alpha_{R,r_1}}p_2^{\alpha_{C,1}/2-\alpha_{C,r_2}}T^{1/4}\right)=o_P\left(\varphi_2^\ast(p_1,p_2,T)\right),\notag \\
&&\max_{1 \leq t \leq T} \left\|\widetilde{\lambda}_{R,1}^{1/2}\widetilde{\boldsymbol V}_{R}^{-1/2}\left( \widehat{\mathbf{R}} - \mathbf{R} \widehat{\mathbf{H}}_R^\ast\mathbf{B}_R^{-1}\right)^{^\intercal} \mathbf{E}_t \mathbf{C} \widehat{\mathbf{H}}_C^\ast\mathbf{B}_C^{-1} \widetilde{\boldsymbol V}_{C}^{-1/2}\right\|_F\\	
&=&O_P\left(p_1^{1/2-\alpha_{R,r_1}/2}p_2^{\alpha_{C,1}/2-\alpha_{C,r_2}}T^{1/4}\right)O_P\left(p_1^{\alpha_{R,1}-\alpha_{R,r_1}} p_1^{1/2-\alpha_{R,1}/2}p_2^{-\alpha_{C,1}/2}T^{-1}\left(1 + p_1^{-\alpha_{R,1}/2}p_2^{1-\alpha_{C,1}/2} \right)\right)\\
&=&O_P\left(p_1^{\alpha_{R,1}/2-\alpha_{R,r_1}}p_2^{\alpha_{C,1}/2-\alpha_{C,r_2}}T^{1/4}\right)O_P\left(p_1^{1-\alpha_{R,r_1}/2}p_2^{-\alpha_{C,1}/2}T^{-1}\left(1 + p_1^{-\alpha_{R,1}/2}p_2^{1-\alpha_{C,1}/2} \right)\right)\\
&=&o_P\left(p_1^{\alpha_{R,1}/2-\alpha_{R,r_1}}p_2^{\alpha_{C,1}/2-\alpha_{C,r_2}}T^{1/4}\right)=o_P\left(\varphi_2^\ast(p_1,p_2,T)\right),\\
&&\max_{1 \leq t \leq T}\left\|\widetilde{\lambda}_{R,1}^{1/2}\widetilde{\boldsymbol V}_{R}^{-1/2}\mathbf{B}_R^{-1}\left(\widehat{\mathbf{H}}_R^\ast\right)^{^\intercal} \mathbf{R}^{^\intercal} \mathbf{E}_t \left( \widehat{\mathbf{C}} - \mathbf{C} \widehat{\mathbf{H}}_C^\ast\mathbf{B}_C^{-1}\right)\widetilde{\boldsymbol V}_{C}^{-1/2}\right\|_F \\
&=&O_P\left(p_1^{\alpha_{R,1}/2-\alpha_{R,r_1}}p_2^{1/2-\alpha_{C,r_2}/2}T^{1/4} \right)O_P\left(p_2^{\alpha_{C,1}-\alpha_{C,r_2}} p_2^{1/2-\alpha_{C,1}/2}p_1^{-\alpha_{R,1}/2}T^{-1}\left(1+p_2^{-\alpha_{C,1}/2}p_1^{1-\alpha_{R,1}/2}\right)\right)\\
&=&O_P\left(p_1^{\alpha_{R,1}/2-\alpha_{R,r_1}}p_2^{\alpha_{C,1}/2-\alpha_{C,r_2}}T^{1/4}\right)O_P\left( p_2^{1-\alpha_{C,r_2}/2}p_1^{-\alpha_{R,1}/2}T^{-1}\left(1+p_2^{-\alpha_{C,1}/2}p_1^{1-\alpha_{R,1}/2}\right)\right)\\
&=&o_P\left(p_1^{\alpha_{R,1}/2-\alpha_{R,r_1}}p_2^{\alpha_{C,1}/2-\alpha_{C,r_2}}T^{1/4}\right)=o_P\left(\varphi_2^\ast(p_1,p_2,T)\right),\notag\\
&&\max_{1 \leq t \leq T} \left\| \widetilde{\lambda}_{R,1}^{1/2}\widetilde{\boldsymbol V}_{R}^{-1/2}\mathbf{B}_R^{-1}\left(\widehat{\mathbf{H}}_R^\ast\right)^{^\intercal} \mathbf{R}^{^\intercal} \mathbf{E}_t \mathbf{C} \widehat{\mathbf{H}}_C^\ast\mathbf{B}_C^{-1} \widetilde{\boldsymbol V}_{C}^{-1/2}\right\|_F\notag\\
&=&O_P\left(p_1^{\alpha_{R,1}/2-\alpha_{R,r_1}}p_2^{\alpha_{C,1}/2-\alpha_{C,r_2}}T^{1/4}\right)=O_P\left(\varphi_2^\ast(p_1,p_2,T)\right). 
\end{eqnarray*}
Combining the above results, we complete the proof of Theorem \ref{thm:4.1}(ii). 

\medskip

(iii) It follows from Lemma \ref{le:B.14} that there exists a positive constant $\epsilon_1$ such that
\begin{equation}
{\sf P}\left(\left\vert\frac{\widehat\lambda_{R,k+1}}{\widehat\lambda_{R,k}}\right\vert>\epsilon_1\times p_1^{\alpha_{R,r_1}-\alpha_{R,1}}\right)\rightarrow1,\ \ k=0,\ldots,r_1-1.\label{eqA.37}
\end{equation}
For $k=r_1$, following the proof of Lemma \ref{le:B.14} and using the condition $p_1^{\alpha_{R,1}-\alpha_{R,r_1}}\xi_R(p_1,p_2,T)=o(1)$, we have
$$
\frac{\widehat\lambda_{R,r_1+1}}{\widehat\lambda_{R,r_1}}=O_P\left(\xi_R(p_1,p_2,T)\right)=o_P\left(p_1^{\alpha_{R,r_1}-\alpha_{R,1}}\right),
$$ 
indicating that
\begin{equation}\label{eqA.38}
{\sf P}\left(\left\vert\frac{\widehat\lambda_{R,r_1+1}}{\widehat\lambda_{R,r_1}}\right\vert \geq \epsilon_1\times p_1^{\alpha_{R,r_1}-\alpha_{R,1}}\right)={\sf P}\left(\left\vert\frac{\widehat\nu_{R,r_1+1}}{\widehat\nu_{R,r_1}}\right\vert \geq \epsilon_1\times p_1^{\alpha_{R,r_1}-\alpha_{R,1}}\right)\rightarrow0.
\end{equation}
Finally, for $k=r_1+1,\ldots,K_1$, as in the proof of Theorem \ref{thm:3.3}, we have 
\begin{equation}\label{eqA.39}
{\sf P}\left(\left\vert\frac{\widehat\lambda_{R,k+1}}{\widehat\lambda_{R,k}}\right\vert> \epsilon_1\right)\rightarrow1
\end{equation}
Combining (\ref{eqA.37})--(\ref{eqA.39}), we prove the consistency property for $\widehat{r}_1$. The consistency of $\widehat{r}_2$ can be proved in exactly the same way. \hfill$\blacksquare$

\bigskip

\noindent{\bf Proof of Theorem \ref{thm:4.2}}.\ \ (i) Note that 
\[
\left(\frac{1}{p_2^{\alpha_{C,1}}}\overline{\boldsymbol\Omega}_R\right)\overline{\mathbf R}\mathbf{B}_{R}^{-2} =\overline{\mathbf R}\check{\boldsymbol V}_R^\ast.
\]
By \eqref{eq2.9} and Lemma \ref{le:B.18}, we have
\[
\left(\check{\mathbf{R}}^\ast-\mathbf{R}\overline{\mathbf{H}}_R^\ast\right)\mathbf{B}_R^{-1}=\frac{1}{p_2^{\alpha_{C,1}}}(\boldsymbol{\Pi}_{1}+\boldsymbol{\Pi}_{2}+\boldsymbol{\Pi}_{3}) \overline{\mathbf{R}} \mathbf{B}_R^{-2} \check{\boldsymbol V}_{R}^{\ast^{-1}}.
\]
Using $\left\|\mathbf{R}\right\|_F = O(p_1^{\alpha_{R,1}/2})$ and Lemma \ref{le:B.17}, we have
\begin{eqnarray}
\left\|\frac{1}{p_2^{\alpha_{C,1}}}(\boldsymbol{\Pi}_{1}+\boldsymbol{\Pi}_{2})\right\|_F &=& \frac{1}{p_2^{\alpha_{C,1}}} \times O(p_1^{\alpha_{R,1}/2}) \times O_P\left(p_1^{1/2} p_2^{\alpha_{C,1}/2}T^{-1/2}\right)\notag\\
&=& O_P\left(p_1^{\frac{1+\alpha_{R,1}}{2}}p_2^{-\frac{\alpha_{C,1}}{2}}T^{-1/2}\right).\notag
\end{eqnarray}
It follows from Lemma \ref{le:B.7} that
$$
\left\|\frac{1}{p_2^{\alpha_{C,1}}}\boldsymbol{\Pi}_{3} \right\|_F=O_P\left(p_1^{\frac{1}{2}}p_2^{1-\alpha_{C,1}} + p_1 p_2^{\frac{1}{2}-\alpha_{C,1}}T^{-\frac{1}{2}}\right).
$$
Combining the above two results with $\left\|\mathbf{B}_R^{-2}\right\|_F = O(p_1^{-\alpha_{R,r_1}})$, $\left\|\overline{\mathbf{R}} \right\|_F = O(1)$ and Lemma \ref{le:B.18}, we complete the proof of (\ref{eq4.7}). The proof of (\ref{eq4.8}) is analogous.

\medskip

(ii) Observe that
{\small\begin{eqnarray*}
\widehat{\mathbf{F}}_t&=&\overline{\lambda}_{R,1}^{1/2}\overline{\boldsymbol V}_{R}^{-1/2}\mathbf{B}_R\overline{\mathbf{H}}_R^{\ast^{-1}}\mathbf{F}_t\left(\overline{\mathbf{H}}_C^{\ast^{-1}}\right)^{^\intercal}\mathbf{B}_C\overline{\boldsymbol V}_{C}^{-1/2}+\overline{\lambda}_{R,1}^{1/2}\overline{\boldsymbol V}_{R}^{-1/2}\left[\overline{\mathbf{R}}^{^\intercal}\left(\mathbf{R}-\overline{\mathbf{R}}\mathbf{B}_R\overline{\mathbf{H}}_R^{\ast^{-1}}\right)\mathbf{F}_t\left(\mathbf{C}-\overline{\mathbf{C}}\mathbf{B}_C\overline{\mathbf{H}}_C^{\ast^{-1}}\right)^{^\intercal}\overline{\mathbf{C}}\right.\\
&&\left.+\overline{\mathbf{R}}^{^\intercal}\left(\mathbf{R}-\overline{\mathbf{R}}\mathbf{B}_R\overline{\mathbf{H}}_R^{\ast^{-1}}\right)\mathbf{F}_t\left(\overline{\mathbf{H}}_C^{\ast^{-1}}\right)^{^\intercal}\mathbf{B}_C + \mathbf{B}_R\overline{\mathbf{H}}_R^{\ast^{-1}} \mathbf{F}_t \left(\mathbf{C}-\overline{\mathbf{C}}\mathbf{B}_C\overline{\mathbf{H}}_C^{\ast^{-1}}\right)^{^\intercal}\overline{\mathbf{C}} \right]\overline{\boldsymbol V}_{C}^{-1/2}\\
&&+ \overline{\lambda}_{R,1}^{1/2}\overline{\boldsymbol V}_{R}^{-1/2}\left[\left( \overline{\mathbf{R}} - \mathbf{R} \overline{\mathbf{H}}_R^\ast\mathbf{B}_R^{-1} \right)^{^\intercal} \mathbf{E}_t\left( \overline{\mathbf{C}} - \mathbf{C} \overline{\mathbf{H}}_C^\ast\mathbf{B}_C^{-1}\right)
+ \left( \overline{\mathbf{R}} - \mathbf{R} \overline{\mathbf{H}}_R^\ast\mathbf{B}_R^{-1}\right)^{^\intercal} \mathbf{E}_t \mathbf{C}\overline{\mathbf{H}}_C^\ast\mathbf{B}_C^{-1} \right.\\
&& + \left. \mathbf{B}_R^{-1}\left(\overline{\mathbf{H}}_R^\ast\right)^{^\intercal} \mathbf{R}^{^\intercal} \mathbf{E}_t \left( \overline{\mathbf{C}} - \mathbf{C} \overline{\mathbf{H}}_C^\ast\mathbf{B}_C^{-1}\right)+ \mathbf{B}_R^{-1}\left(\overline{\mathbf{H}}_R^\ast\right)^{^\intercal} \mathbf{R}^{^\intercal} \mathbf{E}_t \mathbf{C} \overline{\mathbf{H}}_C^\ast\mathbf{B}_C^{-1} \right]\overline{\boldsymbol V}_{C}^{-1/2}
\end{eqnarray*}}
and
$$
\overline{\lambda}_{R,1}^{1/2}\overline{\boldsymbol V}_{R}^{-1/2}\mathbf{B}_R\overline{\mathbf{H}}_R^{\ast^{-1}}\mathbf{F}_t\left(\overline{\mathbf{H}}_C^{\ast^{-1}}\right)^{^\intercal}\mathbf{B}_C\overline{\boldsymbol V}_{C}^{-1/2} = \check{\nu}_{R,1}^{\ast^{1/2}}\check{\boldsymbol V}_R^{\ast^{-1/2}}\overline{{\boldsymbol H}}_R^{\ast^{-1}}{\boldsymbol F}_t\left(\overline{{\boldsymbol H}}_C^{\ast^{-1}}\right)^{^\intercal}\check{\boldsymbol V}_C^{\ast^{-1/2}},
$$
where $\check{\nu}_{R,1}^{\ast} = p_1^{-\alpha_{R,1}}p_2^{-\alpha_{C,1}}\overline{\lambda}_{R,1}$ is the first diagonal element of $\check{\boldsymbol V}_R^{\ast}$, which is positive and bounded {\em w.p.a.1} according to Lemma \ref{le:B.18}.

By Theorem \ref{thm:4.2}(i), we readily have that
\begin{eqnarray}
\left\Vert\overline{\mathbf{R}}\mathbf{B}_R\overline{\mathbf{H}}_R^{\ast^{-1}}-\mathbf{R}\right\Vert_F&=&O_P\left(p_1^{\alpha_{R,1}-\alpha_{R,r_1}/2}\left(p_1^{1/2-\alpha_{R,1}}p_2^{1-\alpha_{C,1}} + p_1^{1-\alpha_{R,1}}p_2^{1/2-\alpha_{C,1}}T^{-1/2} \right)\right),\label{eqA.40}\\
\left\Vert\overline{\mathbf{C}}\mathbf{B}_C\overline{\mathbf{H}}_C^{\ast^{-1}}-\mathbf{C}\right\Vert_F&=&O_P\left(p_2^{\alpha_{C,1}-\alpha_{C,r_2}/2}\left(p_2^{1/2-\alpha_{C,1}}p_1^{1-\alpha_{R,1}} + p_2^{1-\alpha_{C,1}}p_1^{1/2-\alpha_{R,1}}T^{-1/2} \right)\right).\label{eqA.41}
\end{eqnarray}
By (\ref{eqA.40}), (\ref{eqA.41}), $\max_{1\leq t\leq T}\|\mathbf{F}_t\|_F = O_P(T^{1/2})$ and Assumption \ref{ass:4.4}(i), we may show that
\begin{eqnarray*}
&&\max_{1 \leq t \leq T} \left\|\overline{\lambda}_{R,1}^{1/2}\overline{\boldsymbol V}_{R}^{-1/2} \overline{\mathbf{R}}^{^\intercal}\left(\mathbf{R}-\overline{\mathbf{R}}\mathbf{B}_R\overline{\mathbf{H}}_R^{\ast^{-1}}\right)\mathbf{F}_t\left(\mathbf{C}-\overline{\mathbf{C}}\mathbf{B}_C\overline{\mathbf{H}}_C^{\ast^{-1}}\right)^{^\intercal}\overline{\mathbf{C}}\overline{\boldsymbol V}_{C}^{-1/2}\right\|_F \\
&=& O_P\left(p_1^{\alpha_{R,1}-\alpha_{R,r_1}}\left(p_1^{1/2-\alpha_{R,1}}p_2^{1-\alpha_{C,1}} + p_1^{1-\alpha_{R,1}}p_2^{1/2-\alpha_{C,1}}T^{-1/2} \right)\right) O_P(T^{1/2})\\
&&O_P\left(p_2^{\alpha_{C,1}-\alpha_{C,r_2}}\left(p_2^{1/2-\alpha_{C,1}}p_1^{1-\alpha_{R,1}} + p_2^{1-\alpha_{C,1}}p_1^{1/2-\alpha_{R,1}}T^{-1/2} \right)\right)\notag\\
&=& o_P\left(\varpi_1^\ast(p_1,p_2,T)\right),\\
&&\max_{1 \leq t \leq T}\left\| \overline{\lambda}_{R,1}^{1/2}\overline{\boldsymbol V}_{R}^{-1/2}\overline{\mathbf{R}}^{^\intercal}\left(\mathbf{R}-\overline{\mathbf{R}}\mathbf{B}_R\overline{\mathbf{H}}_R^{\ast^{-1}}\right)\mathbf{F}_t\left(\overline{\mathbf{H}}_C^{\ast^{-1}}\right)^{^\intercal}\mathbf{B}_C\overline{\boldsymbol V}_{C}^{-1/2}\right\|_F\\
&=& O_P\left(p_1^{\alpha_{R,1}-\alpha_{R,r_1}}\left(p_1^{1/2-\alpha_{R,1}}p_2^{1-\alpha_{C,1}} + p_1^{1-\alpha_{R,1}}p_2^{1/2-\alpha_{C,1}}T^{-1/2} \right)\right) O_P(T^{1/2})\\
&=&O_P\left(\varpi_1^\ast(p_1,p_2,T)\right),\\
&&\max_{1 \leq t \leq T}\left\|\overline{\lambda}_{R,1}^{1/2}\overline{\boldsymbol V}_{R}^{-1/2} \mathbf{B}_R\overline{\mathbf{H}}_R^{\ast^{-1}} \mathbf{F}_t \left(\mathbf{C}-\overline{\mathbf{C}}\mathbf{B}_C\overline{\mathbf{H}}_C^{\ast^{-1}}\right)^{^\intercal}\overline{\mathbf{C}} \overline{\boldsymbol V}_{C}^{-1/2}\right\| _F\\
&=&O_P\left(p_2^{\alpha_{C,1}-\alpha_{C,r_2}}\left(p_2^{1/2-\alpha_{C,1}}p_1^{1-\alpha_{R,1}} + p_2^{1-\alpha_{C,1}}p_1^{1/2-\alpha_{R,1}}T^{-1/2} \right)\right)O_P(T^{1/2})\\
&=&O_P\left(\varpi_1^\ast(p_1,p_2,T)\right).
\end{eqnarray*}
By Theorem \ref{thm:4.2}(i), Lemma \ref{le:B.16} and Assumption \ref{ass:4.4}(i), we have
\begin{eqnarray*}
&&\max_{1 \leq t \leq T} \left\|\overline{\lambda}_{R,1}^{1/2}\overline{\boldsymbol V}_{R}^{-1/2}\left( \overline{\mathbf{R}} - \mathbf{R} \overline{\mathbf{H}}_R^\ast\mathbf{B}_R^{-1} \right)^{^\intercal} \mathbf{E}_t^\circ\left( \overline{\mathbf{C}} - \mathbf{C} \overline{\mathbf{H}}_C^\ast\mathbf{B}_C^{-1}\right)\overline{\boldsymbol V}_{C}^{-1/2} \right\|_F \\
&=& O_P\left(p_1^{1/2-\alpha_{R,r_1}/2}p_2^{1/2-\alpha_{C,r_2}/2}T^{1/4}\right)O_P\left(p_1^{\alpha_{R,1}-\alpha_{R,r_1}}\left(p_1^{1/2-\alpha_{R,1}}p_2^{1-\alpha_{C,1}}+p_1^{1-\alpha_{R,1}}p_2^{1/2-\alpha_{C,1}}T^{-1/2}\right)\right)\notag\\
&&O_P\left(p_2^{\alpha_{C,1}-\alpha_{C,r_2}}\left(p_2^{1/2-\alpha_{C,1}}p_1^{1-\alpha_{R,1}}+p_2^{1-\alpha_{C,1}}p_1^{1/2-\alpha_{R,1}}T^{-1/2}\right)\right)\notag \\
&=& o_P\left(\varpi_1^{\ast}(p_1,p_2,T) \right),\\
&&\max_{1 \leq t \leq T} \left\|\overline{\lambda}_{R,1}^{1/2}\overline{\boldsymbol V}_{R}^{-1/2} \left( \overline{\mathbf{R}} - \mathbf{R} \overline{\mathbf{H}}_R^\ast\mathbf{B}_R^{-1}\right)^{^\intercal} \mathbf{E}_t^\circ \mathbf{C}\overline{\mathbf{H}}_C^\ast\mathbf{B}_C^{-1}\overline{\boldsymbol V}_{C}^{-1/2}\right\|_F\\
&=&O_P\left(p_1^{1/2-\alpha_{R,r_1}/2}p_2^{\alpha_{C,1}/2-\alpha_{C,r_2}}T^{1/4} p_1^{\alpha_{R,1}-\alpha_{R,r_1}}\left(p_1^{1/2-\alpha_{R,1}}p_2^{1-\alpha_{C,1}}+p_1^{1-\alpha_{R,1}}p_2^{1/2-\alpha_{C,1}}T^{-1/2}\right) \right)\\
&=&o_P\left(\varpi_1^\ast(p_1,p_2,T)\right),\\
&&\max_{1 \leq t \leq T}\left\|\overline{\lambda}_{R,1}^{1/2}\overline{\boldsymbol V}_{R}^{-1/2}\mathbf{B}_R^{-1}\left(\overline{\mathbf{H}}_R^\ast\right)^{^\intercal} \mathbf{R}^{^\intercal} \mathbf{E}_t^\circ \left( \overline{\mathbf{C}} - \mathbf{C} \overline{\mathbf{H}}_C^\ast\mathbf{B}_C^{-1}\right) \overline{\boldsymbol V}_{C}^{-1/2}\right\|_F \\
&=&O_P\left(p_2^{1/2-\alpha_{C,r_2}/2}p_1^{\alpha_{R,1}/2-\alpha_{R,r_1}}T^{1/4} p_2^{\alpha_{C,1}-\alpha_{C,r_2}}\left(p_2^{1/2-\alpha_{C,1}}p_1^{1-\alpha_{R,1}}+p_2^{1-\alpha_{C,1}}p_1^{1/2-\alpha_{R,1}}T^{-1/2}\right) \right)\\
&=&o_P\left(\varpi_1^\ast(p_1,p_2,T)\right),\\
&&\max_{1 \leq t \leq T} \left\| \overline{\lambda}_{R,1}^{1/2}\overline{\boldsymbol V}_{R}^{-1/2}\mathbf{B}_R^{-1}\left(\overline{\mathbf{H}}_R^\ast\right)^{^\intercal} \mathbf{R}^{^\intercal} \mathbf{E}_t^\circ \mathbf{C} \overline{\mathbf{H}}_C^\ast\mathbf{B}_C^{-1}\overline{\boldsymbol V}_{C}^{-1/2}\right\|_F\notag\\
&=&O_P\left(p_1^{\alpha_{R,1}/2-\alpha_{R,r_1}}p_2^{\alpha_{C,1}/2-\alpha_{C,r_2}}T^{1/4}\right)= o_P\left(\varpi_1^{\ast}(p_1,p_2,T) \right). 
\end{eqnarray*}
By Theorem \ref{thm:4.2}(i), Lemma \ref{le:B.20} and Assumption \ref{ass:4.4}(i), we have
\begin{eqnarray*}
&&\max_{1 \leq t \leq T} \left\|\overline{\lambda}_{R,1}^{1/2}\overline{\boldsymbol V}_{R}^{-1/2}\left( \overline{\mathbf{R}} - \mathbf{R} \overline{\mathbf{H}}_R^\ast\mathbf{B}_R^{-1}\right)^{^\intercal} \mathbf{E}_t^\dag \left( \overline{\mathbf{C}} - \mathbf{C} \overline{\mathbf{H}}_C^\ast\mathbf{B}_C^{-1}\right)\overline{\boldsymbol V}_{C}^{-1/2} \right\|_F \\
&=&O_P\left(\frac{s_R^{1/2}s_C^{1/2}}{p_1^{\alpha_{R,r_1}/2}p_2^{\alpha_{C,r_2}/2}}T^{1/4}\right)O_P\left(p_1^{1/2-\alpha_{R,1}}p_2^{1-\alpha_{C,1}}+p_1^{1-\alpha_{R,1}}p_2^{1/2-\alpha_{C,1}}T^{-1/2}\right)O_P\left(\varpi_1^{\ast}(p_1,p_2,T)\right)\\
&=& o_P\left(\varpi_1^{\ast}(p_1,p_2,T) \right),\\
&&\max_{1 \leq t \leq T} \left\|\overline{\lambda}_{R,1}^{1/2}\overline{\boldsymbol V}_{R}^{-1/2}\left(\overline{\mathbf{R}} - \mathbf{R} \overline{\mathbf{H}}_R^\ast \mathbf{B}_R^{-1} \right)^{^\intercal} \mathbf{E}_t^\dag \mathbf{C} \overline{\mathbf{H}}_C^\ast\mathbf{B}_C^{-1}\overline{\boldsymbol V}_{C}^{-1/2}\right\|_F\\	
&=&O_P\left(\frac{s_R^{1/2}s_C^{\alpha_{C,1}/2}}{p_1^{\alpha_{R,r_1}/2}p_2^{\alpha_{C,r_2}}}T^{1/4}\right)O_P\left(\varpi_1^{\ast}(p_1,p_2,T)\right)=o_P\left(\varpi_1^{\ast}(p_1,p_2,T) \right),\\
&&\max_{1 \leq t \leq T}\left\|\overline{\lambda}_{R,1}^{1/2}\overline{\boldsymbol V}_{R}^{-1/2}\mathbf{B}_R^{-1}\overline{\mathbf{H}}_R^{\ast^\intercal} \mathbf{R}^{^\intercal} \mathbf{E}_t^\dag \left( \overline{\mathbf{C}} - \mathbf{C} \overline{\mathbf{H}}_C^\ast\mathbf{B}_C^{-1}\right) \overline{\boldsymbol V}_{C}^{-1/2}\right\|_F \\
&=&O_P\left(\frac{s_C^{1/2}s_R^{\alpha_{R,1}/2}}{p_2^{\alpha_{C,r_2}/2}p_1^{R,r_1}}T^{1/4}\right)O_P\left(\varpi_1^{\ast}(p_1,p_2,T)\right)=o_P\left(\varpi_1^{\ast}(p_1,p_2,T) \right),\\
&&\max_{1 \leq t \leq T} \left\| \overline{\lambda}_{R,1}^{1/2}\overline{\boldsymbol V}_{R}^{-1/2}\mathbf{B}_R^{-1}\overline{\mathbf{H}}_R^{\ast^\intercal} \mathbf{R}^{^\intercal} \mathbf{E}_t^\dag \mathbf{C} \overline{\mathbf{H}}_C^\ast\mathbf{B}_C^{-1}\overline{\boldsymbol V}_{C}^{-1/2}\right\|_F\notag\\
&=&O_P\left(\frac{s_R^{\alpha_{R,1}/2}s_C^{\alpha_{C,1}/2}}{p_1^{\alpha_{R,r_1}}p_2^{\alpha_{C,r_2}}}T^{3/4}\right) = O_P\left(\varpi_2^{\ast}(p_1,p_2,T) \right). 
\end{eqnarray*}
Combining the above results, we complete the proof of (\ref{eq4.9}). 

\medskip

(iii) The proof is similar to that of Theorem \ref{thm:4.1}(iii). We only need to modify the proof of (\ref{eqA.38}). Following the proof of Lemma \ref{le:B.18} and using the condition $p_1^{\alpha_{R,1}-\alpha_{R,r_1}}\xi_R^\dag(p_1,p_2,T)=o(1)$, we have
\begin{eqnarray}
\frac{\overline\lambda_{R,r_1+1}}{\overline\lambda_{R,r_1}}&=&O_P\left(p_1^{1/2-\alpha_{R,r_1}}p_2^{1-\alpha_{C,1}} + p_1^{1-\alpha_{R,r_1}}p_2^{1/2-\alpha_{C,1}} T^{-1/2}\right)\notag\\
&=&O_P\left(\xi_R(p_1,p_2,T\right)=o_P\left(p_1^{\alpha_{R,r_1}-\alpha_{R,1}}\right),\notag
\end{eqnarray} 
leading to
$$
{\sf P}\left(\left\vert\frac{\overline\lambda_{R,r_1+1}}{\overline\lambda_{R,r_1}}\right\vert \geq \epsilon_1\times p_1^{\alpha_{R,r_1}-\alpha_{R,1}}\right)\rightarrow0.
$$
\hfill$\blacksquare$


\section*{Appendix B:\ Technical lemmas}
\renewcommand{\theequation}{B.\arabic{equation}}
\setcounter{equation}{0}

\renewcommand{\thelemma}{B.\arabic{lemma}}
\setcounter{lemma}{0}

\begin{lemma}\label{le:B.1}

Suppose that Assumption \ref{ass:3.2}(i)(ii) is satisfied. Then we have
\begin{eqnarray}
&&{\sf E}\left\Vert \sum_{t=1}^{T}{\boldsymbol E}_t{\boldsymbol E}_{t}^{^\intercal}\right\Vert_F^2=O\left(p_1p_2^2T^2+p_1^2p_2T\right),\label{eqB.1}\\
&&{\sf E}\left\Vert \sum_{t=1}^{T}{\boldsymbol E}_t^{^\intercal}{\boldsymbol E}_{t}\right\Vert_F^2=O\left(p_1^2p_2T^2+p_1p_2^2T\right).\label{eqB.2}
\end{eqnarray}

\end{lemma}

\noindent{\bf Proof of Lemma \ref{le:B.1}}.\ \ We first give the proof of (\ref{eqB.1}). Write ${\boldsymbol E}_{t,\bullet j}$ as the $j$-th column vector of ${\boldsymbol E}_{t}$. Note that
\begin{eqnarray}
{\sf E}\left\Vert \sum_{t=1}^{T}{\boldsymbol E}_t{\boldsymbol E}_{t}^{^\intercal}\right\Vert_F^2&=&{\sf E}\left\Vert \sum_{t=1}^{T}\sum_{j=1}^{p_2}{\boldsymbol E}_{t,\bullet j}{\boldsymbol E}_{t,\bullet j}^{^\intercal}\right\Vert_F^2\notag\\
&=&\sum_{i_1=1}^{p_1}\sum_{i_2=1}^{p_1}{\sf E}\left\vert \sum_{t=1}^{T}\sum_{j=1}^{p_2}e_{t,(i_1,j)}e_{t,(i_2,j)}\right\vert^2\notag\\
&\leq& 2T^2p_2^2\max_{1\leq t\leq T}\sum_{i_1=1}^{p_1}\sum_{i_2=1}^{p_1}\vert\rho_{R,t,(i_1,i_2)}\vert^2+\notag\\
&&2\sum_{i_1=1}^{p_1}\sum_{i_2=1}^{p_1}{\sf E}\left\vert\sum_{t=1}^{T}\sum_{j=1}^{p_2} \left[e_{t,(i_1,j)}e_{t,(i_2,j)}-\rho_{R,t,(i_1,i_2)}\right]\right\vert^2\notag\\
&\leq&2c_1p_1p_2^2T^2+2c_1p_1^2p_2T,\notag
\end{eqnarray}
using Assumption \ref{ass:3.2}(i)(ii). The proof of (\ref{eqB.1}) is completed. The proof of (\ref{eqB.2}) is similar.\hfill$\blacksquare$ 

\medskip

By Assumption \ref{ass:3.1}(i) and the Beveridge-Nelson decomposition, 
\begin{equation}
	{\mathbf f}_t=\sum_{s=1}^t{\mathbf u}_s+{\mathbf f}_0=\overline{\boldsymbol A}\sum_{s=1}^t{\boldsymbol \eta}_s+{\mathbf f}_0+\widetilde{\mathbf u}_0-\widetilde{\mathbf u}_t,\label{eqB.3}
\end{equation}
where $\widetilde{\mathbf u}_t=\sum_{j=0}^\infty\widetilde{\boldsymbol A}_j{\boldsymbol\eta}_{t-j}$ with $\widetilde{\boldsymbol A}_j=\sum_{k=j+1}^\infty{\boldsymbol A}_k$. Letting $\overline{\mathbf f}_t=\overline{\boldsymbol A}\sum_{s=1}^t{\boldsymbol \eta}_s$ and $\widetilde{\mathbf f}_t={\mathbf f}_0+\widetilde{\mathbf u}_0-\widetilde{\mathbf u}_t$, we write
\begin{equation}\label{eqB.4}
	{\boldsymbol F}_t={\sf vec}^{-1}({\mathbf f}_t)={\sf vec}^{-1}(\overline{\mathbf f}_t)+{\sf vec}^{-1}(\widetilde{\mathbf f}_t)=:\overline{\boldsymbol F}_t+\widetilde{\boldsymbol F}_t.
\end{equation}

\medskip

\begin{lemma}\label{le:B.2}

Suppose that Assumptions \ref{ass:3.1}(i)(ii) and \ref{ass:3.2} are satisfied. Then we have
\begin{equation}\label{eqB.5}
{\sf E}\left\|\sum_{t=1}^{T} {\boldsymbol F}_t {\boldsymbol C}^{^\intercal} {\boldsymbol E}_{t}^{^\intercal}\right\|_F^2 = O\left(p_1 p_2^{\alpha_C}T^2\right),
\end{equation}
and
\begin{equation}\label{eqB.6}
{\sf E}\left\|\sum_{t=1}^{T} {\boldsymbol E}_t {\boldsymbol C}^{^\intercal} {\boldsymbol F}_{t}^{^\intercal}\right\|_F^2 = O\left(p_1 p_2^{\alpha_C}T^2\right).
\end{equation}
\end{lemma}

\noindent{\bf Proof of Lemma \ref{le:B.2}}.\ \ We only prove \eqref{eqB.5} as the proof of \eqref{eqB.6} is exactly the same. Recall that ${\boldsymbol C}_{j\bullet}$ is the $j$-th row vector of ${\boldsymbol C}$ and ${\boldsymbol E}_{t,\bullet j}$ is the $j$-th column vector of ${\boldsymbol E}_{t}$. Note that
\begin{eqnarray}
	{\sf E}\left\Vert \sum_{t=1}^{T}{\boldsymbol F}_t{\boldsymbol C}^{^\intercal}{\boldsymbol E}_{t}^{^\intercal}\right\Vert_F^2&=&{\sf E}\left\Vert \sum_{t=1}^{T}{\boldsymbol F}_t\sum_{j=1}^{p_2}{\boldsymbol C}_{j\bullet}^{^\intercal}{\boldsymbol E}_{t,\bullet j}^{^\intercal}\right\Vert_F^2\notag\\
	&=&\sum_{i=1}^{p_1}{\sf E}\left\Vert \sum_{t=1}^{T}\sum_{j=1}^{p_2}e_{t,(i,j)}{\boldsymbol C}_{j\bullet}{\boldsymbol F}_t^{^\intercal}\right\Vert_F^2\notag\\
	&\leq&2\sum_{i=1}^{p_1}{\sf E}\left\Vert \sum_{t=1}^{T}\sum_{j=1}^{p_2}e_{t,(i,j)}{\boldsymbol C}_{j\bullet}\overline{\boldsymbol F}_t^{^\intercal}\right\Vert_F^2+2\sum_{i=1}^{p_1}{\sf E}\left\Vert \sum_{t=1}^{T}\sum_{j=1}^{p_2}e_{t,(i,j)}{\boldsymbol C}_{j\bullet}\widetilde{\boldsymbol F}_t^{^\intercal}\right\Vert_F^2,\notag
\end{eqnarray}
where the last inequality is due to (\ref{eqB.4}) and the triangle inequality. 

By the moment condition in Assumption \ref{ass:3.1}(i), we have $\max_{0\leq t\leq T}{\sf E}\Vert \widetilde{\mathbf u}_t\Vert_F^4<\infty$ and ${\sf E}\Vert {\mathbf f}_0\Vert_F^4<\infty$, indicating that 
\begin{equation}\label{eqB.7}
	\max_{1\leq t\leq T}{\sf E}\Vert \widetilde{\mathbf F}_t\Vert_F^4<\infty.
\end{equation}
By using Assumption \ref{ass:3.2}(iii), Cauchy-Schwarz inequality and (\ref{eqB.7}), we have
$$
\sum_{i=1}^{p_1}{\sf E}\left\|\sum_{t=1}^{T}\sum_{j=1}^{p_2}e_{t,(i,j)}{\boldsymbol C}_{j\cdot}\widetilde{{\boldsymbol F}}_t^{^\intercal} \right\|_F^2 = O\left(p_1p_2^{\alpha_C} T^2\right).
$$
Letting $G_{s,i}=\sum_{t=s}^{T}\sum_{j=1}^{p_2}e_{t,(i,j)}{\boldsymbol C}_{j\cdot}$, since $\{\boldsymbol{\eta}_t\}$ is independent of $\{{\boldsymbol E}_t\}$, we have
\begin{eqnarray*}
	\sum_{i=1}^{p_1}{\sf E}\left\|\sum_{t=1}^{T}\sum_{j=1}^{p_2}e_{t,(i,j)}{\boldsymbol C}_{j\cdot}\overline{{\boldsymbol F}}_t^{^\intercal} \right\|_F^2 &=&\sum_{i=1}^{p_1}{\sf E}\left\|\sum_{s=1}^{T} G_{s,i}[\mathrm{vec}^{-1}(\overline{\mathbf{A}}\boldsymbol{\eta}_s)]^{^\intercal} \right\|_F^2 \\
	&=&\sum_{i=1}^{p_1}\sum_{s=1}^{T}{\sf E}\left\| G_{s,i}[\mathrm{vec}^{-1}(\overline{\mathbf{A}}\boldsymbol{\eta}_s)]^{^\intercal} \right\|_F^2\\
	&=&O\left(p_1p_2^{\alpha_C}T^2\right).
\end{eqnarray*}
The proof of (\ref{eqB.5}) is completed. \hfill$\blacksquare$

\medskip

\begin{lemma}\label{le:B.3}
Suppose that Assumptions \ref{ass:3.1}(i)(ii) and \ref{ass:3.2} are satisfied. Then we have
	\begin{equation}\label{eqB.8}
		\left\|\frac{1}{p_2^{\alpha_C}T^2}\sum_{t=1}^{T} {\boldsymbol F}_t {\boldsymbol C}^{^\intercal}{\boldsymbol C} {\boldsymbol F}_{t}^{^\intercal} - \int_{0}^{1}\mathbf{W}(u)\boldsymbol{\Sigma}_{C}\mathbf{W}(u)^{^\intercal}\mathrm{d}u\right\|_F = o_P\left(1\right),
	\end{equation}
	and
	\begin{equation}\label{eqB.9}
		\left\|\frac{1}{p_1^{\alpha_R}T^2}\sum_{t=1}^{T} {\boldsymbol F}_t {\boldsymbol R}^{^\intercal}{\boldsymbol R} {\boldsymbol F}_{t}^{^\intercal} - \int_{0}^{1}\mathbf{W}(u)\boldsymbol{\Sigma}_{R}\mathbf{W}(u)^{^\intercal}\mathrm{d}u\right\|_F = o_P\left(1\right).
	\end{equation}
	
\end{lemma}

\noindent{\bf Proof of Lemma \ref{le:B.3}}.\ \ By (\ref{eq3.2}) and (\ref{eq3.3}) in Assumption \ref{ass:3.1}(ii) as well as the continuous mapping theorem, it is sufficient to show 
\[
\max_{1\leq t\leq T} T^{-1/2} \left\|{\boldsymbol F}_t - \mathbf{W}(t/T)\right\|_F = o_P(1).
\] 
Write
$$
\max_{1\leq t\leq T} T^{-1/2} \left\|{\boldsymbol F}_t - \mathbf{W}(t/T)\right\|_F\leq \max_{1\leq t\leq T}T^{-1/2}  \left\|\overline{{\boldsymbol F}}_t - \mathbf{W}(t/T)\right\|_F + \max_{1\leq t\leq T}T^{-1/2}  \left\|\widetilde{{\boldsymbol F}}_t\right\|_F,
$$
where $\overline{{\boldsymbol F}}_t$ and $\widetilde{{\boldsymbol F}}_t$ are defined in (\ref{eqB.4}). Since $\overline{{\boldsymbol F}}_t$ is a partial sum of i.i.d random matrix, by using the Gaussian approximation in Theorem 4 of \cite{GZ09}, we have 
$$
\max_{1\leq t\leq T}T^{-1/2}  \left\|\overline{{\boldsymbol F}}_t - \mathbf{W}(t/T)\right\|_F = O_P(T^{-1/4}).
$$
Since ${\sf E}\|\widetilde{{\boldsymbol F}}_t\|_F^4$ is bounded by (\ref{eqB.7}), we have
$$
{\sf E}\left[\max_{1\leq t\leq T}\left\|\widetilde{{\boldsymbol F}}_t\right\|_F^4\right] \leq \sum_{t=1}^{T}{\sf E}\left\|\widetilde{{\boldsymbol F}}_t\right\|_F^4 = O(T),
$$
indicating that 
\[
\max_{1\leq t\leq T}T^{-1/2}  \left\|\widetilde{{\boldsymbol F}}_t\right\|_F = O_P(T^{-1/4}).
\]
The proof is completed.\hfill$\blacksquare$

\medskip

Let $\nu_{R,k}$ be the $k$-th largest eigenvalue of ${\boldsymbol\Sigma}_R^{1/2}[\int_0^1 {\boldsymbol W}(u){\boldsymbol\Sigma}_C{\boldsymbol W}(u)^{^\intercal}du]{\boldsymbol\Sigma}_R^{1/2}$, $k=1,\ldots,r_1$, and $\nu_{C,k}$ the $k$-th largest eigenvalue of ${\boldsymbol\Sigma}_C^{1/2}[\int_0^1 {\boldsymbol W}(u)^{^\intercal}{\boldsymbol\Sigma}_R{\boldsymbol W}(u)du]{\boldsymbol\Sigma}_C^{1/2}$, $k=1,\ldots,r_2$.

\medskip

\begin{lemma}\label{le:B.4}

Suppose that Assumptions \ref{ass:3.1}, \ref{ass:3.2} and \ref{ass:3.3}(i) are satisfied. Then we have
	\begin{equation}\label{eqB.10}
		\left\|\widehat{{\boldsymbol V}}_R - {\boldsymbol V}_R\right\| = o_P\left(1\right)
	\end{equation}
	and
	\begin{equation}\label{eqB.11}
		\left\|\widehat{{\boldsymbol V}}_C - {\boldsymbol V}_C\right\| = o_P\left(1\right),
	\end{equation}
where ${\boldsymbol V}_R ={\sf diag}\{\nu_{R,1},...,\nu_{R,r_1}\}$ and ${\boldsymbol V}_C = {\sf diag}\{\nu_{C,1},...,\nu_{C,r_2}\}$.

\end{lemma}

\noindent{\bf Proof of Lemma \ref{le:B.4}}.\ \ Note that $\widehat{{\boldsymbol V}}_R$ collects the first $r_1$ largest eigenvalues of matrix $(p_1^{\alpha_R}p_2^{\alpha_C}T)^{-1}\widehat{\boldsymbol{\Omega}}_{R}$ and 
\begin{eqnarray*}
\frac{1}{p_1^{\alpha_R}p_2^{\alpha_C}T}\widehat{\boldsymbol{\Omega}}_{R}
	&=&\frac{1}{p_1^{\alpha_R}p_2^{\alpha_C}T^2}\sum_{t=1}^{T}{\boldsymbol R}{\boldsymbol F}_t{\boldsymbol C}^{^\intercal}{\boldsymbol C}{\boldsymbol F}_{t}^{^\intercal}{\boldsymbol R}^{^\intercal} +\frac{1}{p_1^{\alpha_R}p_2^{\alpha_C}T}\boldsymbol{\Lambda}_{1}+\frac{1}{p_1^{\alpha_R}p_2^{\alpha_C}T}\boldsymbol{\Lambda}_{2}+\frac{1}{p_1^{\alpha_R}p_2^{\alpha_C}T}\boldsymbol{\Lambda}_{3}\\
	&=:& \mathbf{I}_1 + \mathbf{I}_2 + \mathbf{I}_3 + \mathbf{I}_4.
\end{eqnarray*}
For the first term $\mathbf{I}_1$, by using $\max_{1\leq t\leq T}T^{-1/2} \left\|{\boldsymbol F}_t - \mathbf{W}(t/T)\right\|_F = O_P(T^{-1/4})$ in the proof of Lemma \ref{le:B.3} and the continuous mapping theorem, we have
\[
\left\|\mathbf{I}_1 - \mathbf{I}_1^*\right\| = o_P(1),\quad \mathbf{I}_1^* = \frac{1}{p_1^{\alpha_R}}{\boldsymbol R}\left[\int_{0}^{1}\mathbf{W}(u)\boldsymbol{\Sigma}_{C} \mathbf{W}(u)^\intercal \mathrm{d}u\right]{\boldsymbol R}^{^\intercal}.
\]
Meanwhile, following the proof of Theorem \ref{thm:3.1} in Appendix A and using Assumption \ref{ass:3.3}(i), we have
$$
\left\|\mathbf{I}_2 + \mathbf{I}_3 + \mathbf{I}_4\right\| = O_P\left(p_1^{\frac{1-\alpha_R}{2}}p_2^{-\frac{\alpha_C}{2}}T^{-1} + p_1^{\frac{1}{2}-\alpha_R}p_2^{1-\alpha_C}T^{-1} + p_1^{1-\alpha_R}p_2^{\frac{1}{2}-\alpha_C}T^{-\frac{3}{2}}\right)=o_P(1).
$$
Hence, using the standard matrix perturbation theory and noting  
\[
\max_{1\leq k\leq r_1}\left\vert \psi_k(\mathbf{I}_1^*)-\nu_{R,k}\right\vert=o_P(1)
\]
by (\ref{eq3.2}) and Assumption \ref{ass:3.1}(iii), we prove \eqref{eqB.10}. In addition, due to the low-rank structure on $\mathbf{I}_1$, its last $p_1-r_1$ eigenvalues are zeros. Consequently, the last $p_1-r_1$ eigenvalues of $(p_1^{\alpha_R}p_2^{\alpha_C}T)^{-1}\widehat{\boldsymbol{\Omega}}_{R}$ converge to zeros at the rate:
$$
O_P\left(p_1^{\frac{1-\alpha_R}{2}}p_2^{-\frac{\alpha_C}{2}}T^{-1} + p_1^{\frac{1}{2}-\alpha_R}p_2^{1-\alpha_C}T^{-1} + p_1^{1-\alpha_R}p_2^{\frac{1}{2}-\alpha_C}T^{-\frac{3}{2}}\right).
$$ 
The proof of \eqref{eqB.11} is analogous and thus omitted. \hfill$\blacksquare$

\medskip

Define
$$
	{\boldsymbol\Delta}_C=\int_0^1 {\boldsymbol W}(u){\boldsymbol\Sigma}_C{\boldsymbol W}(u)^{^\intercal}du\quad \text{and} \quad
	{\boldsymbol\Delta}_R=\int_0^1 {\boldsymbol W}(u)^{^\intercal}{\boldsymbol\Sigma}_R{\boldsymbol W}(u)du.
$$
Let ${\boldsymbol W}_{R,0}$ and ${\boldsymbol W}_{C,0}$ be matrices consisting of the eigenvectors of ${\boldsymbol\Delta}_C^{1/2}{\boldsymbol\Sigma}_R{\boldsymbol\Delta}_C^{1/2}$ and ${\boldsymbol\Delta}_R^{1/2}{\boldsymbol\Sigma}_C{\boldsymbol\Delta}_R^{1/2}$, respectively.

\medskip

\begin{lemma}\label{le:B.5}
Suppose that Assumptions \ref{ass:3.1}, \ref{ass:3.2} and \ref{ass:3.3}(i) are satisfied. The rotation matrices $\widehat{{\boldsymbol H}}_R$ and $\widehat{{\boldsymbol H}}_C$ are invertible w.p.a.1. Furthermore, we have the following convergence results:
	\begin{equation}\label{eqB.12}
		\left\|\widehat{{\boldsymbol H}}_R - {\boldsymbol H}_R\right\| = o_P\left(1\right), \quad {\boldsymbol H}_R = \pmb{\Delta}_C^{1/2}\mathbf{W}_{R,0} {\boldsymbol V}_R^{-1/2},
	\end{equation}
	\begin{equation}\label{eqB.13}
		\left\|\widehat{{\boldsymbol H}}_C - {\boldsymbol H}_C\right\| = o_P\left(1\right), \quad {\boldsymbol H}_C = \pmb{\Delta}_R^{1/2}\mathbf{W}_{C,0} {\boldsymbol V}_C^{-1/2}.
	\end{equation}
\end{lemma}

\noindent{\bf Proof of Lemma \ref{le:B.5}}.\ \ Define $\widehat{\boldsymbol{\Sigma}}_{R} =p_1^{-\alpha_R/2}{\boldsymbol R}^{^\intercal} \widehat{{\boldsymbol R}} = p_1^{-\alpha_R}{\boldsymbol R}^{^\intercal} (p_1^{\alpha_R/2}\widehat{{\boldsymbol R}} )=:p_1^{-\alpha_R}{\boldsymbol R}^{^\intercal} \widetilde{{\boldsymbol R}}$, and
$$
\boldsymbol{\Gamma}_{C} = \frac{1}{p_2^{\alpha_C}T^2}\sum_{t=1}^{T}{\boldsymbol F}_t{\boldsymbol C}^{^\intercal}{\boldsymbol C}{\boldsymbol F}_{t}^{^\intercal}.
$$
It follows from (\ref{eqA.1}) and the proof of Lemma \ref{le:B.4} that
\begin{equation}\label{eqB.14}
	\widehat{{\boldsymbol R}}^{^\intercal} \left(\frac{1}{p_1^{\alpha_R}p_2^{\alpha_C}T^2}\sum_{t=1}^{T}{\boldsymbol R}{\boldsymbol F}_t{\boldsymbol C}^{^\intercal}{\boldsymbol C}{\boldsymbol F}_{t}^{^\intercal}{\boldsymbol R}^{^\intercal}\right) \widehat{{\boldsymbol R}} + o_P(1) = \widehat{{\boldsymbol V}}_{R}.
\end{equation}
Taking limit on both sides of \eqref{eqB.14} and using Lemmas \ref{le:B.3} and \ref{le:B.4} as well as Assumptions \ref{ass:3.1}(ii)(iii), we have
$$
\left(\lim_{p_1\to \infty} \widehat{\boldsymbol{\Sigma}}_{R}\right)^{^\intercal} \boldsymbol{\Delta}_{C}\left(\lim_{p_1\to \infty} \widehat{\boldsymbol{\Sigma}}_{R}\right) = {\boldsymbol V}_R,
$$
indicating that $\widehat{\boldsymbol{\Sigma}}_{R}$ is asymptotically invertible.

We next only prove \eqref{eqB.12} as the proof of \eqref{eqB.13} is similar. Let
$\widetilde{\mathbf{W}}_{R} = \mathbf{W}_{R} \mathbf{D}_R^{-1}$ with $\mathbf{W}_{R} = \boldsymbol{\Gamma}_{C}^{1/2}\widehat{\boldsymbol{\Sigma}}_{R}$ and $\mathbf{D}_R=\left(\mathrm{diag}\{ \mathbf{W}_R^{^\intercal}\mathbf{W}_R\} \right)^{1/2}$, $\widetilde{\boldsymbol{\Omega}}_R = \boldsymbol{\Gamma}_{C}^{1/2}(p_1^{-\alpha_R}{\boldsymbol R}^{^\intercal} {\boldsymbol R})\boldsymbol{\Gamma}_{C}^{1/2}$, 
\[
\widetilde{\boldsymbol{\Omega}}_{*} = \boldsymbol{\Gamma}_{C}^{1/2}\frac{{\boldsymbol R}^{^\intercal}}{p_1^{\alpha_R/2}}\left(\frac{1}{p_1^{\alpha_R}p_2^{\alpha_C}T}\widehat{\boldsymbol{\Omega}}_R-\frac{1}{p_1^{\alpha_R}p_2^{\alpha_C}T^2}\sum_{t=1}^{T}{\boldsymbol R}{\boldsymbol F}_t{\boldsymbol C}^{^\intercal}{\boldsymbol C}{\boldsymbol F}_{t}^{^\intercal}{\boldsymbol R}^{^\intercal} \right)\frac{\widetilde{{\boldsymbol R}}}{p_1^{\alpha_R/2}}.
\]
Note that
\[
\left(\widetilde{\boldsymbol{\Omega}}_R + \widetilde{\boldsymbol{\Omega}}_{*}\mathbf{W}_R^{-1}\right)\widetilde{\mathbf{W}}_R = \widetilde{\mathbf{W}}_R\widehat{{\boldsymbol V}}_R,
\]
indicating that $\widetilde{\mathbf{W}}_R$ is a matrix consisting of the eigenvectors of $\widetilde{\boldsymbol{\Omega}}_R + \widetilde{\boldsymbol{\Omega}}_{*}\mathbf{W}_{R}^{-1}$. On the other hand, as $\widehat{{\boldsymbol H}}_R=\boldsymbol{\Gamma}_{C}\widehat{\boldsymbol{\Sigma}}_{R}\widehat{{\boldsymbol V}}_R^{-1}=\boldsymbol{\Gamma}_{C}^{1/2}(\widetilde{\mathbf{W}}_R\mathbf{D}_R )\widehat{{\boldsymbol V}}_R^{-1}$, using Lemmas \ref{le:B.3} and \ref{le:B.4}, in order to complete the proof of \eqref{eqB.12}, we only need to show 
\begin{equation}\label{eqB.15}
\left\|\mathbf{D}_R^2 - {\boldsymbol V}_R\right\| = o_P(1)\ \ \text{and}\ \ \left\|\widetilde{\mathbf{W}}_R - \mathbf{W}_{R,0}\right\| = o_P(1).
\end{equation}

For the first assertion in (\ref{eqB.15}), we note that
\begin{eqnarray*}
	\left\|\mathbf{D}_R^2 - {\boldsymbol V}_R\right\|&\leq&\left\|\mathbf{D}_R^2 - \widehat{{\boldsymbol V}}_R\right\|+ \left\|\widehat{{\boldsymbol V}}_R - {\boldsymbol V}_R\right\|\\
	& =& \left\|\mathbf{D}_R^2 - \widehat{{\boldsymbol V}}_R\right\|+o_P(1).
\end{eqnarray*}
As in the proof of Lemma \ref{le:B.4},
\begin{eqnarray*}
	\left\|\mathbf{D}_R^2 - \widehat{{\boldsymbol V}}_R\right\|&\leq&\left\| \widehat{{\boldsymbol R}}^{^\intercal}\left(\frac{1}{p_1^{\alpha_R}p_2^{\alpha_C}T^2}\widehat{\boldsymbol{\Omega}}_{R} - \mathbf{I}_1\right) \widehat{{\boldsymbol R}}\right\|_F\\ 
	&=&O_P\left(p_1^{\frac{1-\alpha_R}{2}}p_2^{-\frac{\alpha_C}{2}}T^{-1} + p_1^{\frac{1}{2}-\alpha_R}p_2^{1-\alpha_C}T^{-1} + p_1^{1-\alpha_R}p_2^{\frac{1}{2}-\alpha_C}T^{-\frac{3}{2}}\right)=o_P(1).
\end{eqnarray*}
Let $\mathbf{W}_{R,1}$ be an $r_1\times r_2$ matrix consisting of the eigenvalues of $\widetilde{\boldsymbol{\Omega}}_R$. By Assumption \ref{ass:3.1}(ii) and Lemma \ref{le:B.4}, we have
$$
\left\|\widetilde{\mathbf{W}}_R - \mathbf{W}_{R,0}\right\|\leq\left\|\widetilde{\mathbf{W}}_R - \mathbf{W}_{R,1}\right\| + \left\|\mathbf{W}_{R,1} - \mathbf{W}_{R,0}\right\|=\left\|\widetilde{\mathbf{W}}_R - \mathbf{W}_{R,1}\right\|+o_P(1).
$$
By using \cite{DK70}'s $\sin\theta$ theorem, we have
$$
\left\|\widetilde{\mathbf{W}}_R - \mathbf{W}_{R,1}\right\|\leq C\left\|\widetilde{\boldsymbol{\Omega}}_{*}\right\| \cdot \left\|\mathbf{W}_R^{-1}\right\|=o_P(1),
$$
completing the second assertion in (\ref{eqB.15}).\hfill$\blacksquare$

\medskip

\begin{lemma}\label{le:B.6}

Suppose that Assumptions \ref{ass:3.2} and \ref{ass:3.3}(ii) are satisfied. The following uniform convergence holds:
	\begin{equation}\label{eqB.16}
		\max_{1\leq t\leq T}\left\{\left\|\frac{{\boldsymbol R}^{^\intercal}{\boldsymbol E}_t{\boldsymbol C}}{p_1^{\alpha_R/2}p_2^{\alpha_C/2}}\right\|_F + \left\|\frac{{\boldsymbol E}_t}{p_1^{1/2}p_2^{1/2}}\right\|_F + \left\|\frac{{\boldsymbol E}_t{\boldsymbol C}}{p_1^{1/2}p_2^{\alpha_C/2}}\right\|_F +\left\|\frac{{\boldsymbol R}^{^\intercal}{\boldsymbol E}_t}{p_1^{\alpha_R/2}p_2^{1/2}}\right\|_F \right\} = O_P\left(T^{1/4}\right).
	\end{equation}
\end{lemma}

\noindent{\bf Proof of Lemma \ref{le:B.6}}.\ \ By using Assumption \ref{ass:3.3}(ii), we have 
$$
{\sf E}\left[\max_{1\leq t\leq T}\left\|\frac{{\boldsymbol R}^{^\intercal}{\boldsymbol E}_t{\boldsymbol C}}{p_1^{\alpha/2}p_2^{\beta/2}}\right\|_F^4\right] \leq \sum_{t=1}^{T}{\sf E}\left\|\frac{{\boldsymbol R}^{^\intercal}{\boldsymbol E}_t{\boldsymbol C}}{p_1^{\alpha_R/2}p_2^{\alpha_C/2}}\right\|_F^4 = O(T),
$$
leading to
$$
\max_{1\leq t\leq T}\left\|\frac{{\boldsymbol R}^{^\intercal}{\boldsymbol E}_t{\boldsymbol C}}{p_1^{\alpha_R/2}p_2^{\alpha_C/2}}\right\|_F = O_P\left(T^{1/4}\right).
$$
The other convergence results can be proved in a similar way. \hfill$\blacksquare$

\medskip

\begin{lemma}\label{le:B.7}

Suppose that Assumption \ref{ass:3.6} is satisfied. Then we have
\begin{eqnarray}
&&{\sf E}\left\Vert \sum_{t=1}^{T}(\Delta {\boldsymbol E}_t) (\Delta{\boldsymbol E}_{t})^{^\intercal}\right\Vert_F^2=O\left(p_1p_2^2T^2+p_1^2p_2T\right),\label{eqB.17}\\
&&{\sf E}\left\Vert \sum_{t=1}^{T}(\Delta {\boldsymbol E}_t)^{^\intercal}(\Delta {\boldsymbol E}_{t})\right\Vert_F^2=O\left(p_1^2p_2T^2+p_1p_2^2T\right).\label{eqB.18}
\end{eqnarray}

\end{lemma}

\noindent{\bf Proof of Lemma \ref{le:B.7}}.\ \ Let $\Delta{\boldsymbol E}_{t,\bullet j}$ and $\Delta e_{t,(i,j)}$ be the $j$-th column vector and $(i,j)$-entry of $\Delta{\boldsymbol E}_{t}$, respectively. We only give the proof of (\ref{eqB.17}) as the proof of (\ref{eqB.18}) is analogous. By Assumption \ref{ass:3.6}(i)(ii), we may show that
\begin{equation}\label{eqB.19}
\sum_{i_1=1}^{p_1}\sum_{i_2=1}^{p_1}\vert \rho_{R,t,(i_1,i_2)}^{\Delta e}\vert\leq C p_1,
\ \ 
\max_{1\leq i_1, i_2\leq p_1}{\sf E}\left\vert \sum_{s=t_1}^{t_2}\sum_{j=1}^{p_2} \left[\Delta e_{s,(i_1,j)}\Delta e_{s,(i_2,j)}-\rho_{R,s,(i_1,i_2)}^{\Delta e}\right]\right\vert^2\leq C p_2(t_2-t_1)
\end{equation}
for any $1\leq t\leq T$ and $1\leq t_1<t_2\leq T$, where $\rho_{R,t,(i_1,i_2)}^{\Delta e}=\frac{1}{p_2}\sum_{j=1}^{p_2}{\sf E}[\Delta e_{t,(i_1,j)}\Delta e_{t,(i_2,j)}]$. With (\ref{eqB.19}), we have
\begin{eqnarray}
{\sf E}\left\Vert \sum_{t=1}^{T}(\Delta{\boldsymbol E}_t)(\Delta{\boldsymbol E}_{t})^{^\intercal}\right\Vert_F^2&=&{\sf E}\left\Vert \sum_{t=1}^{T}\sum_{j=1}^{p_2}(\Delta{\boldsymbol E}_{t,\bullet j})(\Delta{\boldsymbol E}_{t,\bullet j})^{^\intercal}\right\Vert_F^2\notag\\
&=&\sum_{i_1=1}^{p_1}\sum_{i_2=1}^{p_1}{\sf E}\left\vert \sum_{t=1}^{T}\sum_{j=1}^{p_2}\Delta e_{t,(i_1,j)}\Delta e_{t,(i_2,j)}\right\vert^2\notag\\
&\leq& 2\sum_{i_1=1}^{p_1}\sum_{i_2=1}^{p_1}{\sf E}\left\vert\sum_{t=1}^{T}\sum_{j=1}^{p_2} \left[\Delta e_{t,(i_1,j)}\Delta e_{t,(i_2,j)}-\rho_{R,t,(i_1,i_2)}^{\Delta e}\right]\right\vert^2+\notag\\
&&2T^2p_2^2\sum_{i_1=1}^{p_1}\sum_{i_2=1}^{p_1}\left\vert\rho_{R,t,(i_1,i_2)}^{\Delta e}\right\vert^2\notag\\
&\leq&C\left( p_1^2p_2T+ p_1p_2^2T^2\right).\notag
\end{eqnarray}
The proof of (\ref{eqB.17}) is completed. \hfill$\blacksquare$ 

\medskip

By Assumption \ref{ass:3.5}(i) and the Granger's representation \citep[e.g.,][]{J95}, 
\begin{equation}
{\sf vec}({\boldsymbol F}_t)={\boldsymbol C}\sum_{s=1}^t {\sf vec}({\boldsymbol V}_s)+{\boldsymbol\alpha}\left({\boldsymbol\beta}^{^\intercal}{\boldsymbol\alpha}\right)^{-1}\ R(L){\boldsymbol\beta}^{^\intercal} {\sf vec}({\boldsymbol V}_t)+{\boldsymbol C}{\sf vec}({\boldsymbol F}_0),\label{eqB.20}
\end{equation}
where ${\boldsymbol C}={\boldsymbol\beta}_\bot \left({\boldsymbol\alpha}_\bot^{^\intercal}{\boldsymbol\beta}_\bot\right)^{-1}{\boldsymbol\alpha}_\bot^{^\intercal}$, $R(L)=\sum_{s=0}^\infty \left({\boldsymbol I}_{k_1k_2}+{\boldsymbol\beta}^{^\intercal}{\boldsymbol\alpha}\right)^s L^s$, and ${\boldsymbol\alpha}_\bot$ and ${\boldsymbol\beta}_\bot$ are the orthogonal complements of ${\boldsymbol\alpha}$ and ${\boldsymbol\beta}$, respectively. It follows from (\ref{eqB.20}) that 
\begin{equation}\label{eqB.21}
{\sf vec}(\Delta {\boldsymbol F}_t)={\boldsymbol C} {\sf vec}({\boldsymbol V}_t)+{\boldsymbol\alpha}\left({\boldsymbol\beta}^{^\intercal}{\boldsymbol\alpha}\right)^{-1}R(L){\boldsymbol\beta}^{^\intercal} {\sf vec}(\Delta {\boldsymbol V}_t),
\end{equation}
which is stationary over $t$.

\medskip

\begin{lemma}\label{le:B.8}

Suppose that Assumptions \ref{ass:3.5}(i)(ii) and \ref{ass:3.6}(i)(ii) are satisfied. Then we have
\begin{equation}\label{eqB.22}
{\sf E}\left\|\sum_{t=1}^{T} (\Delta{\boldsymbol F}_t) {\boldsymbol C}^{^\intercal} (\Delta{\boldsymbol E}_{t})^{^\intercal}\right\|_F^2 = O\left(p_1 p_2^{\alpha_C}T\right),
\end{equation}
and
\begin{equation}\label{eqB.23}
{\sf E}\left\|\sum_{t=1}^{T} (\Delta{\boldsymbol E}_t) {\boldsymbol C}^{^\intercal} (\Delta {\boldsymbol F}_{t})^{^\intercal}\right\|_F^2 = O\left(p_1 p_2^{\alpha_C}T\right).
\end{equation}
\end{lemma}

\noindent{\bf Proof of Lemma \ref{le:B.8}}.\ \ We only prove \eqref{eqB.22} as the proof of \eqref{eqB.23} is similar. By (\ref{eqB.21}) and Assumption \ref{ass:3.5}(ii), 
\begin{equation}\label{eqB.24}
\max_{1\leq t\leq T}{\sf E}\Vert \Delta{\boldsymbol F}_t\Vert_F^4<\infty.
\end{equation}
By Assumption \ref{ass:3.6}(i)(ii), we have
\begin{equation}\label{eqB.25}
\max_{1\leq i\leq p_1}{\sf E}\left\Vert p_2^{-\alpha_C/2}\sum_{j=1}^{p_2}\Delta e_{t,(i,j)}{\boldsymbol C}_{j\bullet}\right\Vert^4\leq C
\end{equation}
for any $1\leq t\leq T$. Combining (\ref{eqB.24}), (\ref{eqB.25}) and Cauchy-Schwarz inequality, we have
\begin{eqnarray}
	{\sf E}\left\Vert \sum_{t=1}^{T} (\Delta{\boldsymbol F}_t) {\boldsymbol C}^{^\intercal} (\Delta{\boldsymbol E}_{t})^{^\intercal}\right\Vert_F^2&=&{\sf E}\left\Vert \sum_{t=1}^{T}\Delta{\boldsymbol F}_t\sum_{j=1}^{p_2}{\boldsymbol C}_{j\bullet}^{^\intercal}\Delta{\boldsymbol E}_{t,\bullet j}^{^\intercal}\right\Vert_F^2\notag\\
	&=&\sum_{i=1}^{p_1}{\sf E}\left\Vert \sum_{t=1}^{T}\sum_{j=1}^{p_2}\Delta e_{t,(i,j)}{\boldsymbol C}_{j\bullet} \Delta {\boldsymbol F}_t^{^\intercal}\right\Vert_F^2\notag\\
	&=&O\left(p_1p_2^{\alpha_C} T\right),\notag
\end{eqnarray}
completing the proof of (\ref{eqB.22}).\hfill$\blacksquare$

\medskip

Let $\nu_{R,k}^\circ$ be the $k$-th largest eigenvalue of  ${\boldsymbol\Sigma}_R^{1/2}{\sf E}\left[ \Delta {\boldsymbol F}_t{\boldsymbol\Sigma}_C \Delta {\boldsymbol F}_t^{^\intercal}\right]{\boldsymbol\Sigma}_R^{1/2}$, $k=1,\ldots,r_1$, and $\nu_{C,k}^\circ$ the $k$-th largest eigenvalue of ${\boldsymbol\Sigma}_C^{1/2} {\sf E}\left[ \Delta {\boldsymbol F}^{^\intercal}_t{\boldsymbol\Sigma}_R \Delta {\boldsymbol F}_t\right] {\boldsymbol\Sigma}_C^{1/2}$, $k=1,\ldots,r_2$. 

\medskip

\begin{lemma}\label{le:B.9}

Suppose that Assumptions \ref{ass:3.1}(ii) and \ref{ass:3.5}--\ref{ass:3.7} are satisfied. Then we have
	\begin{equation}\label{eqB.26}
		\left\|\check{\boldsymbol V}_R - {\boldsymbol V}_{R,\circ}\right\| = o_P\left(1\right)
	\end{equation}
	and
	\begin{equation}\label{eqB.27}
		\left\|\check{\boldsymbol V}_C - {\boldsymbol V}_{C,\circ}\right\| = o_P\left(1\right),
	\end{equation}
where ${\boldsymbol V}_{R,\circ} ={\sf diag}\{\nu_{R,1}^\circ,...,\nu_{R,r_1}^\circ\}$ and ${\boldsymbol V}_{C,\circ} = {\sf diag}\{\nu_{C,1}^\circ,...,\nu_{C,r_2}^\circ\}$.

\end{lemma}

\noindent{\bf Proof of Lemma \ref{le:B.9}}.\ \ Let 
\[
{\boldsymbol\Pi}_\circ=\frac{1}{p_1^{\alpha_R}p_2^{\alpha_C}T}\sum_{t=1}^{T}{\boldsymbol R}(\Delta{\boldsymbol F}_{t}){\boldsymbol C}^{^\intercal}{\boldsymbol C}(\Delta{\boldsymbol F}_{t})^{^\intercal}{\boldsymbol R}^{^\intercal}.
\]
It follows from (\ref{eqA.17})--(\ref{eqA.19}) and the standard matrix perturbation theory that
\begin{equation}\label{eqB.28}
\max_{1\leq k\leq r_1}\left\vert \check{\nu}_{R,k}-\nu_k({\boldsymbol\Pi}_\circ)\right\vert=o_P(1),
\end{equation}
where $\check{\nu}_{R,k}$ denotes the $k$-th largest eigenvalue of $\check{\boldsymbol V}_R$. With (\ref{eqB.28}) and Assumptions \ref{ass:3.1}(ii) and \ref{ass:3.5}(iii), we complete the proof of (\ref{eqB.26}). The proof of (\ref{eqB.27}) is analogous. \hfill$\blacksquare$

\medskip

Define
$$
	{\boldsymbol\Delta}_{C,\circ}={\sf E}\left[ \Delta {\boldsymbol F}_t{\boldsymbol\Sigma}_C \Delta {\boldsymbol F}_t^{^\intercal}\right]\quad \text{and} \quad
	{\boldsymbol\Delta}_{R,\circ}={\sf E}\left[ \Delta {\boldsymbol F}_t^{^\intercal}{\boldsymbol\Sigma}_R \Delta {\boldsymbol F}_t\right].
$$
Let ${\boldsymbol W}_{R,\circ}$ and ${\boldsymbol W}_{C,\circ}$ be matrices consisting of the eigenvectors of ${\boldsymbol\Delta}_{C,\circ}^{1/2}{\boldsymbol\Sigma}_R{\boldsymbol\Delta}_{C,\circ}^{1/2}$ and ${\boldsymbol\Delta}_{R,\circ}^{1/2}{\boldsymbol\Sigma}_C{\boldsymbol\Delta}_{R,\circ}^{1/2}$, respectively.

\medskip

\begin{lemma}\label{le:B.10}
Suppose that Assumptions \ref{ass:3.1}(ii) and \ref{ass:3.5}--\ref{ass:3.7} are satisfied. The rotation matrices $\overline{{\boldsymbol H}}_R$ and $\overline{{\boldsymbol H}}_C$ are invertible w.p.a.1, and have the following convergence properties:
	\begin{equation}\label{eqB.29}
		\left\|\overline{{\boldsymbol H}}_R - {\boldsymbol H}_{R,\circ}\right\| = o_P\left(1\right), \quad {\boldsymbol H}_{R,\circ} = \pmb{\Delta}_{C,\circ}^{1/2}\mathbf{W}_{R,\circ} {\boldsymbol V}_{R,\circ}^{-1/2},
	\end{equation}
	\begin{equation}\label{eqB.30}
		\left\|\overline{{\boldsymbol H}}_C - {\boldsymbol H}_{C,\circ}\right\| = o_P\left(1\right), \quad {\boldsymbol H}_{C,\circ} = \pmb{\Delta}_{R,\circ}^{1/2}\mathbf{W}_{C,\circ} {\boldsymbol V}_{C,\circ}^{-1/2}.
	\end{equation}
\end{lemma}

\noindent{\bf Proof of Lemma \ref{le:B.10}}.\ \ The proof is similar to the proof of Lemma \ref{le:B.5} with some modifications.\hfill$\blacksquare$

\medskip

\begin{lemma}\label{le:B.11}

Suppose that Assumptions \ref{ass:3.6}(ii) and \ref{ass:3.8}(ii)(iii) are satisfied. The following uniform convergence holds:
	\begin{equation}\label{eqB.31}
		\max_{1\leq t\leq T}\left\{\left\|\frac{{\boldsymbol R}^{^\intercal}{\boldsymbol E}_t^\dag{\boldsymbol C}}{s_R^{\alpha_R/2}s_C^{\alpha_C/2}}\right\|_F + \left\|\frac{{\boldsymbol E}_t^\dag}{s_R^{1/2}s_C^{1/2}}\right\|_F + \left\|\frac{{\boldsymbol E}_t^\dag{\boldsymbol C}}{s_R^{1/2}s_C^{\alpha_C/2}}\right\|_F +\left\|\frac{{\boldsymbol R}^{^\intercal}{\boldsymbol E}_t^\dag}{s_R^{\alpha_R/2}s_C^{1/2}}\right\|_F \right\} = O_P\left(T^{3/4}\right).
	\end{equation}
\end{lemma}

\noindent{\bf Proof of Lemma \ref{le:B.11}}.\ \ By using Assumption \ref{ass:3.8}(ii), we have 
$$
{\sf E}\left[\max_{1\leq t\leq T}\left\|\frac{{\boldsymbol R}^{^\intercal}{\boldsymbol E}_t^\dag{\boldsymbol C}}{s_R^{\alpha/2}s_C^{\beta/2}}\right\|_F^4\right] \leq \sum_{t=1}^{T}{\sf E}\left\|\frac{{\boldsymbol R}^{^\intercal}{\boldsymbol E}_t^\dag{\boldsymbol C}}{s_R^{\alpha_R/2}s_C^{\alpha_C/2}}\right\|_F^4\leq c_4\sum_{t=1}^T t^2 = O\left(T^3\right),
$$
indicating that
$$
\max_{1\leq t\leq T}\left\|\frac{{\boldsymbol R}^{^\intercal}{\boldsymbol E}_t^\dag{\boldsymbol C}}{s_R^{\alpha_R/2}s_C^{\alpha_C/2}}\right\|_F = O_P\left(T^{3/4}\right).
$$
The other uniform convergence results can be similarly proved. \hfill$\blacksquare$

\medskip

\begin{lemma}\label{le:B.12}

Suppose that Assumptions \ref{ass:3.1}(i), \ref{ass:3.2}(i)(ii), \ref{ass:4.1} and \ref{ass:4.2}(i) are satisfied. Then we have
\begin{eqnarray}
&&{\sf E}\left\Vert \sum_{t=1}^{T}{\boldsymbol F}_t{\boldsymbol C}^{^\intercal}{\boldsymbol E}_{t}^{^\intercal}\right\Vert_F^2=O_P\left(p_1p_2^{\alpha_{C,1}}T^2\right),\label{eqB.32}\\
&&{\sf E}\left\Vert \sum_{t=1}^{T}{\boldsymbol E}_t{\boldsymbol C}{\boldsymbol F}_{t}^{^\intercal}\right\Vert_F^2=O_P\left(p_1p_2^{\alpha_{C,1}}T^2\right).\label{eqB.33}
\end{eqnarray}

\end{lemma}

\noindent{\bf Proof of Lemma \ref{le:B.12}}.\ \ The proof is similar to the proof of Lemma \ref{le:B.2}. We only prove (\ref{eqB.32}) as the proof of (\ref{eqB.33}) is analogous. Recall that
$$
{\sf E}\left\Vert \sum_{t=1}^{T}{\boldsymbol F}_t{\boldsymbol C}^{^\intercal}{\boldsymbol E}_{t}^{^\intercal}\right\Vert_F^2\leq2\sum_{i=1}^{p_1}{\sf E}\left\Vert \sum_{t=1}^{T}\sum_{j=1}^{p_2}e_{t,(i,j)}{\boldsymbol C}_{j\bullet}\overline{\boldsymbol F}_t^{^\intercal}\right\Vert^2+2\sum_{i=1}^{p_1}{\sf E}\left\Vert \sum_{t=1}^{T}\sum_{j=1}^{p_2}e_{t,(i,j)}{\boldsymbol C}_{j\bullet}\widetilde{\boldsymbol F}_t^{^\intercal}\right\Vert^2.
$$
By Assumption \ref{ass:4.2}(i), (\ref{eqB.7}) and the Cauchy-Schwarz inequality, 
\begin{eqnarray}
	\sum_{i=1}^{p_1}{\sf E}\left\Vert \sum_{t=1}^{T}\sum_{j=1}^{p_2}e_{t,(i,j)}{\boldsymbol C}_{j\bullet}\widetilde{\boldsymbol F}_t^{^\intercal}\right\Vert^2&\leq&\sum_{i=1}^{p_1}\sum_{1\leq t,s\leq T}\left({\sf E}\left\Vert \sum_{j=1}^{p_2}e_{t,(i,j)}{\boldsymbol C}_{j\bullet}\right\Vert^4\right)^{1/4}\left({\sf E}\left\Vert\widetilde{\boldsymbol F}_t\right\Vert^4\right)^{1/4}\notag\\
	&&\left({\sf E}\left\Vert \sum_{j=1}^{p_2}e_{s,(i,j)}{\boldsymbol C}_{j\bullet}\right\Vert^4\right)^{1/4}\left({\sf E}\left\Vert\widetilde{\boldsymbol F}_s\right\Vert^4\right)^{1/4}\notag\\
	&=&O\left(p_1p_2^{\alpha_{C,1}}T^2\right).\label{eqB.34}
\end{eqnarray}

On the other hand, letting $G_{t,i}=\sum_{s=t}^T\sum_{j=1}^{p_2}e_{s,(i,j)}{\boldsymbol C}_{j\bullet}$, we re-write 
\[
\sum_{t=1}^{T}\sum_{j=1}^{p_2}e_{t,(i,j)}{\boldsymbol C}_{j\bullet}\overline{\boldsymbol F}_t^{^\intercal}=\sum_{t=1}^{T}G_{t,i}\left[{\sf vec}^{-1}(\overline{\boldsymbol A}{\boldsymbol \eta}_t)\right]^{^\intercal},
\]
which is a sum of martingale differences since $\{{\boldsymbol \eta}_t\}$ is a sequence of i.i.d. random vectors independent of $\{G_{t,i}\}$ for any $i$. Then, by Assumption \ref{ass:4.2}(i), we readily have that 
\begin{equation}\label{eqB.35}
\sum_{i=1}^{p_1}{\sf E}\left\Vert \sum_{t=1}^{T}\sum_{j=1}^{p_2}e_{t,(i,j)}{\boldsymbol C}_{j\bullet}\overline{\boldsymbol F}_t^{^\intercal}\right\Vert^2=O\left(p_1p_2^{\alpha_{C,1}}T^2\right).
\end{equation}
Using (\ref{eqB.34}) and (\ref{eqB.35}), we complete the proof of (\ref{eqB.32}).\hfill$\blacksquare$

\medskip

\begin{lemma}\label{le:B.13}
	
Suppose that Assumptions \ref{ass:3.1}(i) and \ref{ass:4.1} are satisfied. Then we have
\begin{eqnarray}
&&\left\Vert \frac{1}{p_2^{\alpha_{C,1}}T^2}\sum_{t=1}^{T}{\boldsymbol F}_t{\boldsymbol C}^{^\intercal}{\boldsymbol C}{\boldsymbol F}_{t}^{^\intercal}-\int_0^1 {\boldsymbol W}(u){\boldsymbol\Sigma}_C^{\ast,1}{\boldsymbol W}(u)^{^\intercal}du\right\Vert=o_P(1),\label{eqB.36}\\
&&\left\Vert \frac{1}{p_1^{\alpha_{R,1}}T^2}\sum_{t=1}^{T}{\boldsymbol F}_t^{^\intercal}{\boldsymbol R}^{^\intercal}{\boldsymbol R}{\boldsymbol F}_{t}-\int_0^1 {\boldsymbol W}(u)^{^\intercal}{\boldsymbol\Sigma}_R^{\ast,1}{\boldsymbol W}(u)du\right\Vert=o_P(1).\label{eqB.37}
\end{eqnarray}
where ${\boldsymbol W}(\cdot)$ is an $r_1\times r_2$ matrix of Brownian motions with covariance of ${\sf vec}({\boldsymbol W}(\cdot))$ being $\overline{\boldsymbol A}{\boldsymbol\Sigma}_\eta \overline{\boldsymbol A}^{^\intercal}$, 
\[
{\boldsymbol\Sigma}_C^{\ast,1} = \lim_{p_2\to \infty}\frac{1}{p_2^{\alpha_{C,1}}}{\boldsymbol C}^{^\intercal}{\boldsymbol C}\quad \text{and}\quad {\boldsymbol\Sigma}_R^{\ast,1} = \lim_{p_2\to \infty}\frac{1}{p_1^{\alpha_{R,1}}}{\boldsymbol R}^{^\intercal}{\boldsymbol R}.
\]
	
\end{lemma}

\noindent{\bf Proof of Lemma \ref{le:B.13}}.\ \ The proof is similar to that of Lemma \ref{le:B.3}.\hfill$\blacksquare$

\medskip

Let $\nu_{R,k}^\ast$ be the $k$-th largest eigenvalue of ${\boldsymbol\Sigma}_R^{\ast^{1/2}}[\int_0^1 {\boldsymbol W}(u){\boldsymbol\Sigma}_C^{\ast,1}{\boldsymbol W}(u)^{^\intercal}du]{\boldsymbol\Sigma}_R^{\ast^{1/2}} \mathbf{B}_{R}^{2}$, $k=1,\ldots,r_1$, and $\nu_{C,k}^\ast$ the $k$-th largest eigenvalue of ${\boldsymbol\Sigma}_C^{\ast^{1/2}}[\int_0^1 {\boldsymbol W}(u)^{^\intercal}{\boldsymbol\Sigma}_R^{\ast,1}{\boldsymbol W}(u)du]{\boldsymbol\Sigma}_C^{\ast^{1/2}}\mathbf{B}_C^{2}$, $k=1,\ldots,r_2$.

\medskip 

\begin{lemma}\label{le:B.14}

Suppose that Assumptions \ref{ass:3.1}(i), \ref{ass:3.2}(i)(ii) and \ref{ass:4.1}--\ref{ass:4.3} are satisfied. Then we have
\begin{equation}\label{eqB.38}
\left\Vert \widehat{\boldsymbol V}_R^\ast -{\boldsymbol V}_R^\ast\right\Vert=o_P(1),\ \ \ \left\Vert \widehat{\boldsymbol V}_C^\ast-{\boldsymbol V}_C^\ast\right\Vert=o_P(1),
\end{equation}
where $\widehat{\boldsymbol V}_R^\ast=p_2^{-\alpha_{C,1}}\mathbf{B}_{R}^{-2}\widetilde{\boldsymbol V}_R$, $\widehat{\boldsymbol V}_C^\ast=p_1^{-\alpha_{R,1}}\mathbf{B}_{C}^{-2}\widetilde{\boldsymbol V}_C$, ${\boldsymbol V}_R^\ast=\mathbf{B}_R^{-2}{\sf diag}\{\nu_{R,1}^\ast,\ldots,\nu_{R,r_1}^\ast\}\succ0$ and ${\boldsymbol V}_C^\ast=\mathbf{B}_C^{-2}{\sf diag}\{\nu_{C,1}^\ast,\ldots,\nu_{C,r_2}^\ast\}\succ0$.
	
\end{lemma}

\noindent{\bf Proof of Lemma \ref{le:B.14}}.\ \ To save the space, we only prove the first assertion in (\ref{eqB.38}). It follows from (\ref{eqA.2}) that
\begin{eqnarray*}
\frac{1}{p_2^{\alpha_{C,1}}T}\widehat{\boldsymbol\Omega}_R&=&\frac{1}{p_2^{\alpha_{C,1}}T^2}\sum_{t=1}^{T}{\boldsymbol R}{\boldsymbol F}_{t}{\boldsymbol C}^{^\intercal}{\boldsymbol C}{\boldsymbol F}_{t}^{^\intercal}{\boldsymbol R}^{^\intercal}+\frac{1}{p_2^{\alpha_{C,1}}T}\left({\boldsymbol\Lambda}_{1}+{\boldsymbol\Lambda}_{2}+{\boldsymbol\Lambda}_{3}\right)\notag\\
&=:&{\boldsymbol\Pi}_{R,1}+{\boldsymbol\Pi}_{R,2}+{\boldsymbol\Pi}_{R,3}+{\boldsymbol\Pi}_{R,4}.\notag
\end{eqnarray*}
Using Lemmas \ref{le:B.1} and \ref{le:B.12}, we have
\[
\Vert {\boldsymbol\Pi}_{R,2}+{\boldsymbol\Pi}_{R,3}+{\boldsymbol\Pi}_{R,4}\Vert=O_P\left(p_1p_2^{1/2-\alpha_{C,1}}T^{-3/2} + p_1^{1/2}p_2^{1-\alpha_{C,1}}T^{-1} + p_1^{(\alpha_{R,1}+1)/2}p_2^{-\alpha_{C,1}/2}T^{-1}\right),
\]
which, together with Weyl's inequality, leads to
\begin{equation}\label{eqB.39}
\max_{1\leq k\leq r_1}\left\vert \widetilde\nu_{R,k}^\ast-\widetilde\nu_{R,k}^\dag\right\vert=O_P\left(p_1p_2^{1/2-\alpha_{C,1}}T^{-3/2} + p_1^{1/2}p_2^{1-\alpha_{C,1}}T^{-1} + p_1^{(\alpha_{R,1}+1)/2}p_2^{-\alpha_{C,1}/2}T^{-1}\right),
\end{equation}
where $\widetilde\nu_{R,k}^\ast$ and $\widetilde\nu_{R,k}^\dag$ denote the $k$-th largest eigenvalues of $\frac{1}{p_2^{\alpha_{C,1}}T}\widehat{\boldsymbol\Omega}_R$ and ${\boldsymbol\Pi}_{R,1}$, respectively. For a positive definite matrix $\mathbf{A}$ and a positive definite diagonal matrix $\mathbf{B}$, we have the following eigenvalue bounds: 
$$
\psi_{\mathrm{min}}(\mathbf{A})b_{\mathrm{min}} \leq \psi_{i}(\mathbf{AB})\leq \psi_{\mathrm{max}}(\mathbf{A})b_{\mathrm{max}},
$$
where $b_{\mathrm{min}}$ and $b_{\mathrm{max}}$ are the smallest and largest diagonal entries of $\mathbf{B}$, respectively. Note that
$$
{\boldsymbol\Pi}_{R,1}=\mathbf{B}_R\left(\frac{1}{p_2^{\alpha_{C,1}}T^2}\sum_{t=1}^{T}{\boldsymbol F}_{t}{\boldsymbol C}^{^\intercal}{\boldsymbol C}{\boldsymbol F}_{t}^{^\intercal}\right)\mathbf{B}_R\left(\mathbf{B}_R^{-1}{\boldsymbol R}^{^\intercal}{\boldsymbol R}\mathbf{B}_R^{-1}\right).
$$
By Assumption \ref{ass:4.1}(i) and (\ref{eqB.36}) in Lemma \ref{le:B.13}, the diagonal matrix containing the first $r_1$ eigenvalues of ${\boldsymbol\Pi}_{R,1}$ is of order $\mathbf{B}_R^2$, and
\begin{equation}\label{eqB.40}
 \max_{1\leq k\leq r_1}p_1^{-\alpha_{R,k}}\left\vert \widetilde\nu_{R,k}^\dag- \nu_{R,k}^\ast\right\vert=o_P(1).
\end{equation}
With (\ref{eqB.39}), (\ref{eqB.40}) and the definition of $\widehat{\boldsymbol V}_R^\ast$, noting that
$$
O_P\left(\|\mathbf{B}_R^{-2}\|_F\right)O_P\left(p_1p_2^{1/2-\alpha_{C,1}}T^{-3/2} + p_1^{1/2}p_2^{1-\alpha_{C,1}}T^{-1} + p_1^{(\alpha_{R,1}+1)/2}p_2^{-\alpha_{C,1}/2}T^{-1}\right) = o_P(1).
$$ 
by Assumption \ref{ass:4.3}, we complete the proof of the first assertion in (\ref{eqB.38}).\hfill$\blacksquare$ 

\medskip

Recall that
\begin{eqnarray*}
\widehat{\boldsymbol H}_R^\ast&=&\left(\frac{1}{p_2^{\alpha_{C,1}}T^2}\sum_{t=1}^{T}{\boldsymbol F}_{t}{\boldsymbol C}^{^\intercal}{\boldsymbol C}{\boldsymbol F}_{t}^{^\intercal}\right)\left({\boldsymbol R}^{^\intercal}\widehat{\boldsymbol R}\mathbf{B}_R^{-1}\right)\left(\widehat{\boldsymbol V}_R^\ast\right)^{-1},\notag\\
\widehat{\boldsymbol H}_C^\ast&=&\left(\frac{1}{p_1^{\alpha_{R,1}}T^2}\sum_{t=1}^{T}{\boldsymbol F}_{t}^{^\intercal}{\boldsymbol R}^{^\intercal}{\boldsymbol R}{\boldsymbol F}_{t}\right)\left({\boldsymbol C}^{^\intercal}\widehat{\boldsymbol C}\mathbf{B}_C^{-1}\right)\left(\widehat{\boldsymbol V}_C^\ast\right)^{-1}.
\end{eqnarray*}

\begin{lemma}\label{le:B.15}

Suppose that Assumptions \ref{ass:3.1}(i), \ref{ass:3.2}(i)(ii) and \ref{ass:4.1}--\ref{ass:4.3} are satisfied. The rotation matrices $\widehat{\boldsymbol H}_R^\ast$ and $\widehat{\boldsymbol H}_C^\ast$ are invertible w.p.a.1. 	

\end{lemma}

\noindent{\bf Proof of Lemma \ref{le:B.15}}.\ \ Let $\widehat{\boldsymbol\Sigma}_{R}^\ast={\boldsymbol R}^{^\intercal}\widehat{\boldsymbol R}\mathbf{B}_R^{-1}$. Note that 
\[
\frac{1}{p_2^{\alpha_{C,1}}T}\mathbf{B}_R^{-1}\widehat{\mathbf{R}}^{^\intercal}\widehat{\boldsymbol{\Omega}}_R\widehat{\mathbf{R}}\mathbf{B}_R^{-1} = \widehat{\boldsymbol V}_R^\ast.
\]
 Following the proof of Lemma \ref{le:B.14} and using $\|\widehat{\boldsymbol R} \|_F = O(1)$, we have
\begin{equation}\label{eqB.41}
\left(\mathbf{B}_R^{-1}\widehat{\boldsymbol R}^{^\intercal}{\boldsymbol R}\right)\left(\frac{1}{p_2^{\alpha_{C,1}}T^2}\sum_{t=1}^{T}{\boldsymbol F}_{t}{\boldsymbol C}^{^\intercal}{\boldsymbol C}{\boldsymbol F}_{t}^{^\intercal}\right)\left({\boldsymbol R}^{^\intercal}\widehat{\boldsymbol R}\mathbf{B}_R^{-1}\right)+o_P(1)=\widehat{\boldsymbol V}_R^\ast \stackrel{P}\to {\boldsymbol V}_R^\ast \succ 0.
\end{equation}
Taking limit on both sides of (\ref{eqB.41}) and using Lemmas \ref{le:B.13} and \ref{le:B.14}, we have
\[
\left(\lim_{p_1\rightarrow\infty}\widehat{\boldsymbol\Sigma}_{R}^\ast\right)^{^\intercal}{\boldsymbol\Delta}_{C}^\ast\left(\lim_{p_1\rightarrow\infty}\widehat{\boldsymbol\Sigma}_{R}^\ast\right)={\boldsymbol V}_R^\ast,
\]
where ${\boldsymbol\Delta}_{C}^\ast=\int_0^1 {\boldsymbol W}(u){\boldsymbol\Sigma}_C^{\ast,1}{\boldsymbol W}(u)^{^\intercal}du$. This indicates that $\widehat{\boldsymbol\Sigma}_{R}^\ast$ is invertible {\em w.p.a.1}. From the definition of $\widehat{\boldsymbol H}_R^\ast$ and Lemma \ref{le:B.14}, we may show that $\widehat{\boldsymbol H}_R^\ast$ is invertible {\em w.p.a.1}. Similarly, we can also prove that $\widehat{\boldsymbol H}_C^\ast$ is invertible {\em w.p.a.1}. \hfill$\blacksquare$

\medskip

\begin{lemma}\label{le:B.16}
	
Suppose that Assumptions \ref{ass:3.2}(i) and \ref{ass:4.2} are satisfied. The following uniform convergence holds:
\begin{equation}
\max_{1\leq t\leq T}\left\{\left\Vert \mathbf{B}_R^{-1} {\boldsymbol R}^{^\intercal}{\boldsymbol E}_t{\boldsymbol C}\mathbf{B}_C^{-1}\right\Vert_F+\left\Vert \frac{{\boldsymbol E}_t}{\sqrt{p_1p_2}}\right\Vert_F+\left\Vert p_1^{-1/2}{\boldsymbol E}_t{\boldsymbol C}\mathbf{B}_C^{-1}\right\Vert_F+ \left\Vert p_2^{-1/2}\mathbf{B}_R^{-1}{\boldsymbol R}^{^\intercal}{\boldsymbol E}_t\right\Vert_F\right\}=O_P\left(T^{1/4}\right).\label{eqB.42}
\end{equation}
	
\end{lemma}

\noindent{\bf Proof of Lemma \ref{le:B.16}}.\ \ This proof is similar to that of Lemma \ref{le:B.6}.\hfill$\blacksquare$

\medskip

\begin{lemma}\label{le:B.17}

Suppose that Assumptions \ref{ass:3.5}(i)(ii), \ref{ass:3.6}, \ref{ass:4.1}(i) and \ref{ass:4.4} are satisfied. Then we have
\begin{equation}\label{eqB.43}
{\sf E}\left\|\sum_{t=1}^{T} (\Delta{\boldsymbol F}_t) {\boldsymbol C}^{^\intercal} (\Delta{\boldsymbol E}_{t})^{^\intercal}\right\|_F^2 = O\left(p_1 p_2^{\alpha_{C,1}}T\right),
\end{equation}
and
\begin{equation}\label{eqB.44}
{\sf E}\left\|\sum_{t=1}^{T} (\Delta{\boldsymbol E}_t) {\boldsymbol C}^{^\intercal} (\Delta {\boldsymbol F}_{t})^{^\intercal}\right\|_F^2 = O\left(p_1 p_2^{\alpha_{C,1}}T\right).
\end{equation}

\end{lemma}

\noindent{\bf Proof of Lemma \ref{le:B.17}}.\ \ This proof is similar to that of Lemmas \ref{le:B.8} and \ref{le:B.12}.\hfill$\blacksquare$

\medskip

Let $\nu_{R,k}^\ddag$ be the $k$-th largest eigenvalue of  ${\boldsymbol\Sigma}_R^{\ast^{1/2}}{\sf E}\left[ \Delta {\boldsymbol F}_t{\boldsymbol\Sigma}_C^{\ast,1} \Delta {\boldsymbol F}_t^{^\intercal}\right]{\boldsymbol\Sigma}_R^{\ast^{1/2}} \mathbf{B}_R^2$, $k=1,\ldots,r_1$, and $\nu_{C,k}^\ddag$ the $k$-th largest eigenvalue of ${\boldsymbol\Sigma}_C^{\ast^{1/2}} {\sf E}\left[ \Delta {\boldsymbol F}^{^\intercal}_t{\boldsymbol\Sigma}_R^{\ast,1} \Delta {\boldsymbol F}_t\right] {\boldsymbol\Sigma}_C^{\ast^{1/2}}\mathbf{B}_C^2$, $k=1,\ldots,r_2$. 

\medskip

\begin{lemma}\label{le:B.18}

Suppose that Assumptions \ref{ass:3.5}(i)(ii), \ref{ass:3.6}, \ref{ass:4.1}(i) and \ref{ass:4.4} are satisfied. Then we have
\begin{equation}\label{eqB.45}
\left\Vert \check{\boldsymbol V}_R^\ast-{\boldsymbol V}_R^\ddag\right\Vert=o_P(1),\ \ \ \left\Vert \check{\boldsymbol V}_C^\ast-{\boldsymbol V}_C^\ddag\right\Vert=o_P(1),
\end{equation}
where $\check{\boldsymbol V}_R^\ast$ and $\check{\boldsymbol V}_C^\ast$ are defined in Section \ref{sec4.2}, 
\[
{\boldsymbol V}_R^\ddag=\mathbf{B}_R^{-2}{\sf diag}\{\nu_{R,1}^\ddag,\ldots,\nu_{R,r_1}^\ddag\}\succ0\quad \text{and}\quad {\boldsymbol V}_C^\ddag=\mathbf{B}_C^{-2}{\sf diag}\{\nu_{C,1}^\ddag,\ldots,\nu_{C,r_2}^\ddag\}\succ0.
\]
	
\end{lemma}

\noindent{\bf Proof of Lemma \ref{le:B.18}}.\ \ As in the proofs of Lemmas \ref{le:B.4} and \ref{le:B.14}, we have
\begin{eqnarray*}
\frac{1}{p_2^{\alpha_{C,1}}}\overline{\boldsymbol\Omega}_R=\frac{1}{p_2^{\alpha_{C,1}}T}\sum_{t=1}^{T}{\boldsymbol R}(\Delta{\boldsymbol F}_{t}){\boldsymbol C}^{^\intercal}{\boldsymbol C}(\Delta{\boldsymbol F}_{t})^{^\intercal}{\boldsymbol R}^{^\intercal}+O_P\left(p_1^{1/2}p_2^{1-\alpha_{C,1}} + p_1p_2^{1/2-\alpha_{C,1}} T^{-1/2}\right).
\end{eqnarray*}
As $p_1^{1/2-\alpha_{R,r_1}}p_2^{1-\alpha_{C,1}} + p_1^{1-\alpha_{R,r_1}}p_2^{1/2-\alpha_{C,1}} T^{-1/2} \to 0$ by Assumption \ref{ass:4.4}(i), the remaining proof follows the arguments in the proof of Lemma \ref{le:B.14}.\hfill$\blacksquare$

\medskip

\begin{lemma}\label{le:B.19}

Suppose that Assumptions \ref{ass:3.5}(i)(ii), \ref{ass:3.6} and \ref{ass:4.4} are satisfied. The rotation matrices $\overline{\boldsymbol H}_R^\ast$ and $\overline{\boldsymbol H}_C^\ast$ defined in Section \ref{sec4.2} are invertible w.p.a.1. 	

\end{lemma}

\noindent{\bf Proof of Lemma \ref{le:B.19}}.\ \ This proof is similar to that of Lemma \ref{le:B.5}.\hfill$\blacksquare$

\medskip

\begin{lemma}\label{le:B.20}

Suppose that Assumption \ref{ass:3.6}(ii) and \ref{ass:4.4} are satisfied. The following uniform convergence holds:
\begin{equation}\label{eqB.47}
\max_{1\leq t\leq T}\left\{\left\|\mathbf{S}_{R}^{-1}{\boldsymbol R}^{^\intercal}{\boldsymbol E}_t^\dag{\boldsymbol C}\mathbf{S}_{C}^{-1}\right\|_F + \left\|\frac{{\boldsymbol E}_t^\dag}{s_R^{1/2}s_C^{1/2}}\right\|_F + \left\|s_R^{-1/2}{\boldsymbol E}_t^\dag{\boldsymbol C}\mathbf{S}_{C}^{-1}\right\|_F +\left\|\mathbf{S}_{R}^{-1}{\boldsymbol R}^{^\intercal}{\boldsymbol E}_t^\dag s_C^{-1/2}\right\|_F \right\} = O_P\left(T^{3/4}\right).
\end{equation}
\end{lemma}

\noindent{\bf Proof of Lemma \ref{le:B.20}}.\ \ This proof is similar to that of Lemmas \ref{le:B.6} and \ref{le:B.11}.\hfill$\blacksquare$

\end{document}